\documentclass[format=manuscript,nonacm=true]{acmart}

\pdfoutput=1

\newcommand{\bra}{\ensuremath{\langle}}
\newcommand{\ket}{\ensuremath{\rangle}}
\newcommand{\ty}{\ensuremath{\mathrm{t}}}
\newcommand{\ho}{\ensuremath{\mathrm{h}}}
\newcommand{\se}{\ensuremath{\mathrm{s}}}
\newcommand{\pa}{\ensuremath{\mathrm{p}}}
\newcommand{\nn}{\ensuremath{\mathrm{i}}}
\newcommand{\lo}{\ensuremath{\mathrm{local}}}

\usepackage{todonotes}
\usepackage{mathtools}
\usepackage{algorithm}
\usepackage{algpseudocode}

\algblockdefx{ParallelFor}{EndParallelFor}[1]{\textbf{parallel for}~#1~\textbf{do}}{\textbf{end for}}

\begin{document}
\title{Engineering Boolean Matrix Multiplication for Multiple-Accelerator Shared-Memory Architectures}

\author{Matti Karppa}
\affiliation{
  \institution{Aalto University}
  \country{Finland}
}
\email{firstname.lastname@aalto.fi}

\author{Petteri Kaski}
\affiliation{
  \institution{Aalto University}
  \country{Finland}
}
\email{firstname.lastname@aalto.fi}
\authorsaddresses{}

\begin{abstract}
We study the problem of multiplying two bit matrices with entries
either over the Boolean algebra $(0,1,\vee,\wedge)$ or over the binary
field $(0,1,+,\cdot)$. We engineer high-performance open-source
algorithm implementations for contemporary multiple-accelerator
shared-memory architectures, with the objective of time-and-energy-efficient 
scaling up to input sizes close to the available shared memory capacity. 
For example, given two terabinary-bit square matrices as input, our 
implementations compute the Boolean product in 
approximately 2100 seconds (1.0~Pbop/s at 3.3~pJ/bop
for a total of 2.1~kWh/product) and the binary product in less than
950 seconds (2.4~effective Pbop/s at 1.5~effective pJ/bop for a total
of 0.92~kWh/product) on an NVIDIA DGX-1 with power consumption 
at peak system power (3.5~kW).\\\hspace*{\parindent}
Our contributions are (a) for the binary product, we use
alternative-basis techniques of Karstadt and Schwartz~[\emph{SP\hspace*{-0.3mm}A\hspace*{-0.07mm}A}\,'17] 
to design novel alternative-basis variants of Strassen's recurrence 
for $2\times 2$ block multiplication~[\emph{Numer.~Math.}~13~(1969)] that 
have been optimized for both the number of additions and low working memory, 
(b) structuring the parallel block recurrences and the memory
layout for coalescent and register-localized execution on accelerator
hardware, (c) low-level engineering of the innermost block products
for the specific target hardware, and (d) structuring the top-level
shared-memory implementation to feed the accelerators with data and
integrate the results for input and output sizes beyond the aggregate
memory capacity of the available accelerators.
\end{abstract}






\maketitle

%%%%%%%%%%%%%%%%%%%%%%%%%%%%%%%%%%%%%%%%%%%%%%%%%%%%%%%%%%%%% Document body %%%

\section{Introduction} 

Matrix multiplication is one of the most widely deployed primitives 
in computing. In a combinatorial context, one often encounters the task
of multiplying {\em Boolean} matrices with entries in $\{0,1\}$, and with  
the arithmetic taking place either over the Boolean algebra $(0,1,\vee,\wedge)$
or over the binary field $(0,1,+,\cdot)$. Fast algorithms for the Boolean 
product and the binary product of two bit matrices underlie the fastest known 
algorithms for a wide range of tasks, such as transitive closure, 
context-free parsing, and triangle detection~%
\cite{FischerM:1971,Furman:1970,ItaiR:1978,Valiant:1975,VassilevskaWilliamsW:2010}.

Our interest in this paper is to engineer high-performance open-source 
implementations of Boolean and binary matrix multiplication for shared-memory 
platforms connected to multiple vector-parallel accelerator devices. Such 
platforms are the state of the art in terms of delivering both 
\begin{itemize}
\item[(i)]
speed through extensive parallelism available in the accelerator devices, and 
\item[(ii)]
energy-efficiency for moderately large data through shared memory.%
\footnote{%
For example,
a representative configuration of this type is the NVIDIA DGX-1, with 
eight Tesla V100 SXM2 accelerator devices 
(in total 40960 32-bit cores running at 1530-MHz boost clock), 
512 GiB of shared DDR4-2133 memory,
and 3.5-kW peak power consumption. This configuration can 
execute, at peak, $40960\cdot 1530\cdot 10^6 \cdot 32 = 2.01\cdot 10^{15}$ bit
operations per second ($2.01$ Pbop/s), which at peak power consumption 
translates to $1.8\cdot 10^{-12}$ Joules of energy consumed per bit operation 
($1.8$ pJ/bop).
}{}
\end{itemize}
Configurations of this type are
not only powerful individual systems; indeed, shared-memory platforms with 
multiple accelerator devices are individual compute nodes in larger
distributed-memory systems ranging from a few compute nodes to current 
leading-edge supercomputers.%
\footnote{%
For example, the Summit supercomputer at Oak Ridge National Laboratory---the 
first-ranked supercomputer on the June 2019 Top500-list 
(\url{https://top500.org})---consists of 4608 compute nodes, 
each of which has 512 GiB of shared DDR4 memory and six Tesla V100 SXM2 
accelerator devices.}{} Thus, scalability to distributed-memory systems starts
from carefully engineered shared-memory-accelerated implementations.

To our knowledge, the present paper constitutes the first 
study of Boolean matrix multiplication that seeks to push the envelope towards 
peak hardware performance on shared-memory-accelerated platforms
simultaneously in terms of speed, energy consumption, and input sizes of up 
to one terabinary-bit 
($1048576\times 1048576 = 2^{20}\times 2^{20}=2^{40}$ bits = 1 Tib = 128 GiB) 
per operand matrix.%
\footnote{%
Here we note
that {\em numerical} rather than Boolean matrix multiplication has been 
extensively studied from an implementation engineering perspective both in 
the setting of shared memory and in the setting of distributed memory---we 
postpone a discussion of related work to \S\ref{sect:earlier}.}{}
For example, given two terabinary-bit square matrices as input, our 
present open-source implementation computes the Boolean product in 
approximately 2100 seconds (1.0~Pbop/s at 3.3~pJ/bop
for a total of 2.1~kWh/product) and the binary product in less than
950 seconds (2.4~effective Pbop/s at 1.5~effective pJ/bop for a total
of 0.92~kWh/product) on an NVIDIA DGX-1 with power consumption 
at peak system power (3.5~kW).

\subsection{Engineering Challenges}

Before outlining our specific contributions in more detail, let us introduce 
some of the challenges when engineering for peak performance 
on inputs that nearly occupy the available capacity of shared memory and 
exceed the total memory available on the accelerators in a platform. 

First, the mathematical framework for algorithm design needs fine-grained 
optimization beyond coarse tools offered by traditional asymptotic analysis. 
Algorithms for fast matrix multiplication are a canonical example of this 
phenomenon, where the asymptotically fastest known algorithm 
designs~\cite{CoppersmithW:1990,DavieS:2013,VassilevskaWilliams:2012,LeGall:2014} hide in their asymptotic analysis very large constants, which make these 
designs infeasible to use in practice; we refer to the recent survey by 
Pan~\cite{Pan:2018} for an extensive discussion. For bit matrices, a natural 
first measure for the concrete efficiency of an algorithm design is the 
number of bit operations executed. For example, the elementary algorithm 
to multiply two $n\times n$ bit matrices uses $2n^3-n^2$ bit operations, and 
the Strassen--Winograd algorithm~\cite{Strassen:1969,Winograd:1971} for the 
binary field uses $6n^{\log_2 7}-5n^2$ bit operations when $n$ is a power 
of two. Already here one observes a leading constant that is three times 
the leading constant for the elementary algorithm, which necessitates
$n\geq 512$ before the Strassen--Winograd algorithm uses fewer bit operations
than the elementary algorithm. The Strassen--Winograd algorithm is known 
to be an optimal implementation of matrix multiplication relying on 
recursive $2\times 2$ multiplications~\cite{Bshouty:1995,Probert:1976}, 
assuming one works in the standard basis. Recently, Karstadt and 
Schwartz~\cite{KarstadtS:2017} introduce an alternative-basis framework 
that reduces the number of bit operations of Strassen's algorithm to 
$5n^{\log_2 7}-4n^2$, assuming the input and output are represented in
an alternative basis. (The transformation of an $n\times n$ bit matrix 
between the standard basis and the alternative basis uses $n^2\log_2 n$ 
bit operations.) 

Second, accelerator hardware is designed to give peak performance when 
executing an identical instruction on each entry of a fixed-length vector 
of words. Thus, all recurrences of bit operations must be designed to 
support vectorization and coalesced memory accesses across words of bits. 
Due to pipelining and long latencies in accelerator hardware
(cf.~Mei and Chu~\cite{MeiC:2017} and Volkov~\cite{Volkov:2016}), 
recurrences should make use of the available lowest-latency memory 
(registers) and expose sufficient parallelism to enable a 
large number of threads simultaneously in execution compared with the number 
of cores for effective latency hiding. Furthermore, the bottom layer of 
recursion needs to be engineered for compatibility with the available 
low-level instruction set, such as vector-shuffles and custom bit operations 
given by a truth table. While the word- and vector-level hardware is at
the moment rapidly evolving towards integrating increasingly domain-specific 
hardware units---for example, units that that perform small-size numerical 
matrix multiplication---the 
high-level architecture with vectorization, pipelining, and long latencies 
is likely to remain more stable over time. In particular, this requires 
designs and engineering for the on-accelerator memory hierarchy that 
saturates the bandwidth of the dedicated hardware units.

Third, assuming each accelerator can be engineered to perform close to the
peak available bandwidth in terms of bit operations per second, the host-level 
algorithm must be engineered to make effective use of the accelerators. 
This in particular means feeding the accelerators with input and 
aggregating their output in comparatively low-bandwidth shared memory 
across a low-bandwidth interconnect between the accelerators and the host. 
Our objective of engineering for inputs that are near the 
capacity of shared memory presents a further challenge in that 
at the host level we must use designs with a low memory footprint. We expect
also these traits to withstand the test of time and thus warrant the present
engineering effort.

\subsection{Our Contributions}

Let us now proceed to our specific contributions to address the aforementioned
engineering challenges. 

\medskip
\noindent
{\em Fine-grained optimization of the base design.}
First, we proceed with a fine-grained optimization of Strassen's 
algorithm~\cite{Strassen:1969} using the alternative-basis framework of 
Karstadt and Schwartz~\cite{KarstadtS:2017} for the binary field. 
Essentially, we investigate all possible alternative bases over the binary 
field, and optimize 
(i) the number of Boolean operations for the alternative-basis multiplication, 
(ii) the weight distribution of linear combinations in the alternative basis, 
and 
(iii) the number of Boolean operations for changing between bases.
Further desirable properties include in-place-computability and 
self-invertibility of the basis changes, as well as the ability to make 
chains of matrix multiplications in the alternative basis. 

\begin{theorem}[Main, Self-Inverse and In-Place Basis Changes]
\label{thm:main}
There exists an alternative-basis bilinear algorithm that multiplies 
two $2\times 2$ input matrices over the binary field using 
\begin{itemize}
\item[$($\hspace*{-0.3mm}a$)$] $3$ additions to preprocess each input, 
\item[$($b$)$] $7$ noncommutative multiplications, and 
\item[$($c$)$] $6$ additions for postprocessing to obtain the output. 
\end{itemize}
Moreover, the weight distribution in the alternative bases is $1,1,1,1,2,2,2$. 
The basis changes between the standard basis and the alternative bases each
use\/ $2$ additions, are self-inverse, and can be computed in-place. 
\end{theorem}

Applied recursively in the alternative basis, Theorem~\ref{thm:main} gives 
binary matrix multiplication using $5n^{\log_2 7}-4n^2$ bit operations 
when $n$ is a power of two. Changing between the standard basis and the 
alternative bases takes $\frac{1}{2}n^2\log_2 n$ bit operations and can be 
done in-place in memory, which presents a minor improvement over the original 
Karstadt--Schwartz design, but comes at the cost of losing the 
chain-multiplication property, which is desirable in many applications. The
next theorem summarizes a new design that supports chain multiplication but
is slightly less efficient than Theorem~\ref{thm:main} in terms of its weight 
distributions, which results in slightly less efficiency when 
aggregating solutions of sub-instances in tight working memory at the host. 

\begin{theorem}[Main, Chain-Multiplication]
\label{thm:main-chaining}
There exists an alternative-basis bilinear algorithm that multiplies 
two $2\times 2$ input matrices over the binary field using 
\begin{itemize}
\item[$($\hspace*{-0.3mm}a$)$] $3$ additions to preprocess each input, 
\item[$($b$)$] $7$ noncommutative multiplications, and 
\item[$($c$)$] $6$ additions for postprocessing to obtain the output. 
\end{itemize}
Moreover, the weight distributions are $1,1,1,1,2,2,2$ and $1,1,1,1,2,3,3$ 
for taking linear combinations of the operands and the results of 
noncommutative multiplications, respectively. 
The basis changes between the standard basis and the alternative basis each
use\/ $2$ additions, can be computed in-place, and support chain-multiplication
in the alternative basis. 
\end{theorem}

The proofs of Theorem~\ref{thm:main} and Theorem~\ref{thm:main-chaining} are
presented in \S\ref{sect:alt-basis} together with a development of the
alternative-basis framework using Kronecker products and Yates's 
algorithm~\cite{Yates1937} that enables easy parallelization on vectorized 
accelerator hardware.

\medskip
\noindent
{\em Engineering for performance on accelerator hardware.}
Second, beyond optimizing the base design, we engineer an implementation
suitable for vectorized accelerator hardware with extensive 
bit-operation-and-memory bandwidth but
long latencies. Here the key engineering principle is to expose sufficient 
parallelism to saturate the compute cores with work and hide latency, but not 
to exceed the available on-accelerator memory. It is well known 
(e.g.~\cite{BallardDHLS:2012a,LipshitzBDS:2012})
that bilinear block-recursive algorithm designs for matrix multiplication 
enable a tradeoff between 
\begin{enumerate}
\item
parallel processing (executing the recursive block 
multiplications in parallel, or ``breadth-first'', using independent memory 
for each recursive case) and 
\item
serial processing (by processing the recursive 
multiplications serially one after another, or ``depth-first'', 
reusing memory). 
\end{enumerate}
We observe that this tradeoff also 
applies to alternative-basis algorithms, and illustate its application with 
a recursive design whose upper levels proceed serially through recursive 
calls to lower levels executed in parallel, relying on the recurrences 
underlying Theorems~\ref{thm:main} and~\ref{thm:main-chaining}.
The lower parallel levels consist of (i) parallel preprocessing for both
inputs, (ii) parallel low-level optimized $64\times 64$ 
bit-matrix-multiplication, and (iii) parallel postprocessing to recover 
the output. We fine-tune the performance of these levels for the target 
hardware by (a) merging consecutive levels so that the intermediate results 
are stored in per-thread registers and (b) working with the widest available 
per-thread load and store instructions for communicating between the
per-thread registers and the on-accelerator memory. 

For an $n\times n$ binary product for large enough $n$, these engineering
considerations together with Theorem~\ref{thm:main} and~\ref{thm:main-chaining}
enable us to obtain on a single Tesla V100 SXM2 accelerator an empirical 
effective bit-operation bandwidth that exceeds the theoretical boost-clock 
peak bandwidth for bit operations. Here by {\em effective} bandwidth we 
mean $\frac{2n^3-n^2}{T}$ bit operations per second, where $2n^3-n^2$ is the 
number of bit operations to compute an $n\times n$ binary product with the 
elementary algorithm, and $T$ is the measured wall-clock running time to 
compute an $n\times n$ binary product. We postpone a detailed review of 
empirical performance to \S\ref{sect:experiments}.

\medskip
\noindent
{\em Host-level implementation with a low memory footprint.}
Third, we engineer host-level subroutines that in parallel feed multiple 
accelerators with recursive subproblems and aggregate the results obtained 
to the host-level buffers. Here we follow a strategy of using multiple groups 
of threads on the host CPUs, where each group contains exactly one thread for 
each accelerator. Each group is responsible for a 
specific task in a pipeline of tasks, such as preparing subproblems from the 
host-level input, solving subproblems on accelerators, and integrating the 
results of subproblems into the host-level output. The threads coordinate 
their work through standard synchronization primitives such as mutexes and 
blocking. To maintain a low working-memory footprint in host memory, we 
use $q$-fold Kronecker products of the decomposition matrices 
underlying Theorems~\ref{thm:main} and~\ref{thm:main-chaining} to produce 
subproblems and aggregate the sub-results. This strategy in particular 
benefits from optimization of the weights of the decomposition matrices 
discussed above. 

For the binary product, this strategy suffices to obtain aggregate effective 
bandwidth that exceeds the aggregate peak boost-clock bandwidth of the 
accelerators on inputs whose size is beyond the aggregate memory capacity
of the accelerators. In particular, we consider it a very satisfactory 
engineering outcome as well as benchmark that we can deliver the binary 
product of two one-terabinary-bit square matrices in less than one thousand 
seconds and with less than one kilowatt-hour of energy consumption.

\medskip
\noindent
{\em Open source.}
We release our present experimental algorithm implementations as open source% 
\footnote{Cf.~\url{https://github.com/mkarppa/matmul}.}{}
to enable further engineering and to communicate precisely the lowest-level 
design decisions for current microarchitectures.

\subsection{Earlier and Related Work}
\label{sect:earlier}

From a theory perspective, the study of fast algorithms for Boolean matrix 
multiplication has proceeded along essentially two lines of study.
The first line of study either relies on the binary field directly or embeds 
the Boolean algebra to a ring, employing techniques for fast matrix 
multiplication over rings to obtain the result. 
Asymptotically the fastest known such algorithms run in time 
$O(n^{\omega+o(1)})$, where $2\leq\omega<2.3728639$ is the exponent of matrix 
multiplication 
\cite{CoppersmithW:1990,DavieS:2013,VassilevskaWilliams:2012,LeGall:2014}.

While such algorithms are the fastest known in terms of asymptotic efficiency, 
practical algorithms for fast matrix multiplication over rings rely on 
recursive tensor techniques using small base tensors 
(e.g.~\cite{BallardDHLS:2012a,BensonB:2015,CenkH:2017,DAlbertoBN:2011,GraysonG:1996,HuangRMG:2017,HuangSHG:2016,KarstadtS:2017,KumarHJS:1993,LipshitzBDS:2012,LuoD:1995}) 
or trilinear aggregation--cancellation techniques \cite{Pan:1984,Pan:1978} 
for a small number of independent matrix 
multiplications~(e.g.~\cite{Kaporin:1999,Kaporin:2004}). These practical
studies differ from our present work in that they consider numerical 
(floating-point) matrix multiplication for CPU-based shared- or 
distributed-memory systems, whereas we optimize for bit matrices and a 
multiple-accelerator shared-memory system. Furthermore, we rely on an
alternative-basis approach optimized for bit matrices, whereas most of
the earlier work---with the exception of Karstadt and 
Schwartz~\cite{KarstadtS:2017}---operates in the standard basis. 

A second line of study on Boolean matrix multiplication seeks combinatorial 
rather than algebraic techniques to obtain competitive 
algorithms~\cite{Arlazarov:1970,BansalW:2012,Chan:2015,VassilevskaWilliamsW:2010,Yu:2018}. Currently, the fastest such algorithm runs 
asymptotically in time $\hat O(n^3/\log^4 n)$~\cite{Yu:2018}. However,
we are not aware of engineering work to bring algorithms in this second 
line of study to the computing practice on contemporary parallel architectures.

\section{Alternative-Basis Matrix-Multiplication}
\label{sect:alt-basis}

This section develops the mathematical framework for alternative-basis
matrix multiplication over the binary field $(0,1,+,\cdot)$.
We start by recalling the Strassen--Winograd standard-basis design, and 
then proceed to introduce two novel alternative-basis designs for use with 
the binary field as well as establishing our main theorems 
(Theorem~\ref{thm:main} and Theorem~\ref{thm:main-chaining}). 
In essence, both new designs are variants of the design in 
Karstadt and Schwartz~\cite{KarstadtS:2017}, with some further optimization 
of the designs in particular as pertains to the additive complexity and 
in-place computability of the basis changes. The rest of the section develops 
the mathematical framework of fast alternative-basis matrix multiplication 
with no claim on originality, apart perhaps from an expositionary choice 
to work with Kronecker products and Yates's
algorithm~\cite{Yates1937} to highlight the symmetries and the easily
vector-parallelizable sum-product-sum-layered structure of the framework. 
As discussed in the introduction, we expect this framework to withstand the 
test of time with further designs and evolving computing hardware.

\subsection{The Strassen--Winograd Recurrences}

We start by recalling the classical Strassen--Winograd design that works
in the standard basis. It will be convenient to first state all algorithms 
as straight-line programs to highlight their additive complexity. 
Toward this end, suppose we are to multiply two matrices 
\begin{equation}
\label{eq:ab}
A=\left[\begin{array}{cc}A_{00}&A_{01}\\A_{10}&A_{11}\end{array}\right]
\quad\text{and}\quad
B=\left[\begin{array}{cc}B_{00}&B_{01}\\B_{10}&B_{11}\end{array}\right]
\end{equation}
with entries over the binary field $(0,1,+,\cdot)$. 

The Strassen-Winograd design is as follows. 
First, compute the linear combinations
\begin{equation}
\label{eq:sw-input}
  \begin{aligned}
    \!T_0 &\leftarrow A_{10} + A_{11} \, ,&\!\!\!
    T_1 &\leftarrow A_{01} \, ,&\!\!\!
    T_2 &\leftarrow A_{01} + A_{11} \, ,&\!\!
    T_3 &\leftarrow A_{10} + T_2 \, ,&\!\!
    T_4 &\leftarrow A_{00} + T_3 \, ,&\!\!\!
    T_5 &\leftarrow A_{10} \, ,&\!\!\!
    T_6 &\leftarrow A_{00} \, ,&\!\!\! \\
    \!S_0 &\leftarrow B_{10} + B_{11} \, ,&\!\!\!
    S_1 &\leftarrow B_{10} \, ,&\!\!\!
    S_2 &\leftarrow B_{01} + B_{11} \, ,&\!\!
    S_3 &\leftarrow B_{10} + S_2 \, ,&\!\!
    S_4 &\leftarrow B_{01} \, ,&\!\!\!
    S_5 &\leftarrow B_{00} + S_3\, ,&\!\!\!
    S_6 &\leftarrow B_{00} \, .&\!\!\!
  \end{aligned}
\end{equation}
Then, multiply the linear combinations to obtain the products
\begin{equation}
\label{eq:sw-mul}
    Q_0\leftarrow T_0S_0\,,\quad
    Q_1\leftarrow T_1S_1\,,\quad
    Q_2\leftarrow T_2S_2\,,\quad
    Q_3\leftarrow T_3S_3\,,\quad
    Q_4\leftarrow T_4S_4\,,\quad
    Q_5\leftarrow T_5S_5\,,\quad
    Q_6\leftarrow T_6S_6\,.\quad
\end{equation}
Finally, compute the linear combinations of products
\begin{equation}
\label{eq:sw-output}
\begin{aligned}
    U_0 &\leftarrow Q_1 + Q_3 \, ,&\quad 
    U_1 &\leftarrow Q_2 + U_0 \, ,&\quad 
    U_2 &\leftarrow Q_4 + U_0 \, ,&\quad&  &\\
    C_{00} &\leftarrow Q_1 + Q_6 \, ,&\quad
    C_{01} &\leftarrow Q_0 + U_2 \, ,&\quad
    C_{10} &\leftarrow Q_5 + U_1 \, ,&\quad
    C_{11} &\leftarrow Q_0 + U_1 \, .&
\end{aligned}
\end{equation}
In total, this straight-line program makes exactly 15 additions and 
7 multiplications. Furthermore, a direct calculation shows that 
the straight-line program correctly computes the product
\begin{equation}
\label{eq:c-equals-ab}
\begin{split}
C=\left[\begin{array}{cc}
C_{00}&C_{01}\\C_{10}&C_{11}\end{array}\right]
=\left[\begin{array}{cc}
A_{00}B_{00}+A_{01}B_{10}&
A_{00}B_{01}+A_{01}B_{11}\\
A_{10}B_{00}+A_{11}B_{10}&
A_{10}B_{01}+A_{11}B_{11}\\
\end{array}\right]=AB\,.
\end{split}
\end{equation}

\subsection{An Alternative-Basis Algorithm with Self-Inverse Basis Changes}

\label{sect:mb}

Let us now turn to alternative-basis designs. We continue to assume that the 
input is given in $2\times 2$ block form, as in~\eqref{eq:ab}. 
Our first new alternative-basis algorithm is as follows. 
First, change basis by computing 
\begin{equation}
  \label{eq:mb-fwdbasisrecurrence}
  \begin{aligned}
    \hat A_{00} &\leftarrow A_{00}\, ,& \quad
    \hat A_{01} &\leftarrow A_{01}\, ,& \quad
    \hat A_{10} &\leftarrow A_{10}\, ,& \quad
    \hat A_{11} &\leftarrow A_{01}+A_{10}+A_{11}\, ,&\\
    \hat B_{00} &\leftarrow B_{00}\, ,& \quad
    \hat B_{01} &\leftarrow B_{01}\, ,& \quad
    \hat B_{10} &\leftarrow B_{10}\, ,& \quad
    \hat B_{11} &\leftarrow B_{01}+B_{10}+B_{11}\, ,&\\
  \end{aligned}
\end{equation}
We observe that the same basis change is applied for both matrices
$A$ and $B$. Next, compute the linear combinations
\begin{equation}
  \label{eq:mb-tsrecurrence}
  \begin{aligned}
    \hat T_0 &\leftarrow \hat A_{00} \, ,&\!\!
    \hat T_1 &\leftarrow \hat A_{01} \, ,&\!\!
    \hat T_2 &\leftarrow \hat A_{10} \, ,&\!\!
    \hat T_3 &\leftarrow \hat A_{11} \, ,&\!\!
    \hat T_4 &\leftarrow \hat A_{00} + \hat A_{11} \, ,&\!\!
    \hat T_5 &\leftarrow \hat A_{01} + \hat A_{11} \, ,&\!\!
    \hat T_6 &\leftarrow \hat A_{10} + \hat A_{11} \, ,\\
    \hat S_0 &\leftarrow \hat B_{00} \, ,&\!\!
    \hat S_1 &\leftarrow \hat B_{10} \, ,&\!\!
    \hat S_2 &\leftarrow \hat B_{00} + \hat B_{11} \, ,&\!\!
    \hat S_3 &\leftarrow \hat B_{11} \, ,&\!\!
    \hat S_4 &\leftarrow \hat B_{01} \, ,&\!\!
    \hat S_5 &\leftarrow \hat B_{01} + \hat B_{11}\, ,&\!\!
    \hat S_6 &\leftarrow \hat B_{10} + \hat B_{11}\, .
  \end{aligned}
\end{equation}
Multiply the linear combinations to obtain the products
\begin{equation}
   \label{eq:mb-qmul}
    \hat Q_0\leftarrow \hat T_0\hat S_0\,,\quad
    \hat Q_1\leftarrow \hat T_1\hat S_1\,,\quad
    \hat Q_2\leftarrow \hat T_2\hat S_2\,,\quad
    \hat Q_3\leftarrow \hat T_3\hat S_3\,,\quad
    \hat Q_4\leftarrow \hat T_4\hat S_4\,,\quad
    \hat Q_5\leftarrow \hat T_5\hat S_5\,,\quad
    \hat Q_6\leftarrow \hat T_6\hat S_6\,.\quad
\end{equation}
Compute the linear combinations of products
\begin{equation}
  \label{eq:mb-qrecurrence}
    \hat C_{00} \leftarrow \hat Q_0 + \hat Q_1 \, ,\quad
    \hat C_{01} \leftarrow \hat Q_4 + \hat Q_6 \, ,\quad
    \hat C_{10} \leftarrow \hat Q_2 + \hat Q_5 \, ,\quad
    \hat C_{11} \leftarrow \hat Q_1 + \hat Q_3 + \hat Q_5 + \hat Q_6 \, .
\end{equation}
Finally, transform $\hat C$ back to the standard basis with
\begin{equation}
  \label{eq:mb-revbasisrecurrence}
    C_{00} \leftarrow \hat C_{00} \, , \qquad
    C_{01} \leftarrow \hat C_{01} + \hat C_{11} \, , \qquad
    C_{10} \leftarrow \hat C_{10} + \hat C_{11} \, , \qquad
    C_{11} \leftarrow \hat C_{11} \, . 
\end{equation}
A direct calculation shows that the product \eqref{eq:c-equals-ab} is correctly
evaluated. This establishes Theorem~\ref{thm:main} with the exception of the
claim on weights, which can be easily verified from \eqref{eq:mb-albega}
in what follows.

Compared with the design of Karstadt and Schwartz~\cite{KarstadtS:2017}, 
we observe that the basis transformations 
\eqref{eq:mb-fwdbasisrecurrence} and \eqref{eq:mb-revbasisrecurrence} 
over the binary field use only two additions per matrix, whereas their 
design uses three additions per matrix, but works over an arbitrary ring.
Furthermore, our transformations 
\eqref{eq:mb-fwdbasisrecurrence} and \eqref{eq:mb-revbasisrecurrence} 
admit straightforward in-place computation as well as 
in-place inversion---indeed, both \eqref{eq:mb-fwdbasisrecurrence} 
and \eqref{eq:mb-revbasisrecurrence} are easily verified 
to be self-inverses over the binary field.

\smallskip
\noindent
{\em Remark.}
A drawback of the design above is that it does not enable alternative-basis 
chaining of matrix multiplications in the sense that the 
transformations \eqref{eq:mb-fwdbasisrecurrence} and 
\eqref{eq:mb-revbasisrecurrence} are not inverses of {\em each other}, 
which would be advantageous in applications that seek chain-multiplication. 
Our next design removes this drawback but loses the appealing self-inverse 
property.

\subsection{An Alternative-Basis Algorithm with Chaining}

\label{sect:ab}

Our second algorithm retains the arithmetic advantage in basis changes 
over the Karstadt--Schwartz~\cite{KarstadtS:2017} design and works in 
an alternative basis for the matrix ring. For a matrix in the standard basis
\[
X=\left[\begin{array}{cc}X_{00}&X_{01}\\X_{10}&X_{11}\end{array}\right]\,,
\]
we transform to the alternative basis by computing
\begin{equation}
\label{eq:ab-fwdbasisrecurrence}
  \hat X_{00}\leftarrow X_{00}\,,\quad
  \hat X_{01}\leftarrow X_{01}\,,\quad
  \hat X_{11}\leftarrow X_{01}+X_{11}\,,\quad
  \hat X_{10}\leftarrow \hat X_{11}+X_{10}\,.\quad
\end{equation}
The inverse transform from the alternative basis to the standard basis 
is given by 
\begin{equation}
\label{eq:ab-revbasisrecurrence}
  X_{00}\leftarrow \hat X_{00}\,,\quad
  X_{01}\leftarrow \hat X_{01}\,,\quad
  X_{10}\leftarrow \hat X_{10}+\hat X_{11}\,,\quad
  X_{11}\leftarrow \hat X_{01}+\hat X_{11}\,.\quad
\end{equation}
The algorithm now proceeds as follows. Given two matrices $A$ and $B$ as 
input, we transform both to the alternative basis 
using \eqref{eq:ab-fwdbasisrecurrence} to obtain the matrices 
$\hat A$ and $\hat B$.
Then, compute the linear combinations
\begin{equation}
  \label{eq:ab-tsrecurrence}
  \begin{aligned}
    \hat T_0 &\leftarrow  \hat A_{00} \, ,&\!\!
    \hat T_1 &\leftarrow  \hat A_{01} \, ,&\!\!
    \hat T_2 &\leftarrow  \hat A_{10} \, ,&\!\!
    \hat T_3 &\leftarrow  \hat A_{11} \, ,&\!\!
    \hat T_4 &\leftarrow  \hat A_{00} + \hat A_{10} \, ,&\!\!
    \hat T_5 &\leftarrow  \hat A_{01} + \hat A_{10} \, ,&\!\! 
    \hat T_6 &\leftarrow  \hat A_{10} + \hat A_{11} \, ,\\
    \hat S_0 &\leftarrow  \hat B_{00} \, ,&\!\!
    \hat S_1 &\leftarrow  \hat B_{10} + \hat B_{11} \, ,&\!\!
    \hat S_2 &\leftarrow  \hat B_{10} \, ,&\!\!
    \hat S_3 &\leftarrow  \hat B_{11} \, ,&\!\!
    \hat S_4 &\leftarrow  \hat B_{01} \, ,&\!\!
    \hat S_5 &\leftarrow  \hat B_{01} + \hat B_{10}\, ,&\!\!
    \hat S_6 &\leftarrow  \hat B_{00} + \hat B_{10}\, .
  \end{aligned}
\end{equation}
Multiply the linear combinations to obtain the products
\begin{equation}
   \label{eq:ab-qmul}
    \hat Q_0\leftarrow \hat T_0\hat S_0\,,\quad
    \hat Q_1\leftarrow \hat T_1\hat S_1\,,\quad
    \hat Q_2\leftarrow \hat T_2\hat S_2\,,\quad
    \hat Q_3\leftarrow \hat T_3\hat S_3\,,\quad
    \hat Q_4\leftarrow \hat T_4\hat S_4\,,\quad
    \hat Q_5\leftarrow \hat T_5\hat S_5\,,\quad
    \hat Q_6\leftarrow \hat T_6\hat S_6\,.\quad
\end{equation}
Compute the linear combinations of products
\begin{equation}
  \label{eq:ab-qrecurrence}
    \hat R \leftarrow  \hat Q_1+\hat Q_2+\hat Q_4\,,\quad
    \hat C_{00} \leftarrow  \hat Q_0 + \hat Q_1\,,\quad
    \hat C_{01} \leftarrow  \hat R + \hat Q_5\,,\quad
    \hat C_{10} \leftarrow  \hat R + \hat Q_6\,,\quad
    \hat C_{11} \leftarrow  \hat Q_3 + \hat Q_4\,.
\end{equation}
Finally, transform the matrix $\hat C$ back to the standard basis 
using \eqref{eq:ab-revbasisrecurrence} to obtain the product matrix $C$.
A direct calculation shows that $C=AB$. Furthermore, since the 
transformations \eqref{eq:ab-fwdbasisrecurrence} and 
\eqref{eq:ab-revbasisrecurrence} are mutual inverses, multiplications
$(\hat A,\hat B)\mapsto \hat C$ in the alternative basis may be chained without 
transforming back to the standard basis in between multiplications;
this property will become immediate from our exposition in 
\S\ref{sect:fast-ab} and \S\ref{sect:new-matrix}.
This establishes Theorem~\ref{thm:main-chaining} with the exception of the
claim on weights, which can be easily verified from \eqref{eq:ab-albega}
in what follows.

\subsection{Arrays, Vectors, and Matrices}

This section develops preliminaries and notational conventions for working 
with binary matrices used throughout the rest of this paper.
The basic data structure we work with is an {\em array} of $m$ entries,
indexed by the set $[m]=\{0,1,\ldots,m-1\}$. 
A {\em tensor} is an array with an associated {\em shape} 
$m_1\times m_2\times\cdots\times m_d$ for nonnegative integers
$m_1,m_2,\ldots,m_d$ with $m=m_1m_2\cdots m_d$. In this case we say
that the tensor has $d$ {\em modes} and that the mode $\ell=1,2,\ldots,d$
has {\em length} $m_\ell$. A tensor with one mode is called a {\em vector}
and a tensor with two modes is called a {\em matrix}.

We index the entries of a tensor $T$ of shape 
$m_1\times m_2\times\cdots\times m_d$ using tuples
$(i_1,i_2,\ldots,i_d)\in[m_1]\times[m_2]\times\cdots\times[m_d]$,
with the convention that the tuple refers to the entry indexed by
\begin{equation}
\label{eq:major-linearization}
i=i_1m_2m_3\cdots m_d+i_2m_3m_4\cdots m_d+\ldots+i_{d-1}m_d+i_d
\end{equation}
in the underlying array of length $m$. In other words, in linearizing
the tuple $(i_1,i_2,\ldots,i_d)\in[m_1]\times[m_2]\times\cdots\times[m_d]$
to a linear index $i\in [m]$, the first index $i_1\in[m_1]$ is 
the most significant, the next index $i_2\in[m_2]$ is the next most 
significant, and so forth. We write $T_{(i_1,i_2,\ldots,i_d)}$ for the 
entry of $T$ indexed by $(i_1,i_2,\ldots,i_d)$. To lighten the notation,
we may also write simply $T_{i_1i_2\cdots i_d}$ assuming the indexing 
is immediate from the context. When working with modes of composite length
$m_\ell=m_{\ell,1}m_{\ell,2}\cdots m_{\ell,k}$ for nonnegative 
integers $m_{\ell,1},m_{\ell,2},\ldots,m_{\ell,k}$, for convenience
we often choose to index such a mode with tuples 
in $[m_{\ell,1}]\times[m_{\ell,2}]\times\cdots\times [m_{\ell,k}]$,
with the understanding that we follow the first-index-major convention 
\eqref{eq:major-linearization} to arrive at the linear index in $[m_\ell]$.

We use the following notation for subtensors of tensors indexed along the
most significicant modes. For a tensor $T$ of shape
$m_1\times m_2\times\cdots\times m_d$ and
a tuple $(i_1,i_2,\ldots,i_\ell)\in[m_1]\times [m_2]\times\cdots\times [m_\ell]$ 
of indices, we write $T_{i_1i_2\cdots i_\ell}$ for the
tensor of shape $m_{\ell+1}\times m_{\ell+2}\times\cdots\times m_d$
with entries 
$(T_{i_1i_2\cdots i_\ell})_{i_{\ell+1}i_{\ell+2}\cdots i_d}$
for all $(i_{\ell+1},i_{\ell+2},\ldots,i_d)\in[m_{\ell+1}]\times [m_{\ell+2}]\times\cdots\times [m_d]$. In particular, we find it convenient to use the same 
notation for entries and subarrays of a tensor $T$, as the structure of the 
indexing tuple will indicate whether an entry or a subarray is meant. 

In what follows let us assume that all arrays have their entries in 
the binary field $(0,1,+,\cdot)$ unless indicated otherwise. 
For an $s\times t$ matrix $\mu$, write write $\mu^\top$ for the 
$t\times s$ transpose of $\mu$ with entries defined for all
$i\in [s]$ and $j\in[t]$ by the rule $\mu^\top_{ij}=\mu_{ji}$.
For an $s\times t$ matrix $\mu$ and a $t\times u$ matrix $\nu$,
we write $\mu\nu$ for the $s\times u$ {\em product} matrix with entries
defined for all $i\in[s]$ and $k\in[u]$ 
by $(\mu\nu)_{ik}=\sum_{j\in [t]}\mu_{ij}\nu_{jk}$.
We write $I=I_n$ for the $n\times n$ identity matrix. 
For an $s\times t$ matrix $\mu$ and a $p\times q$ matrix $\nu$, the
{\em Kronecker product} $\mu\otimes\nu$ is the $sp\times tq$ matrix
with entries defined for all $i\in[s]$, $j\in[t]$, $k\in [p]$, and $\ell\in[q]$
by the rule $(\mu\otimes\nu)_{ikj\ell}=\mu_{ij}\nu_{k\ell}$.
For an $s\times t$ matrix $\mu$, a $t\times u$ matrix $\sigma$,
a $p\times q$ matrix $\nu$, and a $q\times r$ matrix $\tau$, let us
recall the {\em composition rule} for Kronecker products
\begin{equation}
\label{eq:composition-rule}
(\mu\sigma)\otimes(\nu\tau)=(\mu\otimes\nu)(\sigma\otimes\tau)\,.
\end{equation}
Let us also recall the {\em transposition rule} 
\begin{equation}
\label{eq:transposition-rule}
(\mu\otimes\nu)^\top=\mu^\top\otimes\nu^\top\,.
\end{equation}
For two tensors $\sigma$ and $\tau$ of identical shape, 
the {\em entrywise product} $\sigma\odot\tau$ has the same shape,
with entries defined by the rule $(\sigma\odot\tau)_i=\sigma_i\tau_i$
for all indices $i$.

\subsection{The Triple Product Property and Matrix Multiplication}

This section recalls the triple product of matrix multiplication and its 
closure under taking of Kronecker products.
For an $su\times r$ matrix $\zeta$, an $r\times st$ matrix $\xi$, 
and an $r\times tu$ matrix $\eta$, we say that the three-tuple 
$(\zeta|\xi,\eta)$ has the {\em triple product property} with parameters 
$\bra s,t,u\ket_r$ if for all $i,i'\in[s]$, $j,j'\in [t]$, and 
$k,k'\in [u]$ it holds that
\begin{equation}
\label{eq:triple-prod}
\sum_{h\in[r]}
\zeta_{i'k'h}
\xi_{hij}
\eta_{hj'k}
=\begin{cases}
1 & \text{if $i=i'$, $j=j'$, and $k=k'$\,,}\\
0 & \text{otherwise}\,.
\end{cases}
\end{equation}
For example, the Strassen--Winograd design 
\eqref{eq:sw-input}, \eqref{eq:sw-mul}, \eqref{eq:sw-output} 
gives rise to the matrices 
\begin{equation}
\label{eq:sw-xietazeta}
\xi=
\begin{bmatrix}
    0 & 0 & 1 & 1 \\
    0 & 1 & 0 & 0 \\
    0 & 1 & 0 & 1 \\
    0 & 1 & 1 & 1 \\
    1 & 1 & 1 & 1 \\
    0 & 0 & 1 & 0 \\
    1 & 0 & 0 & 0 
\end{bmatrix}\,,\quad
\eta=
\begin{bmatrix}
    0 & 0 & 1 & 1 \\
    0 & 0 & 1 & 0 \\
    0 & 1 & 0 & 1 \\
    0 & 1 & 1 & 1 \\
    0 & 1 & 0 & 0 \\
    1 & 1 & 1 & 1 \\
    1 & 0 & 0 & 0 
\end{bmatrix}\,,\quad\text{and}\quad
\zeta=
\begin{bmatrix}
     0 & 1 & 0 & 0 & 0 & 0 & 1 \\
     1 & 1 & 0 & 1 & 1 & 0 & 0 \\
     0 & 1 & 1 & 1 & 0 & 1 & 0 \\
     1 & 1 & 1 & 1 & 0 & 0 & 0 
\end{bmatrix}\,,\quad
\end{equation}
which are readily verified to satisfy the triple product property
with $s=t=u=2$ and $r=7$.

Let $(\zeta|\xi,\eta)$ satisfy the triple product rule with parameters
$\bra s,t,u\ket_r$. Let $A$ be an $s\times t$ matrix and $B$ be 
a $t\times u$ matrix. 
Define an $s\times u$ matrix $C$ for all $i'\in[s]$ and $k'\in[u]$ 
by the rule
\begin{equation}
\label{eq:triple-mm}
\begin{split}
C_{i'k'}&=
\sum_{h\in[r]}
\zeta_{i'k'h}
\sum_{i\in[s]}
\sum_{j\in[t]}
\xi_{hij}A_{ij}
\sum_{j'\in[t]}
\sum_{k\in[u]}
\eta_{hj'k}B_{j'k}\\
&=
\sum_{i\in[s]}
\sum_{j\in[t]}
\sum_{j'\in[t]}
\sum_{k\in[u]}
A_{ij}
B_{j'k}
\sum_{h\in[r]}
\zeta_{i'k'h}
\xi_{hij}
\eta_{hj'k}\\
&=\sum_{j\in[t]}A_{i'j}B_{jk'}\,,
\end{split}
\end{equation}
where the second equality follows by changing the order of summation 
and the last equality follows from the triple product 
property~\eqref{eq:triple-prod}. Viewing $A$, $B$, and $C$ as vectors of 
length $st$, $tu$, and $su$, respectively, 
from \eqref{eq:triple-mm} it immediately follows that
\begin{equation}
\label{eq:triple-mm-alg}
C=\zeta(\xi A\odot \eta B)\,.
\end{equation}
In other words, we observe that a three-tuple $(\zeta|\xi,\eta)$ with the
triple product property reduces multiplication of the matrices $A$ and $B$ to
(i) taking linear combinations $\xi A$ and $\eta B$ of entries of $A$ and $B$ 
independently, (ii) multiplying these linear combinations pointwise
to obtain $\xi A\odot \eta B$, and
(iii) taking linear combinations $\zeta(\xi A\odot \eta B)$
to obtain the product matrix $C$ of $A$ and $B$.

The triple product property gains its computational power from closure under
Kronecker products and the composition rule. Indeed, let
$(\zeta_\ell|\xi_\ell,\eta_\ell)$ satisfy the triple product property
with parameters $\bra s_\ell,t_\ell,u_\ell\ket_{r_\ell}$ 
for $\ell=1,2,\ldots,d$.
Let $A$ be a vector of length $s_1t_1s_2t_2\cdots s_d t_d$,
let $B$ be a vector of length $t_1u_1t_2u_2\cdots t_d u_d$, and 
let $C$ be a vector of length $s_1t_us_2u_2\cdots s_d u_d$.
Similarly to \eqref{eq:triple-mm}, for all
$(i_1',i_2',\ldots,i_d')\in [s_1]\times[s_2]\times\cdots\times [s_d]$
and
$(k_1',k_2',\ldots,k_d')\in [u_1]\times[u_2]\times\cdots\times [u_d]$,
by the triple product property \eqref{eq:triple-prod} we now have
\begin{equation}
\label{eq:triple-mm-kron}
\begin{split}
&C_{i_1'k_1'i_2'k_2'\cdots i_d'k_d'}=\\[1mm]
&\ \ \ =
\!\!\!
\sum_{\substack{h_1\in[r_1]\\h_2\in[r_2]\\[-1.5mm]\vdots\\h_d\in[r_d]}}
\!\!\!
\zeta^{(1)}_{i_1'k_1'h_1}
\cdots
\zeta^{(d)}_{i_d'k_d'h_d}
\!\!\!
\sum_{\substack{i_1\in[s_1]\\i_2\in[s_2]\\[-1.5mm]\vdots\\i_d\in[s_d]}}\!
\sum_{\substack{j_1\in[t_1]\\j_2\in[t_2]\\[-1.5mm]\vdots\\j_d\in[t_d]}}
\!\!\!
\xi^{(1)}_{h_1i_1j_1}
\cdots
\xi^{(d)}_{h_di_dj_d}
A_{i_1j_1i_2j_2\cdots i_dj_d}
\!\!\!
\sum_{\substack{j_1'\in[t_1]\\j_2'\in[t_2]\\[-1.5mm]\vdots\\j_d'\in[t_d]}}\!
\sum_{\substack{k_1\in[u_1]\\k_2\in[u_2]\\[-1.5mm]\vdots\\k_d\in[u_d]}}
\!\!\!
\eta^{(1)}_{h_1j_1'k_1}
\cdots
\eta^{(d)}_{h_dj_d'k_d}
B_{j_1'k_1j_2'k_2\cdots j_d'k_d}\!\!\!\!\!\!\!\!\!\!\\
&\ \ \ =
\!\!\!
\sum_{\substack{j_1\in[t_1]\\j_2\in[t_2]\\[-1.5mm]\vdots\\j_d\in[t_d]}}
\!\!\!
A_{i_1'j_1i_2'j_2\cdots i_d'j_d}
B_{j_1k_1'j_2k_2'\cdots j_dk_d'}\,.
\end{split}
\end{equation}
Similarly to \eqref{eq:triple-mm-alg}, 
from \eqref{eq:triple-mm-kron} we observe that
\begin{equation}
\label{eq:triple-mm-kron-alg}
C=\bigl(\zeta^{(1)}\otimes\cdots\otimes\zeta^{(d)}\bigr)
  \bigl(\bigl(\xi^{(1)}\otimes\cdots\otimes\xi^{(d)}\bigr)A\odot 
    \bigl(\eta^{(1)}\otimes\cdots\otimes\eta^{(d)}\bigr)B\bigr)\,.
\end{equation}
In other words, to multiply $A$ and $B$, it suffices to first
compute the two matrix-vector products
\begin{equation}
\label{eq:mm-lin-comb-input}
\begin{split}
 T\leftarrow \bigl(\xi^{(1)}\otimes\cdots\otimes\xi^{(d)}\bigr)A\,,\qquad
 S\leftarrow \bigl(\eta^{(1)}\otimes\cdots\otimes\eta^{(d)}\bigr)B\,,
\end{split}
\end{equation}
then multiply the resulting vectors elementwise 
\begin{equation}
\label{eq:mm-elementwise}
 Q\leftarrow T\odot S\,,
\end{equation}
and finally compute the matrix-vector product
\begin{equation}
\label{eq:mm-lin-comb-prod}
 C\leftarrow \bigl(\zeta^{(1)}\otimes\cdots\otimes\zeta^{(d)}\bigr) Q
\end{equation}
to recover the product $C$ of $A$ and $B$ as a vector of length 
$s_1u_1s_2u_2\cdots s_du_d$.

\subsection{Yates's Algorithm and Fast Matrix Multiplication}

This section develops fast matrix multiplication in the above framework 
by reduction to Yates's
algorithm. The computational bottleneck of the matrix-multiplication algorithm 
$(A,B)\mapsto C$ given 
by \eqref{eq:mm-lin-comb-input}, \eqref{eq:mm-elementwise}, and
\eqref{eq:mm-lin-comb-prod} occurs with matrix-vector multiplications
of the following form. For $\ell=1,2,\ldots,d$,
let $\mu^{(\ell)}$ be a matrix of shape $b_\ell\times a_\ell$, and 
let $U$ be a vector of length $a_1a_2\cdots a_d$.
We want to compute the vector $V$ of length $b_1b_2\cdots b_d$ with
\[
V=\bigl(\mu^{(1)}\otimes\mu^{(2)}\otimes\cdots\otimes\mu^{(d)}\bigr)U\,.
\]
The key idea is to rely on the composition rule \eqref{eq:composition-rule}
to implement multiplication with 
$\mu^{(1)}\otimes\mu^{(2)}\otimes\cdots\otimes\mu^{(d)}$ one component
matrix $\mu^{(\ell)}$ at a time, 
via a sequence of sparse matrices $\bar\mu^{[\ell]}$ defined in what follows. 
Let $\pi:\{1,2,\ldots,d\}\rightarrow\{1,2,\ldots,d\}$ be an arbitrary
permutation that encodes the order in which the matrices
$\mu^{(\ell)}$ will be applied.%
\footnote{%
At first reading, it may be convenient 
to assume that $\pi$ is the identity permutation.}
For all $k,\ell\in\{1,2,\ldots,d\}$, let 
\[
m_{k,\ell}^\pi=m_{k,\ell}^\pi(b,a)=\begin{cases}
a_k & \text{if $k\in \pi\bigl(\bigl\{1,2,\ldots,\pi^{-1}(\ell)\bigr\}\bigr)$};\\
b_k & \text{otherwise}. 
\end{cases}
\]
Let us recall that we write $I_n$ for an $n\times n$ identity matrix. 
From the composition rule \eqref{eq:composition-rule} of Kronecker products,
we obtain the decomposition 
\begin{equation}
\label{eq:yates-decomp}
  \mu^{(1)}\otimes\mu^{(2)}\otimes\cdots\otimes\mu^{(d)}
  = \bar\mu^{[\pi(1)]}\bar\mu^{[\pi(2)]} \cdots \bar\mu^{[\pi(d)]}
\end{equation}
with 
\begin{equation}
  \bar\mu^{[\ell]}
  =I_{m^\pi_{1,\ell}}\otimes I_{m^\pi_{2,\ell}}\otimes\cdots\otimes I_{m^\pi_{\ell-1,\ell}}
    \otimes\mu^{(\ell)}\otimes
   I_{m^\pi_{\ell+1,\ell}}\otimes I_{m^\pi_{\ell+2,\ell}}\otimes\cdots\otimes I_{m^\pi_{d,\ell}}
\end{equation}
for $\ell=1,2,\ldots,d$.
In essence, each matrix $\bar\mu^{[\ell]}$ implements 
$\prod^d_{k=1,\,k\neq\ell}m^\pi_{k,\ell}$ independent
matrix-vector multiplications with the matrix $\mu^{(\ell)}$. 
Accordingly, $\bar\mu^{[\ell]}$ is sparse with at most 
$a_\ell b_\ell\prod^d_{k=1,\,k\neq\ell}m^\pi_{k,\ell}$ nonzero entries. 
Furthermore, if matrix-vector multiplication with the $b_\ell\times a_\ell$ 
matrix $\mu^{(\ell)}$ has
a straight-line-program implementation consisting of 
$P_\ell\leq b_\ell(a_\ell-1)$ binary additions, the total number of binary 
additions in the algorithm
\begin{equation}
\label{eq:yates}
\begin{split}
U^{[d]}\leftarrow U\,,\quad
U^{[d-1]}\leftarrow \bar\mu^{[\pi(d)]}U^{[d]}\,,\quad
U^{[d-2]}\leftarrow \bar\mu^{[\pi(d-1)]}U^{[d-1]}\,,\quad
U^{[0]}\leftarrow \bar\mu^{[\pi(1)]}U^{[1]}\,,\quad
&V\leftarrow U^{[0]}\,,
\end{split}
\end{equation}
is 
\begin{equation}
\label{eq:yates-cost}
\sum_{\ell=1}^d 
P_\ell\prod^d_{\substack{k=1\\k\neq\ell}}m^\pi_{k,\ell}\,.
\end{equation}
The algorithm \eqref{eq:yates} is known 
as {\em Yates's algorithm}~\cite{Yates1937} for multiplying 
a Kronecker-product-structured matrix with a given vector. 

Yates's algorithm is known to admit highly efficient parallelization. Indeed, 
since each {\em layer} $U^{[\ell-1]}\leftarrow\bar\mu^{[\pi(\ell)]}U^{[\ell]}$
in Yates's algorithm consists of a large number of independent matrix-vector 
multiplications with the same matrix~$\mu^{(\pi(\ell))}$, 
Yates's algorithm admits immediate vector-parallelization 
(single-instruction-multiple-data parallelization). Furthermore, multiple 
consecutive layers may be aggregated into one layer to optimize use of local 
storage such as per-scalar-thread registers in vectorized parallel execution.
Here the ability to permute layers arbitrarily yields considerable
freedom to optimize both the arithmetic cost as well as the 
vectorized execution and the use of local storage in applications.

For matrix multiplication, using Yates's algorithm with permutations 
$\pi_\zeta,\pi_\xi,\pi_\eta:\{1,2,\ldots,d\}\rightarrow\{1,2,\ldots,d\}$
to implement the matrix-vector multiplications in \eqref{eq:mm-lin-comb-input} 
and \eqref{eq:mm-lin-comb-prod},
we obtain an algorithm design that multiplies an
$s_1s_2\cdots s_d\times t_1t_2\cdots t_d$ matrix with a 
$t_1t_2\cdots t_d\times u_1u_2\cdots u_d$ matrix using exactly
\begin{equation}
\label{eq:yates-mm-add}
\sum_{\ell=1}^d 
P_\ell^{\zeta}\prod^d_{\substack{k=1\\k\neq\ell}}m_{k,\ell}^{\pi_\zeta}(s\odot u,r)
+\!\sum_{\ell=1}^d 
P_\ell^{\xi}\prod^d_{\substack{k=1\\k\neq\ell}}m_{k,\ell}^{\pi_\xi}(r,s\odot t)
+\!\sum_{\ell=1}^d 
P_\ell^{\eta}\prod^d_{\substack{k=1\\k\neq\ell}}m_{k,\ell}^{\pi_\eta}(r,t\odot u)
\end{equation}
binary additions and
$r_1r_2\cdots r_d$
binary multiplications, 
where $P_\ell^\zeta$, $P_\ell^\xi$, and $P_\ell^\eta$
are the number of additions in a straight-line program that multiplies
the matrix $\zeta^{(\ell)}$, $\xi^{(\ell)}$, and $\eta^{(\ell)}$, 
respectively, with a vector, $\ell=1,2,\ldots,d$. 

\subsection{Fast Alternative-Basis Matrix Multiplication}
\label{sect:fast-ab}

Let us now review the key idea of Karstadt and Schwartz~\cite{KarstadtS:2017}
to change basis to reduce the total additive cost within the previous framework.
For $\ell=1,2,\ldots,d$, let $(\zeta^{(\ell)}|\xi^{(\ell)},\eta^{(\ell)})$ 
satisfy the triple product property with parameters 
$\bra s_\ell,t_\ell,u_\ell\ket_{r_\ell}$.
For each $\ell=1,2,\ldots,d$, let us now decompose the matrices 
$\zeta^{(\ell)},\xi^{(\ell)},\eta^{(\ell)}$ into
\begin{equation}
\label{eq:mix-basis}
\zeta^{(\ell)}=\chi^{(\ell)}\gamma^{(\ell)}\,,\quad
\xi^{(\ell)}=\alpha^{(\ell)}\phi^{(\ell)}\,,\quad
\eta^{(\ell)}=\beta^{(\ell)}\psi^{(\ell)}\,,\quad
\end{equation}
where $\phi^{(\ell)}$, $\psi^{(\ell)}$, and $\chi^{(\ell)}$ are arbitrary
invertible matrices. Applying the decomposition \eqref{eq:mix-basis} 
together with the composition rule \eqref{eq:composition-rule} of Kronecker 
products, the key observation of Karstadt and Schwartz~\cite{KarstadtS:2017}
in the present framework is that the multiplication identity 
\eqref{eq:triple-mm-kron-alg} decomposes as
\begin{equation}
\label{eq:triple-mm-kron-alg-ab}
\begin{split}
C&=\bigl(\zeta^{(1)}\otimes\cdots\otimes\zeta^{(d)}\bigr)
   \biggl(\bigl(\xi^{(1)}\otimes\cdots\otimes\xi^{(d)}\bigr)A\odot 
    \bigl(\eta^{(1)}\otimes\cdots\otimes\eta^{(d)}\bigr)B\biggr)\\
 &=\bigl(\chi^{(1)}\gamma^{(1)}\otimes\cdots\otimes\chi^{(d)}\gamma^{(d)}\bigr)
   \biggl(\bigl(\alpha^{(1)}\phi^{(1)}\otimes\cdots\otimes\alpha^{(d)}\phi^{(d)}\bigr)A\odot 
    \bigl(\beta^{(1)}\psi^{(1)}\otimes\cdots\otimes\beta^{(d)}\psi^{(d)}\bigr)B\biggr)\\
 &=\bigl(\chi^{(1)}\otimes\cdots\otimes\chi^{(d)}\bigr)
   \bigl(\gamma^{(1)}\otimes\cdots\otimes\gamma^{(d)}\bigr)
   \biggl(
     \bigl(\alpha^{(1)}\otimes\cdots\otimes\alpha^{(d)}\bigr)
     \bigl(\phi^{(1)}\otimes\cdots\otimes\phi^{(d)}\bigr)A\\
  &\phantom{=\bigl(\chi^{(1)}\otimes\cdots\otimes\chi^{(d)}\bigr)
   \bigl(\gamma^{(1)}\otimes\cdots\otimes\gamma^{(d)}\bigr)}\!\!
   \odot 
     \bigl(\beta^{(1)}\otimes\cdots\otimes\beta^{(d)}\bigr)
     \bigl(\psi^{(1)}\otimes\cdots\otimes\psi^{(d)}\bigr)B
   \biggr)\,.
\end{split}
\end{equation}
From \eqref{eq:triple-mm-kron-alg-ab} we can immediately extract
the following {\em alternative-basis} multiplication algorithm, which we
state in a form that relies on Yates's algorithm for multiplying with each
Kronecker-product-structured matrix in \eqref{eq:triple-mm-kron-alg-ab}.

Let $A$ be a vector of length $s_1t_1s_2t_2\cdots s_d t_d$ and 
let $B$ be a vector of length $t_1u_1t_2u_2\cdots t_d u_d$.
Let $\pi_\phi,\pi_\psi,\pi_\chi,\pi_\alpha,\pi_\beta,\pi_\gamma:\{1,2,\ldots,d\}\rightarrow\{1,2,\ldots,d\}$ be arbitrary permutations.
To multiply $A$ and $B$, first change basis for both inputs by computing
\begin{equation}
\label{eq:mb-mm-input-basis}
\begin{split}
 \hat A\leftarrow \bar\phi^{[\pi_\phi(1)]}\cdots\bar\phi^{[\pi_\phi(d)]}A\,,\qquad
 \hat B\leftarrow \bar\psi^{[\pi_\psi(1)]}\cdots\bar\psi^{[\pi_\psi(d)]}B\,.
\end{split}
\end{equation}
Then compute linear combinations in the new bases
\begin{equation}
\label{eq:mb-mm-lin-comb-input}
\begin{split}
 \hat T\leftarrow \bar\alpha^{[\pi_\alpha(1)]}\cdots\bar\alpha^{[\pi_\alpha(d)]}\hat A\,,\qquad
\hat S\leftarrow \bar\beta^{[\pi_\beta(1)]}\cdots\bar\beta^{[\pi_\beta(d)]}\hat B\,.
\end{split}
\end{equation}
Multiply the resulting vectors elementwise 
\begin{equation}
\label{eq:mb-mm-elementwise}
 \hat Q\leftarrow \hat T\odot \hat S\,.
\end{equation}
Take linear combinations of the products
\begin{equation}
\label{eq:mb-mm-lin-comb-prod}
 \hat C\leftarrow \bar\gamma^{[\pi_\gamma(1)]}\cdots\bar\gamma^{[\pi_\gamma(d)]}\hat Q\,.
\end{equation}
Finally change basis 
\begin{equation}
\label{eq:mb-mm-output-basis}
 C\leftarrow \bar\chi^{[\pi_\chi(1)]}\cdots\bar\chi^{[\pi_\chi(d)]}\hat C
\end{equation}
to recover the product $C$ of $A$ and $B$ as a vector of length 
$s_1u_1s_2u_2\cdots s_du_d$.

The key insight of Karstadt and Schwartz~\cite{KarstadtS:2017} is that the
additive straight-line-program cost of the matrix-vector multiplications 
with the basis-transformed matrices 
$\alpha^{(\ell)}$, $\beta^{(\ell)}$, and $\gamma^{(\ell)}$
may be strictly lower than the corresponding cost for the matrices 
$\xi^{(\ell)}$, $\eta^{(\ell)}$, and $\zeta^{(\ell)}$.
Furthermore, one may engineer the basis changes 
$\phi^{(\ell)}$, $\psi^{(\ell)}$, and $\chi^{(\ell)}$
to such an effect. 

In more precise terms, implementing each of the transformations 
\eqref{eq:mb-mm-input-basis}, \eqref{eq:mb-mm-lin-comb-input}, 
\eqref{eq:mb-mm-lin-comb-prod}, and \eqref{eq:mb-mm-output-basis} 
using Yates's algorithm \eqref{eq:yates} with the identity permutation, we 
obtain that the alternative-basis algorithm uses exactly
\begin{equation}
\label{eq:mb-basis-change-cost}
\sum_{\ell=1}^d 
P_\ell^{\chi}\prod^d_{\substack{k=1\\k\neq\ell}}m_{k,\ell}^{\pi_\chi}(s\odot u,s\odot u)
+\!\sum_{\ell=1}^d 
P_\ell^{\phi}\prod^d_{\substack{k=1\\k\neq\ell}}m_{k,\ell}^{\pi_\phi}(s\odot t,s\odot t)
+\!\sum_{\ell=1}^d 
P_\ell^{\psi}\prod^d_{\substack{k=1\\k\neq\ell}}m_{k,\ell}^{\pi_\psi}(t\odot u,t\odot u)
\end{equation}
binary additions for the basis changes, 
\begin{equation}
\label{eq:mb-linear-comb-cost}
\sum_{\ell=1}^d 
P_\ell^{\gamma}\prod^d_{\substack{k=1\\k\neq\ell}}m_{k,\ell}^{\pi_\gamma}(s\odot u,r)
+\!\sum_{\ell=1}^d 
P_\ell^{\alpha}\prod^d_{\substack{k=1\\k\neq\ell}}m_{k,\ell}^{\pi_\alpha}(r,s\odot t)
+\!\sum_{\ell=1}^d 
P_\ell^{\beta}\prod^d_{\substack{k=1\\k\neq\ell}}m_{k,\ell}^{\pi_\beta}(r,t\odot u)
\end{equation}
binary additions for the linear combinations, and
\begin{equation}
\label{eq:mb-binary-mul-cost}
r_1r_2\cdots r_d
\end{equation}
binary multiplications. 

In the diagonal case with 
$r=r_\ell$,
$s=s_\ell$, 
$t=t_\ell$, 
$u=u_\ell$, 
$\alpha=\alpha^{(\ell)}$,
$\beta=\beta^{(\ell)}$,
$\gamma=\gamma^{(\ell)}$,
$\phi=\phi^{(\ell)}$,
$\psi=\psi^{(\ell)}$, and 
$\chi=\chi^{(\ell)}$ for $\ell=1,2,\ldots,d$, and $s=t=u$ with $r>s^2$, 
we obtain that the alternative-basis algorithm uses 
\begin{equation}
\label{eq:diag-basis-change-cost}
\bigl(P^\chi+P^\phi+P^\psi\bigr)s^{2(d-1)}d
\end{equation}
binary additions for the basis changes, 
\begin{equation}
\label{eq:diag-linear-comb-cost}
\bigl(P^\gamma+P^\alpha+P^\beta\bigr)\sum_{\ell=1}^d s^{2(\ell-1)}r^{d-\ell}
=\bigl(P^\gamma+P^\alpha+P^\beta\bigr)\frac{r^d-s^{2d}}{r-s^2}
\end{equation}
binary additions for the linear combinations,
and $r^d$ binary multiplications. In particular,
from \eqref{eq:diag-basis-change-cost} and \eqref{eq:diag-linear-comb-cost}
we see that optimizing the additive costs 
$P^\gamma$, $P^\alpha$, and $P^\beta$ optimizes the leading constant
of the dominant cost for large $d$. In the non-diagonal case, 
more fine-grained optimization may be performed by optimizing the
components of the decomposition and the permutations to minimize
\eqref{eq:mb-basis-change-cost} and \eqref{eq:mb-linear-comb-cost}.

A useful subfamily of alternative-basis algorithms are the 
alternative-basis algorithms that satisfy 
$\phi^{(\ell)}=\psi^{(\ell)}$ and 
$\chi^{(\ell)}=\bigl(\phi^{(\ell)}\bigr)^{-1}$ 
for all $\ell=1,2,\ldots,d$. We say that such algorithms support
{\em matrix chain multiplication} or {\em chaining} in the alternative basis. 
Indeed, for such an algorithm, one may chain multiplications 
$(\hat A,\hat B)\mapsto \hat C$ in the alternative basis 
using \eqref{eq:mb-mm-lin-comb-input}, \eqref{eq:mb-mm-elementwise}, 
and \eqref{eq:mb-mm-lin-comb-prod} only, without changing back to the 
standard basis in between multiplications.

\subsection{The Two New Algorithms in Matrix Form}
\label{sect:new-matrix}

Let us first express the new alternative-basis algorithm in \S\ref{sect:mb} 
in matrix form. The change-of-basis matrices are
\begin{equation}
\label{eq:mb-basis}
\phi = \psi = 
\begin{bmatrix}
    1 & 0 & 0 & 0 \\
    0 & 1 & 0 & 0 \\
    0 & 0 & 1 & 0 \\
    0 & 1 & 1 & 1
\end{bmatrix}\,,\qquad
\chi = 
\begin{bmatrix}
    1 & 0 & 0 & 0 \\
    0 & 1 & 0 & 1 \\
    0 & 0 & 1 & 1 \\
    0 & 0 & 0 & 1
\end{bmatrix}\,.
\end{equation}
We readily observe that $\phi=\psi=\chi^\top$ and that $\phi=\phi^{-1}$
and $\chi=\chi^{-1}$ over the binary field. 
This self-inverse property and the ease of in-place 
computation of matrix-vector multiplication with \eqref{eq:mb-basis} 
are appealing properties when working with inputs with tight space constraints. 
From \eqref{eq:mb-fwdbasisrecurrence} we observe that we can take 
$P^\phi=P^\psi=2$ and from \eqref{eq:mb-revbasisrecurrence} we observe that 
we can take $P^{\chi}=2$.
The alternative-basis decomposition is defined by the matrices
\begin{equation}
\label{eq:mb-albega}
\alpha=
\begin{bmatrix}
    1 & 0 & 0 & 0 \\
    0 & 1 & 0 & 0 \\
    0 & 0 & 1 & 0 \\
    0 & 0 & 0 & 1 \\
    1 & 0 & 0 & 1 \\
    0 & 1 & 0 & 1 \\
    0 & 0 & 1 & 1 
\end{bmatrix}\,,\quad
\beta=
\begin{bmatrix}
    1 & 0 & 0 & 0 \\
    0 & 0 & 1 & 0 \\
    1 & 0 & 0 & 1 \\
    0 & 0 & 0 & 1 \\
    0 & 1 & 0 & 0 \\
    0 & 1 & 0 & 1 \\
    0 & 0 & 1 & 1 
\end{bmatrix}\,,\quad\text{and}\quad
\gamma=
\begin{bmatrix}
     1 & 1 & 0 & 0 & 0 & 0 & 0 \\
     0 & 0 & 0 & 0 & 1 & 0 & 1 \\
     0 & 0 & 1 & 0 & 0 & 1 & 0 \\
     0 & 1 & 0 & 1 & 0 & 1 & 1 
\end{bmatrix}\,.\quad
\end{equation}
We readily verify that matrix-vector multiplication with the matrices 
$\alpha$ and $\beta$ can be implemented using the 
recurrences~\eqref{eq:mb-tsrecurrence} so that $P^\alpha=P^\beta=3$, and 
matrix-vector multiplication with the matrix $\gamma$ can be implemented 
with $P^{\gamma}=6$ using~\eqref{eq:mb-qrecurrence}. 
We observe that the three-tuple 
$(\chi\gamma|\alpha\phi,\beta\psi)$ with $\phi$, $\alpha$, $\beta$, and 
$\gamma$ given by \eqref{eq:mb-basis} and \eqref{eq:mb-albega} satisfies 
the triple product property \eqref{eq:triple-prod} 
with parameters $\bra 2,2,2\ket_7$.

Let us now turn to the alternative-basis algorithm in \S\ref{sect:ab}.
The change-of-basis matrices are
\begin{equation}
\label{eq:ab-basis}
\phi = \psi = 
\begin{bmatrix}
    1 & 0 & 0 & 0 \\
    0 & 1 & 0 & 0 \\
    0 & 1 & 1 & 1 \\
    0 & 1 & 0 & 1
\end{bmatrix}\quad\text{and}\quad
\chi = 
\begin{bmatrix}
    1 & 0 & 0 & 0 \\
    0 & 1 & 0 & 0 \\
    0 & 0 & 1 & 1 \\
    0 & 1 & 0 & 1
\end{bmatrix}\,,
\end{equation}
We observe that $\chi=\phi^{-1}=\psi^{-1}$, so we have an 
alternative-basis algorithm that supports chaining. 
From \eqref{eq:ab-fwdbasisrecurrence}
we observe that $P^\phi=P^\psi=2$. From \eqref{eq:ab-revbasisrecurrence}
we observe that $P^{\chi}=2$. The alternative-basis decomposition 
is defined by the matrices
\begin{equation}
\label{eq:ab-albega}
\alpha=
\begin{bmatrix}
    1 & 0 & 0 & 0 \\
    0 & 1 & 0 & 0 \\
    0 & 0 & 1 & 0 \\
    0 & 0 & 0 & 1 \\
    1 & 0 & 1 & 0 \\
    0 & 1 & 1 & 0 \\
    0 & 0 & 1 & 1 
\end{bmatrix}\,,\quad
\beta=
\begin{bmatrix}
    1 & 0 & 0 & 0 \\
    0 & 0 & 1 & 1 \\
    0 & 0 & 1 & 0 \\
    0 & 0 & 0 & 1 \\
    0 & 1 & 0 & 0 \\
    0 & 1 & 1 & 0 \\
    1 & 0 & 1 & 0 
\end{bmatrix}\,,\quad\text{and}\quad
\gamma=
\begin{bmatrix}
     1 & 1 & 0 & 0 & 0 & 0 & 0 \\
     0 & 1 & 1 & 0 & 1 & 1 & 0 \\
     0 & 1 & 1 & 0 & 1 & 0 & 1 \\
     0 & 0 & 0 & 1 & 1 & 0 & 0 
\end{bmatrix}\,.\quad
\end{equation}
We readily verify that matrix-vector multiplication with the matrices 
$\alpha$ and $\beta$ can be implemented using the recurrences 
\eqref{eq:ab-tsrecurrence} so that $P^\alpha=P^\beta=3$, and 
matrix-vector multiplication with the matrix $\gamma$ can be implemented 
using \eqref{eq:ab-qrecurrence} so that $P^{\gamma}=6$.
Finally, we have that the three-tuple $(\chi\gamma|\alpha\phi,\beta\psi)$ 
with $\phi$, $\psi$, $\chi$, $\alpha$, $\beta$, and $\gamma$ given 
by \eqref{eq:ab-basis} and \eqref{eq:ab-albega} satisfies the triple 
product property \eqref{eq:triple-prod} with parameters $\bra 2,2,2\ket_7$.

\section{Implementation Engineering}

This section describes our algorithm engineering effort in more
detail, with a focus on implementing the binary product of two matrices; 
our implementation of Boolean products is described more concisely at the
end of this section. We strive for generality of exposition in terms of the 
target platform and in terms of enabling the use of future advances in 
specific alternative-basis decompositions beyond our new decompositions 
presented in ~\S\ref{sect:mb} and \S\ref{sect:ab}.

The high-level exposition withstanding, our goal with the present engineering
effort is that a careful low-level implementation of our engineering framework
will be able to utilize specific hardware configurations across a range of 
configurations efficiently, which we seek to demonstrate in our experiments 
across a range of current platforms in the subsequent section. Furthermore, 
we would like to highlight that detailed low-level parameterization and 
specific optimizations for current target platforms can always be found in 
the accompanying open-source release. 

\subsection{The Family of Target Platforms}

Let us recall from the introduction that we seek generality towards a family 
of target platforms that consist of a single shared-memory compute node 
(the {\em host}) equipped with $N$ independent and identical vectorized 
accelerator devices (the {\em accelerators}), each joined to the host by 
a low-bandwidth interconnect compared with the bandwidth available at 
each device. 

While the detailed parameters of such a configuration are 
expected to vary and evolve over time, this overall topology of the 
configuration---that is, a large-capacity shared-memory host joined by a 
low-bandwidth interconnect to $N$ independent high-bandwidth accelerators 
with restricted memory capacity and vectorized parallelism based on 
thread arrays---can perhaps be expected to remain more stable over time and 
thus warrant engineering attention with the goal of providing a design that 
can be fine-tuned to the varying parameters of each specific configuration. 
A modern typical configuration of this type is a one-or-more-CPU-based 
shared-memory host joined by a peripheral component interconnect to $N$ GPU 
accelerators. 

A further specific goal in our engineering effort is to enable 
consideration of input sizes that are close to the shared-memory capacity
of the host. Accordingly, our present design at the host level has been 
structured for a low shared-memory footprint in terms of working memory, 
but with the understanding that, memory permitting, a faster design can be 
obtained by following a series-parallel strategy at the host, similarly 
to the present framework employed at each accelerator.

One further design choice in the present framework is that we assume that
the accelerators work independently and asynchronously, apart from 
synchronization enforced by the host. This choice scopes out platforms where
the accelerators may be joined to each other by a fast interconnect; indeed, 
making use of such an interconnect would require at least partial 
synchronization between the accelerators.

\subsection{Accelerators and Thread Arrays}

Let us now review our more detailed conventions for working with components
of a target platform, starting with the accelerators. We assume the 
accelerator devices are vector processors designed to execute a large number
of threads in parallel, working asynchronously with one or more arrays of data 
residing in accelerator memory. Each of these data arrays has a shape, which 
makes it convenient to assume that the array of threads working on the data 
has a shape to structure the workload. 

More precisely, we say that a parallel workload of $L=L_1L_2\cdots L_r$ 
asynchronous threads for positive integers $L_1,L_2,\ldots,L_r$ is 
a {\em thread array} of {\em volume} $L$ and {\em shape}
\[
L_1\times L_2\times\cdots\times L_r\,.
\]
This assumption enables us to index a thread $t=0,1,\ldots,L-1$ inside 
a thread array of volume $L$ alternatively by its linear index $t$ or by 
the unique tuple
$(t_1,t_2,\ldots,t_r)\in[L_1]\times [L_2]\times\cdots\times [L_r]$ with 
\begin{equation}
\label{eq:thread-linearization}
t=
t_1L_2L_3\cdots L_r+
t_2L_3L_4\cdots L_r+
\ldots+
t_{r-1}L_r+
t_r
\end{equation}
for compatibility with our convention for linearization of 
data arrays~\eqref{eq:major-linearization}. Workloads based on thread arrays 
are also essentially immediately translatable to modern GPU platforms.

With current and envisaged future platforms in mind, a key
property for a thread-array based design is that of {\em coalescence}. 
We say that a thread array of shape $L_1\times L_2\times\cdots\times L_r$ is 
{\em coalescent} up to $c$ modes if any two threads threads whose tuple-indices
agree in all but least significant $c$ modes $L_{r-c+1},L_{r-c+2},\ldots,L_r$ 
execute an identical stream of instructions; furthermore, whenever an 
instruction in the stream is a memory access, either a single address or 
consecutive addresses of words in accelerator memory are accessed across 
consecutive threads in the $c$ least significant modes. Here it is important
to note that we do not assume that the threads in a thread array execute 
synchronously, although the underlying hardware in most cases has a vectorized
structure that is optimized for synchronized execution of the workload 
in groups of $V$ coalescent threads for a hardware-dependent parameter $V$.

Further engineering principles for accelerators include the following:
\begin{enumerate}
\item
work with a large enough $L$ to expose sufficient parallelism to saturate
the accelerator hardware;
\item
seek coalescent execution by careful design and ordering of the modes of 
the thread array and the modes of the data tensors; and
\item
make sure each thread in the array works with enough local data to make use
of low-latency and high-bandwidth storage available to each thread or groups
of threads, for example, in the form of per-thread registers and 
per-thread-group cache memory.
\end{enumerate}

\subsection{The Host and Coordinating the Work}

The role of the host is to coordinate the work of the $N$ accelerators through multithreading and synchronization primitives at the host. Furthermore, our assumption is that the input and the output require memory capacity in excess of what is available at the accelerators, thus requiring the host to prepare workloads for the accelerators and aggregate the results obtained from the accelerators. We also recall that we are seeking a design that has a low working-memory footprint at the host to enable processing of large inputs. 

Our engineering principles for the host-side implementation include the following:
\begin{enumerate}
\item
design for efficient use of the host-side memory hierarchy; 
each host-thread works with at least one cache line of data at a time, 
utilizing the cache hierarchy available;
\item
the $N$ accelerators are coordinated asynchronously in parallel so that each accelerator has its own pipeline implemented using host-side threads, with synchronization on buffers and submatrices of the output to avoid data races;
\item
each of the $N$ accelerator-pipelines involves four stages, 
implemented using dedicated host-threads: (i) prepare left input to 
accelerator in host memory, (ii) prepare right input to accelerator in host 
memory, (iii) upload input from host memory to accelerator, compute at accelerator, download result from accelerator to host memory, and (iv) aggregate the result in host memory to output;
\item
the accelerators must be supplied with extensive-enough workloads if possible to hide the running time of the other stages of the pipeline behind the accelerator-side compute on the workloads;
\item
the accelerator memory and the available low bandwidth between the host and the accelerators constrain the size of the workloads, which we ease by adopting a series/parallel approach on the accelerators; and
\item
the memory budget at the host is tight due to our goal of scaling to large inputs and the need to buffer of workloads for $N$ accelerators in host memory---we use low-memory-footprint designs to prepare and aggregate the workloads, including in-place basis changes and other primitives help with the memory budget.
\end{enumerate}

\subsection{The High-Level Design}

We are now ready for a high-level exposition of the algorithm design engineered
for our target platforms. Once the high-level description is done, we
proceed to review more detailed implementation considerations. We will use the 
alternative-basis framework of \S\ref{sect:alt-basis}.

Let $d$ be a positive integer that represents the {\em depth} of the 
algorithm design in terms of the number of three-tuples of matrices that 
satisfy the triple-product property \eqref{eq:triple-prod} used to structure 
the arithmetic. The exact implementation of this arithmetic will vary somewhat
from the ideal design \eqref{eq:mb-mm-input-basis}, 
\eqref{eq:mb-mm-lin-comb-input}, \eqref{eq:mb-mm-elementwise}, 
\eqref{eq:mb-mm-lin-comb-prod}, and
\eqref{eq:mb-mm-output-basis} given in \S\ref{sect:alt-basis}, in particular 
due to storage and bandwidth considerations both at the host and at the
accelerators, and due to the need to supply independent and asynchronous 
parallel workloads to each of the $N$ accelerators. Each accelerator will
in turn work both serially (recursively) and in parallel with its specific
subproblems delegated to it by the host. 

We will split $d$ according to the high-level {\em layers} of the design.
We work with four layers: 
\begin{enumerate}
\item
{\em host} layer, 
\item
{\em accelerator serial} layer, 
\item
{\em accelerator parallel} layer, and 
\item
{\em accelerator inner} layer. 
\end{enumerate}
It will be convenient to indicate the layers in our notation with the
symbols ``$\ho$'' (host), ``$\se$'' (accelerator serial), ``$\pa$'' (accelerator parallel), and ``$\nn$'' (inner). 
Accordingly, we have
\[
d=d^{(\ho)}+d^{(\se)}+d^{(\pa)}+d^{(\nn)}\,,
\]
where the nonnegative integers $d^{(\ho)}$, $d^{(\se)}$, $d^{(\pa)}$, 
and $d^{(\nn)}$ indicate the number of decomposition matrices employed at 
each high-level layer. 

For each type of layer $\ty\in\{\ho,\se,\pa,\nn\}$ and each level
$\ell=1,2,\ldots,d^{(\ty)}$ inside a layer, let the three-tuple
\[
\bigl(\chi^{(\ty,\ell)}\gamma^{(\ty,\ell)}\big|
 \alpha^{(\ty,\ell)}\phi^{(\ty,\ell)},
 \beta^{(\ty,\ell)}\psi^{(\ty,\ell)}\bigr)
\]
satisfy the triple product property with parameters 
$\bigl\bra s_\ell^{(\ty)},t_\ell^{(\ty)},u_\ell^{(\ty)}\bigr\ket_{r_\ell^{(\ty)}}$,
where $\chi^{(\ty,\ell)}$, $\phi^{(\ty,\ell)}$, and $\psi^{(\ty,\ell)}$ are
invertible matrices.

Our task is to multiply a given matrix $A'$ of shape $s\times t$
with a given matrix $B'$ of shape $t\times u$ to yield as output
the product matrix $C'=A'B'$ of shape $s\times u$, where
\begin{equation}
\label{eq:input-modes}
\begin{array}{r@{\ \ }l@{\ \ }c@{\,}l@{\,}c@{\,}l@{\,}c@{\,}l@{\,}c@{\,}l}
s&=
&s^{(\ho)}_1\cdots s^{(\ho)}_{d^{(\ho)}}&\cdot
&s^{(\se)}_1\cdots s^{(\se)}_{d^{(\se)}}&\cdot
&s^{(\pa)}_1\cdots s^{(\pa)}_{d^{(\pa)}}&\cdot
&s^{(\nn)}_1\cdots s^{(\nn)}_{d^{(\nn)}}\,,\\
t&=
&t^{(\ho)}_1\cdots t^{(\ho)}_{d^{(\ho)}}&\cdot
&t^{(\se)}_1\cdots t^{(\se)}_{d^{(\se)}}&\cdot
&t^{(\pa)}_1\cdots t^{(\pa)}_{d^{(\pa)}}&\cdot
&t^{(\nn)}_1\cdots t^{(\nn)}_{d^{(\nn)}}\,,\\
u&=
&u^{(\ho)}_1\cdots u^{(\ho)}_{d^{(\ho)}}&\cdot
&u^{(\se)}_1\cdots u^{(\se)}_{d^{(\se)}}&\cdot
&u^{(\pa)}_1\cdots u^{(\pa)}_{d^{(\pa)}}&\cdot
&u^{(\nn)}_1\cdots u^{(\nn)}_{d^{(\nn)}}\,.
\end{array}
\end{equation}

The following subsections will expose the more detailed algorithm design.
We will first present the flow of computation from the host to the accelerators
and back to the host one layer at a time, and then present in detail how the 
host coordinates its own work and the work of the accelerators through 
host-side threading and appropriate synchronization primitives. Finally,
we will parameterize the algorithm implementation.

Our expositionary focus will be on the implementation of the alternative-basis 
phase \eqref{eq:mb-mm-lin-comb-input}, \eqref{eq:mb-mm-elementwise}, and
\eqref{eq:mb-mm-lin-comb-prod} of the algorithm, with less attention and
optimization effort devoted to the pre-and-postprocessing phases of data 
permutation and the basis changes at the host; the latter can be found in 
the accompanying source code.

\smallskip
\noindent
{\em Linearization and indexing conventions recalled.}
At this point it is convenient to recall that we tacitly employ the 
first-index-major linearization convention \eqref{eq:major-linearization} 
when linearizing bit arrays to words of memory; that is, changing the most 
significant (leftmost) indices in a tuple of indices causes the largest stride 
in memory addressing, and the least significant (rightmost) indices will 
index bits inside a word of memory as appropriate. We follow a similar 
convention when linearizing thread arrays on 
accelerators~\eqref{eq:thread-linearization}; that is, hardware 
vectorization occurs with the least significant (rightmost) indices of a tuple 
indexing a thread in an array of asynchronous threads that implements 
a workload on an accelerator. We assume that the accelerator hardware 
computes with vectors of $V$ synchronous threads with consecutive linear 
indices and that each thread can load and store $W$-bit words.

\subsection{Input Permutation at the Host}

To obtain coalesecent execution of the layered design on vectorized 
accelerator hardware, it will be convenient to permute the input 
matrices $A'$ and $B'$ from the $s\times t$ and $t\times u$ layouts 
with \eqref{eq:input-modes} into the interleaved-layout 
vectors $A$ and $B$ of lengths
\begin{equation}
\label{eq:a-interleaved}
s^{(\ho)}_1t^{(\ho)}_1\cdots s^{(\ho)}_{d^{(\ho)}}t^{(\ho)}_{d^{(\ho)}}
s^{(\se)}_1t^{(\se)}_1\cdots s^{(\se)}_{d^{(\se)}}t^{(\se)}_{d^{(\se)}}
s^{(\pa)}_1t^{(\pa)}_1\cdots s^{(\pa)}_{d^{(\pa)}}t^{(\pa)}_{d^{(\pa)}}
s^{(\nn)}_1t^{(\nn)}_1\cdots s^{(\nn)}_{d^{(\nn)}}t^{(\nn)}_{d^{(\nn)}}
\end{equation}
and
\begin{equation}
\label{eq:b-interleaved}
t^{(\ho)}_1u^{(\ho)}_1\cdots t^{(\ho)}_{d^{(\ho)}}u^{(\ho)}_{d^{(\ho)}}
t^{(\se)}_1u^{(\se)}_1\cdots t^{(\se)}_{d^{(\se)}}u^{(\se)}_{d^{(\se)}}
t^{(\pa)}_1u^{(\pa)}_1\cdots t^{(\pa)}_{d^{(\pa)}}u^{(\pa)}_{d^{(\pa)}}
t^{(\nn)}_1u^{(\nn)}_1\cdots t^{(\nn)}_{d^{(\nn)}}u^{(\nn)}_{d^{(\nn)}}\,,
\end{equation}
respectively. This permutation is executed in host memory, using appropriate
parallelization, vectorization primitives, and cache-blocking at the host
for efficiency. For uniform implementation of Boolean and binary matrix
multiplication, our low-level implementation also transposes the right-hand 
side operand at this point by transposing the roles of $t$ and $u$ 
in \eqref{eq:b-interleaved}. The innermost transpose is implemented for
$64\times 64$ submatrices recursively using word operations on 32-bit words,
cf.~\cite{Warren:2013}.

\subsection{Input Basis Change at the Host}

After input permutation, we execute basis changes 
$\hat A\leftarrow\bar\phi A$ and $\hat B\leftarrow\bar\psi B$ at the host.
Both basis changes are executed in parallel using Yates's 
algorithm \eqref{eq:yates} on $d$ levels. The innermost $d^{(\nn)}$ levels
employ the identity basis change and thus are omitted; indeed, we work
with the standard-basis cubic multiplication algorithm in the accelerator 
inner layer in \S\ref{sect:acc-in}, so no basis changes will be required.  

\subsection{Instance Generation at the Host}

We now proceed to describe the sub-instances that the host generates and
forwards to the $N$ accelerators for processing. In particular, we assume that
the alternative-basis inputs $\hat A$ and $\hat B$ have been prepared and
reside in shared memory for parallel access by multiple host-threads working
in parallel to keep all the $N$ accelerators saturated with work; we postpone 
the precise description of the host-side threading and synchronization to 
\S\ref{sect:host-workload}.

In essence, the sub-instances are generated from $\hat A$ and $\hat B$
as the sub-arrays obtained by aggregating the linear combinations at 
the $d^{(\ho)}$ first levels of the base 
design \eqref{eq:mb-mm-lin-comb-input} at once, 
which results in increased arithmetic cost compared with 
\eqref{eq:mb-mm-lin-comb-input} implemented using Yates's algorithm, 
but keeps the working memory needs low at the host, as per our engineering goal
to scale up to large inputs on a single host. 

In precise terms, the sub-instance indexed by 
$(h_1,h_2,\ldots,h_{d^{(\ho)}})\in[r_1^{(\ho)}]\times[r_2^{(\ho)}]\times\cdots\times [r_{d^{(\ho)}}^{(\ho)}]$ consists of the vectors
\begin{equation}
\label{eq:igen}
\begin{split}
\hat T^{[\ho]}_{h_1h_2\cdots h_{d^{(\ho)}}}
&\leftarrow\sum_{\substack{
          (i_1,i_2,\ldots,i_{d^{(\ho)}})\in [s_1^{(\ho)}]\times[s_2^{(\ho)}]\times\cdots\times [s_{d^{(\ho)}}^{(\ho)}]\\
          (j_1,j_2,\ldots,j_{d^{(\ho)}})\in [t_1^{(\ho)}]\times[t_2^{(\ho)}]\times\cdots\times [t_{d^{(\ho)}}^{(\ho)}]}}
    \alpha^{(\ho,1)}_{h_1i_1j_1}
    \alpha^{(\ho,2)}_{h_2i_2j_2}
    \cdots
    \alpha^{(\ho,d^{(\ho)})}_{h_{d^{(\ho)}}i_{d^{(\ho)}}j_{d^{(\ho)}}}
    \hat A_{i_1j_1i_2j_2\cdots i_{d^{(\ho)}}j_{d^{(\ho)}}}\,,\\
\hat S^{[\ho]}_{h_1h_2\cdots h_{d^{(\ho)}}}
&\leftarrow\sum_{\substack{
          (j_1,j_2,\ldots,j_{d^{(\ho)}})\in [t_1^{(\ho)}]\times[t_2^{(\ho)}]\times\cdots\times [t_{d^{(\ho)}}^{(\ho)}]\\
          (k_1,k_2,\ldots,k_{d^{(\ho)}})\in [u_1^{(\ho)}]\times[u_2^{(\ho)}]\times\cdots\times [u_{d^{(\ho)}}^{(\ho)}]}}
    \beta^{(\ho,1)}_{h_1j_1k_1}
    \beta^{(\ho,2)}_{h_2j_2k_2}
    \cdots
    \beta^{(\ho,d^{(\ho)})}_{h_{d^{(\ho)}}j_{d^{(\ho)}}k_{d^{(\ho)}}}
    \hat B_{j_1k_1j_2k_2\cdots j_{d^{(\ho)}}k_{d^{(\ho)}}}\,.
\end{split}
\end{equation}
We compute these vectors by aggregating the sums of subvectors 
in \eqref{eq:igen} as they are stated, taking care to not consider subvectors
whose associated $\alpha$-product (respectively, $\beta$-product) 
in \eqref{eq:igen} is zero. At low level, subroutines for performing 
the XOR and memory copy operations on subvectors have been handcrafted to 
make use of the 256-bit AVX2-registers in CPU hardware. The routines also 
perform reads and writes of 512 bits at a time, which corresponds to the 
length of cache lines on the target CPU. We leave these low-level 
implementation details to the accompanying source code.

In total thus $r_1^{(\ho)}r_2^{(\ho)}\cdots r_{d^{(\ho)}}^{(\ho)}$ 
sub-instances are generated and forwarded for solution by the $N$ accelerators.
Such solving of a sub-instance
$(\hat T^{[\ho]}_{h_1h_2\cdots h_{d^{(\ho)}}},\hat S^{[\ho]}_{h_1h_2\cdots h_{d^{(\ho)}}})$ by an accelerator
consists of (i) uploading the sub-instance to one of the accelerators, 
(ii) having the accelerator compute the solution 
$\hat Q^{[\ho]}_{h_1h_2\cdots h_{d^{(\ho)}}}$, and (iii) downloading 
the solution from the accelerator to shared memory at the host. 
We next describe how each accelerator proceeds to solve a given sub-instance, 
and then describe how each downloaded solution
$\hat Q^{[\ho]}_{h_1h_2\cdots h_{d^{(\ho)}}}$ is aggregated to 
the master solution $\hat C$ at the host. 

\subsection{Accelerator Serial Layer}

Suppose the sub-instance $(\hat T^{[\ho]}_{h_1h_2\cdots h_{d^{(\ho)}}},\hat S^{[\ho]}_{h_1h_2\cdots h_{d^{(\ho)}}})$ has been uploaded to an accelerator. 
The accelerator then proceeds to solve the instance using three layers.
First, a serial layer works with the instance recursively, at the bottom 
of the recursion switching to parallel work with three parts, 
the expanding parallel layer, the inner layer, and the compressing parallel 
layer. We start with a description of the serial (recursive) layer. 
Let the uploaded instance
\[
\begin{split}
\hat T^{[\se,0]}&=\hat T^{[\ho]}_{h_1h_2\cdots h_{d^{(\ho)}}}\,,\\
\hat S^{[\se,0]}&=\hat S^{[\ho]}_{h_1h_2\cdots h_{d^{(\ho)}}}
\end{split}
\]
be the input to the serial layer. The serial layer consists of $d^{(\se)}$ levels.

Each recursive invocation at level $\ell=1,2,\ldots,d^{(\se)}$ can be indexed
by a tuple 
$(h_1,h_2,\ldots,h_{\ell-1})\in[r_1^{(\se)}]\times[r_2^{(\se)}]\times\cdots\times [r_{\ell-1}^{(\se)}]$. In particular, the initial invocation at $\ell=1$ is indexed by the empty tuple and with input consisting of the vectors $\hat T^{[\se,0]}$ and $\hat S^{[\se,0]}$. More generally, 
the input to invocation $(h_1,h_2,\cdots,h_{\ell-1})$ consists of vectors 
$\hat T^{[\se,\ell-1]}_{h_1h_2\cdots h_{\ell-1}}$ and 
$\hat S^{[\se,\ell-1]}_{h_1h_2\cdots h_{\ell-1}}$.
For $h_\ell\in[r_{\ell}^{(\se)}]$, the invocation 
computes the vectors
\begin{equation}
\label{eq:rec-in}
\begin{split}
\hat T^{[\se,\ell]}_{h_1h_2\cdots h_\ell}
\leftarrow
\sum_{\substack{i_\ell\in[s_\ell^{(\se)}]\\j_\ell\in[t_\ell^{(\se)}]\\}}
\alpha^{(\se,\ell)}_{h_\ell i_\ell j_\ell}
\hat T^{[\se,\ell-1]}_{h_1h_2\cdots h_{\ell-1} i_\ell j_\ell}\,,
\qquad
\hat S^{[\se,\ell]}_{h_1h_2\cdots h_\ell}
\leftarrow
\sum_{\substack{j_\ell\in[t_\ell^{(\se)}]\\k_\ell\in[u_\ell^{(\se)}]\\}}
\beta^{(\se,\ell)}_{h_\ell i_\ell j_\ell}
\hat S^{[\se,\ell-1]}_{h_1h_2\cdots h_{\ell-1} i_\ell j_\ell}\,,
\end{split}
\end{equation}
and makes the recursive invocation with index $(h_1,h_2,\ldots,h_\ell)$ 
and input consisting of the vectors 
$\hat T^{[\se,\ell]}_{h_1h_2\cdots h_{\ell}}$ and 
$\hat S^{[\se,\ell]}_{h_1h_2\cdots h_{\ell}}$, obtaining as return value 
and storing the solution $\hat Q^{[\se,\ell]}_{h_1h_2\cdots h_{\ell}}$.
At the bottom level of recursion when $\ell=d^{(\se)}$, the expanding parallel
layer (described in the following section) is invoked with the input, and
the return value is obtained from the output of that layer. 
Once all $r_\ell^{(\se)}$ return values
$\hat Q^{[\se,\ell]}_{h_1h_2\cdots h_{\ell}}$ for $h_\ell\in[r_\ell^{(\se)}]$
are available, the invocation computes and returns the value
$\hat Q^{[\se,\ell-1]}_{h_1h_2\cdots h_{\ell-1}}$ defined for all 
$i_\ell\in [s_\ell^{(\se)}]$ and $j_\ell\in [t_\ell^{(\se)}]$ by
\begin{equation}
\label{eq:rec-out}
\hat Q^{[\se,\ell-1]}_{h_1h_2\cdots h_{\ell-1}i_\ell j_\ell}
\leftarrow
\sum_{h_\ell\in[r_\ell^{(\se)}]}
\gamma^{(\se,\ell)}_{i_\ell j_\ell h_\ell}
\hat Q^{[\se,\ell-1]}_{h_1h_2\cdots h_{\ell}}\,.
\end{equation}
The array-arithmetic in \eqref{eq:rec-in} and \eqref{eq:rec-out} is 
implemented with thread arrays on the accelerator so that each thread 
produces one $W$-bit word of output using optimized straight-line programs 
for matrix-vector multiplication with the $\alpha$, $\beta$, and $\gamma$ 
matrices.

\subsection{Accelerator Expanding Parallel Layer}
\label{sect:acc-exp}

We now turn to the layer that implements Yates's algorithm with each of 
the $\alpha$ and $\beta$ matrices, as follows. Each input to the expanding 
parallel layer arrives from the bottom level of recursion in the serial
layer. Let us write $\hat T^{[\pa,0]}$ and $\hat S^{[\pa,0]}$ 
for the input vectors to the parallel layer. These vectors have lengths
\[
s_1^{(\pa)}t_1^{(\pa)}s_2^{(\pa)}t_2^{(\pa)}\cdots s_{d^{(\pa)}}^{(\pa)}t_{d^{(\pa)}}^{(\pa)}
s_1^{(\nn)}t_1^{(\nn)}s_2^{(\nn)}t_2^{(\nn)}\cdots s_{d^{(\nn)}}^{(\nn)}t_{d^{(\nn)}}^{(\nn)}
\quad
\text{and}\quad
t_1^{(\pa)}u_1^{(\pa)}t_2^{(\pa)}u_2^{(\pa)}\cdots t_{d^{(\pa)}}^{(\pa)}u_{d^{(\pa)}}^{(\pa)}
t_1^{(\nn)}u_1^{(\nn)}t_2^{(\nn)}u_2^{(\nn)}\cdots t_{d^{(\nn)}}^{(\nn)}u_{d^{(\nn)}}^{(\nn)}\,,
\]
respectively.
We describe the processing of the 
input $\hat T^{[\pa,0]}$ only, with the understanding that the processing
of $\hat S^{[\pa,0]}$ is similar but utilizes the $\beta$ matrices in place
of the $\alpha$ matrices. 

Let us write $\bar\alpha^{[p]}$ for the matrix
\[
\bar\alpha^{[\pa]}=
\alpha^{(\pa,1)}
\otimes
\alpha^{(\pa,2)}
\otimes
\cdots
\otimes
\alpha^{(\pa,d^{(\pa)})}
\otimes
I_{s_1^{(\nn)}t_1^{(\nn)}s_2^{(\nn)}t_2^{(\nn)}\cdots s_{d^{(\nn)}}^{(\nn)}t_{d^{(\nn)}}^{(\nn)}}
\]
and decompose the matrix using Yates's decomposition \eqref{eq:yates-decomp}
with respect to the reversal permutation as
\[
\bar\alpha^{[\pa]}=
\bar\alpha^{[\pa,d^{(\pa)}]}
\bar\alpha^{[\pa,d^{(\pa)}-1]}
\cdots
\bar\alpha^{[\pa,1]}\,.
\]
The parallel layer proceeds to compute $\bar\alpha^{[\pa]}\hat T^{[\pa,0]}$
by a sequence of matrix-vector multiplications 
$\hat T^{[\pa,\ell]}\leftarrow \bar\alpha^{[\pa,\ell]}T^{[\pa,\ell-1]}$
for $\ell=1,2,\ldots,d^{(\pa)}$. Each multiplication in the sequence
is implemented with
vectorization using $W$-bit words% 
\footnote{We tacitly assume in what follows that $W$ divides the product
$s_1^{(\nn)}t_1^{(\nn)}s_2^{(\nn)}t_2^{(\nn)}\cdots s_{d^{(\nn)}}^{(\nn)}t_{d^{(\nn)}}^{(\nn)}$. Indeed, $W$ is in most cases a power of two, and the product can be similarly chosen to be a power of two by careful design of the inner layer.}{}
and parallelization via a thread array of 
\[
r_1^{(\pa)}r_2^{(\pa)}\cdots r_{\ell-1}^{(\pa)}
s_{\ell+1}^{(\pa)}t_{\ell+1}^{(\pa)}s_{\ell+2}^{(\pa)}t_{\ell+2}^{(\pa)}\cdots s_{d^{(\pa)}}^{(\pa)}t_{d^{(\pa)}}^{(\pa)}
s_1^{(\nn)}t_1^{(\nn)}s_2^{(\nn)}t_2^{(\nn)}\cdots s_{d^{(\nn)}}^{(\nn)}t_{d^{(\nn)}}^{(\nn)}\,/\,W
\]
threads working asynchronously, so that each thread loads $s_\ell^{(\pa)}t_\ell^{(\pa)}$ words from the array $\hat T^{[\pa,\ell-1]}$ to its local registers, word-bit-parallel multiplies in registers with $W$ copies of the matrix $\alpha^{(\pa,\ell)}$ using an optimized straight-line program, and saves $r_\ell^{(\pa)}$ words to the array $\hat T^{[\pa,\ell]}$. We illustrate the design in pseudocode with Algorithm~\ref{alg:acc-exp-parallel}, where we assume that $w$ is a nonnegative integer with $W=s_{d^{(\nn)}-w+1}^{(\nn)}t_{d^{(\nn)}-w+1}^{(\nn)}\cdots s_{d^{(\nn)}}^{(\nn)}t_{d^{(\nn)}}^{(\nn)}$, and that the parameterization of the inner layer is such that such a $w$ exists; this will be the case in what follows. 

The output of the layer consists of the expanded vectors 
$\hat T^{[\pa,d^{(\pa)}]}$ and $\hat S^{[\pa,d^{(\pa)}]}$
of lengths 
\[
r_1^{(\pa)}r_2^{(\pa)}\cdots r_{d^{(\pa)}}^{(\pa)}
s_1^{(\nn)}t_1^{(\nn)}s_2^{(\nn)}t_2^{(\nn)}\cdots s_{d^{(\nn)}}^{(\nn)}t_{d^{(\nn)}}^{(\nn)}
\qquad
\text{and}\qquad
r_1^{(\pa)}r_2^{(\pa)}\cdots r_{d^{(\pa)}}^{(\pa)}
t_1^{(\nn)}u_1^{(\nn)}t_2^{(\nn)}u_2^{(\nn)}\cdots t_{d^{(\nn)}}^{(\nn)}u_{d^{(\nn)}}^{(\nn)}\,,
\]
respectively, which are in turn given as input to the inner layer.

\begin{algorithm*}
  \small
  \caption{The accelerator expanding parallel layer illustrated in pseudocode. This pseudocode illustrates the procedure implemented using a thread array for level $\ell=1,2,\ldots,d^{(\pa)}$ that takes as input $\hat T^{[\pa,\ell-1]}$ and yields the output $\hat T^{[\pa,\ell]}$. Each thread in the array reads $s_\ell^{(\pa)}t_\ell^{(\pa)}$ words to its local registers, multiplies with the matrix $\alpha^{(\pa,\ell)}$ in registers, and writes $r_\ell^{(\pa)}$ words. This procedure illustrates operation for the left input only; the procedure for the right input is similar.}
  \label{alg:acc-exp-parallel}
  \begin{algorithmic}[1]
    \Procedure{AcceleratorExpandingParallelLeft}{$\hat T^{[\pa,\ell-1]}$, $\ell$}
      \ParallelFor{{\bf thread}

\hspace*{5mm}$(h_1^{(\pa)},\ldots,h_{\ell-1}^{(\pa)},
                                  i_{\ell+1}^{(\pa)},j_{\ell+1}^{(\pa)},
                                  \ldots,
                                  i_{d^{(\pa)}}^{(\pa)},j_{d^{(\pa)}}^{(\pa)},
                                  i_1^{(\nn)},j_1^{(\nn)},
                                  \ldots,
                                  i_{d^{(\nn)}-w}^{(\nn)},j_{d^{(\nn)}-w}^{(\nn)})\in $

\hspace*{1cm}$[r_1^{(\pa)}]\times\cdots\times[r_{\ell-1}^{(\pa)}]\times [s_{\ell+1}^{(\pa)}]\times[t_{\ell+1}^{(\pa)}]\times\cdots\times[s_{d^{(\pa)}}^{(\pa)}]\times[t_{d^{(\pa)}}^{(\pa)}]\times[s_1^{(\nn)}]\times[t_1^{(\nn)}]\times\cdots\times[s_{d^{(\nn)}-w}^{(\nn)}]\times[t_{d^{(\nn)}-w}^{(\nn)}]$

\hspace*{-2mm}}
        \For{$(i_\ell^{(\pa)},j_\ell^{(\pa)})\in[s_\ell^{(\pa)}]\times[t_\ell^{(\pa)}]$}
          \State $\hat T_{i_\ell^{(\pa)}j_\ell^{(\pa)}}^{[\lo,\ell-1]}\leftarrow \hat T^{[\pa,\ell-1]}_{h_1^{(\pa)}\cdots h_{\ell-1}^{(\pa)}i_{\ell}^{(\pa)}j_{\ell}^{(\pa)}i_{\ell+1}^{(\pa)}j_{\ell+1}^{(\pa)}\cdots i_{d^{(\pa)}}^{(\pa)}j_{d^{(\pa)}}^{(\pa)}i_1^{(\nn)}j_1^{(\nn)}\cdots i_{d^{(\nn)}-w}^{(\nn)}j_{d^{(\nn)}-w}^{(\nn)}}$
        \EndFor
        \State $\hat T^{[\lo,\ell]}\leftarrow \alpha^{(\pa,\ell)}\hat T^{[\lo,\ell-1]}$\hspace*{1cm}[[ Implemented with word-bit-operations using an optimized 

\hspace*{4.8cm}
straight-line program for multiplying the $r_\ell^{(\pa)}\times s_\ell^{(\pa)}t_\ell^{(\pa)}$ matrix $\alpha^{(\pa,\ell)}$ 

\hspace*{4.8cm}
with the $s_\ell^{(\pa)}t_\ell^{(\pa)}$ vector $\hat T^{[\lo,\ell-1]}$, 
such as \eqref{eq:mb-tsrecurrence} or \eqref{eq:ab-tsrecurrence}. ]]
        \For{$h_\ell^{(\pa)}\in[r_\ell^{(\pa)}]$}
          \State $\hat T^{[\pa,\ell]}_{h_1^{(\pa)}\cdots h_{\ell-1}^{(\pa)}h_\ell^{(\pa)}i_{\ell+1}^{(\pa)}j_{\ell+1}^{(\pa)}\cdots i_{d^{(\pa)}}^{(\pa)}j_{d^{(\pa)}}^{(\pa)}i_1^{(\nn)}j_1^{(\nn)}\cdots i_{d^{(\nn)}-w}^{(\nn)}j_{d^{(\nn)}-w}^{(\nn)}}\leftarrow \hat T_{h_\ell^{(\pa)}}^{[\lo,\ell]}$
        \EndFor
      \EndParallelFor
    \EndProcedure
  \end{algorithmic}
\end{algorithm*}

\subsection{Accelerator Inner Layer}
\label{sect:acc-in}

The inner layer is the most performance-critical part of the design
since it works with the expanded vectors and thus with the most data
aggregated over the execution of the algorithm. The inner layer takes as
input two vectors $\hat T^{[\nn]}$ and~$\hat S^{[\nn]}$. The output of the
layer is the product
\[
\hat Q^{[\nn]}
=\bar\gamma^{[\nn]}\bigl(
   \bar\alpha^{[\nn]}\hat T^{[\nn]}\odot
   \bar\beta^{[\nn]}\hat S^{[\nn]}\bigr)\,,
\]
where
\[
\begin{split}
\bar\alpha^{[\nn]}&=
I_{r^{(\pa)}_1r^{(\pa)}_2\cdots r^{(\pa)}_{d^{(\pa)}}}
\otimes
\alpha^{(\nn,1)}
\otimes
\alpha^{(\nn,2)}
\otimes
\cdots
\otimes
\alpha^{(\nn,d^{(\nn)})}\,,
\\
\bar\beta^{[\nn]}&=
I_{r^{(\pa)}_1r^{(\pa)}_2\cdots r^{(\pa)}_{d^{(\pa)}}}
\otimes
\beta^{(\nn,1)}
\otimes
\beta^{(\nn,2)}
\otimes
\cdots
\otimes
\beta^{(\nn,d^{(\nn)})}\,,
\\
\bar\gamma^{[\nn]}&=
I_{r^{(\pa)}_1r^{(\pa)}_2\cdots r^{(\pa)}_{d^{(\pa)}}}
\otimes
\gamma^{(\nn,1)}
\otimes
\gamma^{(\nn,2)}
\otimes
\cdots
\otimes
\gamma^{(\nn,d^{(\nn)})}\,.
\end{split}
\]
As the inner layer works with the least significant bit positions of the
bit vectors, in most cases the implementation of the inner layer is
hardware-specific and amounts to making the best possible use of the available
bit- and word-operations in the instruction set as well as the available 
vectorization and associated vector-shuffling instructions. 

For example, 
with $W$-bit words and length-$V$ hardware vectorization of thread arrays
with $M^2=WV$, one possibility to implement the inner 
layer is to perform $r^{(\pa)}_1r^{(\pa)}_2\cdots r^{(\pa)}_{d^{(\pa)}}$
independent $M\times M$ binary matrix multiplications in parallel using
the elementary cubic algorithm implemented with word operations 
and a thread array of $r^{(\pa)}_1r^{(\pa)}_2\cdots r^{(\pa)}_{d^{(\pa)}}V$
threads, so that each thread works with one $W$-bit-word-size fragment 
from each of the $M\times M$ operand matrices and the result matrix, 
with hardware $V$-vector shuffle instructions used to communicate words 
between threads. This is essentially how our low-level implementation of the 
inner layer is structured, though some optimizations remain to be discussed 
in \S\ref{sect:opt}. This part of the framework is also the most sensitive 
to changes in underlying hardware, and so is perhaps the least likely to 
withstand the test of time.

\subsection{Accelerator Compressing Parallel Layer}
\label{sect:acc-comp}

Once the inner layer is complete, its output 
$\hat Q^{[\nn]}$ is given as input to a further parallel layer that 
implements Yates's algorithm with the $\gamma$ matrices, as follows.
Let us write $\hat Q^{[\pa,d^{(\pa)}]}$ for the input vector to the 
compressing parallel layer. Similarly to the expanding layer, 
let us write $\bar\gamma^{[p]}$ for the matrix
\[
\bar\gamma^{[\pa]}=
\gamma^{(\pa,1)}
\otimes
\gamma^{(\pa,2)}
\otimes
\cdots
\otimes
\gamma^{(\pa,d^{(\pa)})}
\otimes
I_{s_1^{(\nn)}u_1^{(\nn)}s_2^{(\nn)}u_2^{(\nn)}\cdots s_{d^{(\nn)}}^{(\nn)}u_{d^{(\nn)}}^{(\nn)}}\,.
\]
and decompose the matrix using Yates's decomposition \eqref{eq:yates-decomp}
with respect to the identity permutation as
\[
\bar\gamma^{[\pa]}=
\bar\gamma^{[\pa,1]}
\bar\gamma^{[\pa,2]}
\cdots
\bar\gamma^{[\pa,d^{(\pa)}]}\,.
\]
The parallel layer proceeds to compute 
$\bar\gamma^{[\pa]}\hat Q^{[\pa,d^{(\pa)}]}$
by a sequence of matrix-vector multiplications 
$\hat Q^{[\pa,\ell-1]}\leftarrow \bar\gamma^{[\pa,\ell]}Q^{[\pa,\ell]}$
for $\ell=d^{(\pa)},d^{(\pa)}-1,\ldots,1$. 
Each multiplication in the sequence is implemented with
vectorization using $W$-bit words and 
parallelization via a thread array of 
\[
r_1^{(\pa)}r_2^{(\pa)}\cdots r_{\ell-1}^{(\pa)}
s_{\ell+1}^{(\pa)}u_{\ell+1}^{(\pa)}s_{\ell+2}^{(\pa)}u_{\ell+2}^{(\pa)}\cdots s_{d^{(\pa)}}^{(\pa)}u_{d^{(\pa)}}^{(\pa)}
s_1^{(\nn)}u_1^{(\nn)}s_2^{(\nn)}u_2^{(\nn)}\cdots s_{d^{(\nn)}}^{(\nn)}u_{d^{(\nn)}}^{(\nn)}\,/\,W
\]
threads working asynchronously, so that each thread loads $r_\ell^{(\pa)}$ words from the array $\hat Q^{[\pa,\ell]}$ to its local registers, word-bit-parallel multiplies in registers with $W$ copies of the matrix $\gamma^{(\pa,\ell)}$ using an optimized straight-line program, and stores $s_\ell^{(\pa)}u_\ell^{(\pa)}$  words to the array $\hat Q^{[\pa,\ell-1]}$. We illustrate the design in pseudocode with Algorithm~\ref{alg:acc-comp-parallel}, where, analogously to Algorithm~\ref{alg:acc-exp-parallel}, we assume that $w$ is a nonnegative integer with $W=s_{d^{(\nn)}-w+1}^{(\nn)}u_{d^{(\nn)}-w+1}^{(\nn)}\cdots s_{d^{(\nn)}}^{(\nn)}u_{d^{(\nn)}}^{(\nn)}$. 

The output of the layer consists of the compressed vector
$\hat Q^{[\pa,0]}$ of length
\[
s_1^{(\pa)}u_1^{(\pa)}s_2^{(\pa)}u_2^{(\pa)}\cdots s_{d^{(\pa)}}^{(\pa)}u_{d^{(\pa)}}^{(\pa)}
s_1^{(\nn)}u_1^{(\nn)}s_2^{(\nn)}u_2^{(\nn)}\cdots s_{d^{(\nn)}}^{(\nn)}u_{d^{(\nn)}}^{(\nn)}\,,
\]
which is in turn given as output to the serial layer.

\begin{algorithm*}
  \small
  \caption{The accelerator compressing parallel layer illustrated in pseudocode. This pseudocode illustrates the procedure implemented using a thread array for level $\ell=d^{(\pa)},d^{(\pa)}-1,\ldots,1$ that takes as input $\hat Q^{[\pa,\ell]}$ and yields the output $\hat Q^{[\pa,\ell-1]}$ . Each thread in the array reads $r_\ell^{(\pa)}$ words to its local registers, multiplies with the matrix $\gamma^{(\pa,\ell)}$ in registers, and writes $s_\ell^{(\pa)}u_\ell^{(\pa)}$ words.}
  \label{alg:acc-comp-parallel}
  \begin{algorithmic}[1]
    \Procedure{AcceleratorCompressingParallel}{$\hat Q^{[\pa,\ell]}$, $\ell$}
      \ParallelFor{{\bf thread}

\hspace*{5mm}$(h_1^{(\pa)},\ldots,h_{\ell-1}^{(\pa)},
                                  i_{\ell+1}^{(\pa)},k_{\ell+1}^{(\pa)},
                                  \ldots,
                                  i_{d^{(\pa)}}^{(\pa)},k_{d^{(\pa)}}^{(\pa)},
                                  i_1^{(\nn)},k_1^{(\nn)},
                                  \ldots,
                                  i_{d^{(\nn)}-w}^{(\nn)},k_{d^{(\nn)}-w}^{(\nn)})\in $

\hspace*{1cm}$[r_1^{(\pa)}]\times\cdots\times[r_{\ell-1}^{(\pa)}]\times [s_{\ell+1}^{(\pa)}]\times[u_{\ell+1}^{(\pa)}]\times\cdots\times[s_{d^{(\pa)}}^{(\pa)}]\times[u_{d^{(\pa)}}^{(\pa)}]\times[s_1^{(\nn)}]\times[u_1^{(\nn)}]\times\cdots\times[s_{d^{(\nn)}-w}^{(\nn)}]\times[u_{d^{(\nn)}-w}^{(\nn)}]$

\hspace*{-2mm}}
        \For{$h_\ell^{(\pa)}\in[r_\ell^{(\pa)}]$}
          \State $\hat Q_{h_\ell^{(\pa)}}^{[\lo,\ell]}\leftarrow \hat Q^{[\pa,\ell]}_{h_1^{(\pa)}\cdots h_{\ell-1}^{(\pa)}h_{\ell}^{(\pa)}i_{\ell+1}^{(\pa)}k_{\ell+1}^{(\pa)}\cdots i_{d^{(\pa)}}^{(\pa)}k_{d^{(\pa)}}^{(\pa)}i_1^{(\nn)}k_1^{(\nn)}\cdots i_{d^{(\nn)}-w}^{(\nn)}k_{d^{(\nn)}-w}^{(\nn)}}$
        \EndFor
        \State $\hat Q^{[\lo,\ell-1]}\leftarrow \gamma^{(\pa,\ell)}\hat Q^{[\lo,\ell]}$\hspace*{1cm}[[ Implemented with word-bit-operations using an optimized 

\hspace*{4.85cm}
straight-line program for multiplying the $s_\ell^{(\pa)}u_\ell^{(\pa)}\times r_\ell^{(\pa)}$ matrix $\gamma^{(\pa,\ell)}$ 

\hspace*{4.85cm}
with the $r_\ell^{(\pa)}$ vector $\hat Q^{[\lo,\ell]}$, 
such as \eqref{eq:mb-qrecurrence} or \eqref{eq:ab-qrecurrence}. ]]
        \For{$(i_\ell^{(\pa)},k_\ell^{(\pa)})\in[s_\ell^{(\pa)}]\times[u_\ell^{(\pa)}]$}
          \State $\hat Q^{[\pa,\ell-1]}_{h_1^{(\pa)}\cdots h_{\ell-1}^{(\pa)}i_\ell^{(\pa)}k_\ell^{(\pa)}i_{\ell+1}^{(\pa)}k_{\ell+1}^{(\pa)}\cdots i_{d^{(\pa)}}^{(\pa)}k_{d^{(\pa)}}^{(\pa)}i_1^{(\nn)}k_1^{(\nn)}\cdots i_{d^{(\nn)}-w}^{(\nn)}k_{d^{(\nn)}-w}^{(\nn)}}\leftarrow \hat Q_{i_\ell^{(\pa)}k_\ell^{(\pa)}}^{[\lo,\ell-1]}$
        \EndFor
      \EndParallelFor
    \EndProcedure
  \end{algorithmic}
\end{algorithm*}

\subsection{Instance Aggregation at the Host}

The aggregation of a solution obtained at an accelerator proceeds as follows
at the host, again with a focus on obtaining a design with a low working memory
footprint in host memory at the price of increased arithmetic cost. 
Before any solutions are accepted, the output vector $\hat C$ is 
initialized to all-$0$ values. Suppose that the solution 
$\hat Q^{[\ho]}_{h_1h_2\cdots h_{d^{(\ho)}}}$ for 
$(h_1,h_2,\ldots,h^{d^{(\ho)}})\in
 [r^{(\ho)}_1]\times[r^{(\ho)}_2]\times\cdots\times[r^{(\ho)}_{d^{(\ho)}}]$ 
has been downloaded from an accelerator to host memory. 
We then execute the following aggregation procedure. For each
$(i_1,i_2,\ldots,i_{d^{(\ho)}})\in[s_1^{(\ho)}]\times[s_2^{(\ho)}]\times\cdots\times [s_{d^{(\ho)}}^{(\ho)}]$ 
and each
$(k_1,k_2,\ldots,k_{d^{(\ho)}})\in[u_1^{(\ho)}]\times[u_2^{(\ho)}]\times\cdots\times [u_{d^{(\ho)}}^{(\ho)}]$,
whenever it holds that 
$\gamma^{(\ho,1)}_{i_1k_1h_1}\gamma^{(\ho,2)}_{i_2k_2h_2}\cdots\gamma^{(\ho,d^{(\ho)})}_{i_{d^{(\ho)}}k_{d^{(\ho)}}h_{d^{(\ho)}}}=1$, 
then aggregate the solution to the output by
\begin{equation}
\label{eq:iagg}
  \hat C_{i_1k_1i_2k_2\cdots i_{d^{(\ho)}}k_{d^{(\ho)}}}
  \leftarrow
  \hat C_{i_1k_1i_2k_2\cdots i_{d^{(\ho)}}k_{d^{(\ho)}}}
  +
  \hat Q^{[\ho]}_{h_1h_2\cdots h_{d^{(\ho)}}}\,.
\end{equation}
When multiple host-threads execute aggregation in parallel, appropriate 
synchronization primitives need to be employed to avoid conflicts between
threads when executing \eqref{eq:iagg}. This will be described in detail in
\S\ref{sect:host-workload}.

\subsection{Output Basis Change at the Host}

After the alternative-basis output $\hat C$ has been aggregated, we execute the
basis change $C\leftarrow\bar\chi \hat C$ at the host, in parallel using 
Yates's algorithm \eqref{eq:yates} on $d$ levels. In our implementation, 
the innermost $d^{(\nn)}$ levels employ the identity basis change 
(since the accelerator inner-layer uses the standard-basis cubic multiplication
algorithm) and thus are omitted. 

\subsection{Output Permutation at the Host}

The last layer of the framework permutes the interleaved-layout 
vector $C$ of length
\begin{equation}
\label{eq:c-interleaved}
s^{(\ho)}_1u^{(\ho)}_1\cdots s^{(\ho)}_{d^{(\ho)}}u^{(\ho)}_{d^{(\ho)}}
s^{(\se)}_1u^{(\se)}_1\cdots s^{(\se)}_{d^{(\se)}}u^{(\se)}_{d^{(\se)}}
s^{(\pa)}_1u^{(\pa)}_1\cdots s^{(\pa)}_{d^{(\pa)}}u^{(\pa)}_{d^{(\pa)}}
s^{(\nn)}_1u^{(\nn)}_1\cdots s^{(\nn)}_{d^{(\nn)}}u^{(\nn)}_{d^{(\nn)}}
\end{equation}
to an $s\times u$ output matrix $C'$ with $s$ and $u$ as 
in \eqref{eq:input-modes}. This completes the description of the layers of 
the algorithm design. 

\subsection{Further Optimization by Merging Levels and Shifting Levels Between Layers}
\label{sect:opt}

The previous framework can be further optimized at the accelerators by making 
more extensive use of the registers local to each thread in a thread array, 
assuming such registers are available in sufficient quantity. This section
documents two techniques towards this end.

\smallskip
\noindent
{\em Merging levels.}
Assuming the 
parameters $r_\ell^{(\pa)},s_\ell^{(\pa)},t_\ell^{(\pa)},u_\ell^{(\pa)}$
of the expanding levels $\ell=1,2,\ldots,d^{(\pa)}$ in 
the expanding layer (cf.~\S\ref{sect:acc-exp}) are small enough, one may utilize
the per-thread registers and compute two consecutive levels, 
$\ell$ and $\ell+1$, instead of only one level $\ell$. 
In this case one works with a thread array of 
\[
r_1^{(\pa)}r_2^{(\pa)}\cdots r_{\ell-1}^{(\pa)}
s_{\ell+2}^{(\pa)}t_{\ell+2}^{(\pa)}s_{\ell+3}^{(\pa)}t_{\ell+3}^{(\pa)}\cdots s_{d^{(\pa)}}^{(\pa)}t_{d^{(\pa)}}^{(\pa)}
s_1^{(\nn)}t_1^{(\nn)}s_2^{(\nn)}t_2^{(\nn)}\cdots s_{d^{(\nn)}}^{(\nn)}t_{d^{(\nn)}}^{(\nn)}\,/\,W
\]
threads, so that each thread loads $s_\ell^{(\pa)}t_\ell^{(\pa)}s_{\ell+1}^{(\pa)}t_{\ell+1}^{(\pa)}$ words from accelerator memory to its local registers, then multiplies word-bit-parallel in registers with $W$ copies of the matrix $\alpha^{(\pa,\ell)}\otimes\alpha^{(\pa,\ell+1)}$ using optimized straight-line programs for both component matrices (that is, one implements with register arithmetic Yates's algorithm using straight-line programs for $\alpha^{(\pa,\ell)}$ and $\alpha^{(\pa,\ell+1)}$ for the levels $\ell$ and $\ell+1$ computed in registers), and finally saves $r_\ell^{(\pa)}r_{\ell+1}^{(\pa)}$ words to accelerator memory. Similar merging of levels may be used for the compressing layer (cf.~\S\ref{sect:acc-comp}). Our low-level implementation uses this two-level merging for pairs of levels closest to the inner layer. In particular one wants to optimize at levels closest to the inner layer because these layers process the longest vectors and thus do the most work. If more per-thread registers are available, more consecutive levels can be merged and computed in registers.

\smallskip
\noindent
{\em Shifting levels between layers.}
The inner layer can use per-thread registers more efficiently
by shifting expanding/compressing levels to the inner layer from the
expanding/compressing layers. Assuming that the parameters of the 
expansion/compression are 
$r_\ell^{(\pa)},s_\ell^{(\pa)},t_\ell^{(\pa)},u_\ell^{(\pa)}$,
and that the inner layer originally implements $M\times M$ standard-basis 
matrix multiplication (cf.~\S\ref{sect:acc-in}) using one $W$-bit-word
per thread, the inner layer with shifting (i) first loads 
$s_\ell t_\ell+t_\ell u_\ell$ words to each thread, 
(ii) expands in registers using straight-line programs for
$\alpha^{(\pa,\ell)}$ and $\beta^{(\pa,\ell)}$
to $r_\ell+r_\ell$ words of operands per thread, 
(iii) executes in registers $r_\ell$ independent $M\times M$ matrix 
multiplications (essentially repeating the original inner layer $r_\ell$ 
times in registers) to obtain $r_\ell$ words of expanded results, 
and (iv) finally compresses to $s_\ell u_\ell$ words of results per thread
using a straight-line program for $\gamma^{(\pa,\ell)}$ 
and writes these results to accelerator memory. Our low-level implementation
follows this strategy of shifting one level of expansion/compression to the
inner layer to make more efficient use of per-thread registers and to perform
less accelerator-memory transactions at the inner layer.

\subsection{Coordinating the Workload}
\label{sect:host-workload}

This section describes how the entire workload for the host and 
the $N$ accelerators is coordinated at the host using host-threads and 
appropriate synchronization.

Let us start by recalling that the top-level alternative-basis workload 
$(\hat A,\hat B)\mapsto\hat C$ 
consists of $r_1^{(\ho)}r_2^{(\ho)}\cdots r_{d^{(\ho)}}^{(\ho)}$ sub-instances
that need to be 
(i) generated from $\hat A$ and $\hat B$ on the host, 
(ii) solved in an accelerator, and 
(iii) the solution aggregated to~$\hat C$ on the host. 
This suggests processing the workload with $N$ 
essentially independent pipelines implemented by one or more host-threads each
so that each pipeline is responsible for keeping exactly one accelerator 
busy with work. Synchronization between pipelines is also required because 
the aggregation steps~\eqref{eq:iagg} of different pipelines can access and 
update the same subarray of $\hat C$, which can lead to data races unless 
synchronization is used to ensure serial updates to each subarray. 

Our strategy is to implement the pipeline for each accelerator $\ell\in[N]$ 
using three buffers $T_\ell,S_\ell,Q_\ell$ in host memory and four 
host-threads: one thread to prepare the left input $T_\ell$ in host memory,
one thread to prepare the right input $S_\ell$ in host memory, 
one thread to multiply $T_\ell,S_\ell$ on accelerator $\ell$ to obtain 
the result $Q_\ell$ (including uploading the input and downloading the result),
and one thread to aggregate the result $Q_\ell$ to the result $\tilde C$.

Pseudocode for coordinating work at the host is given in 
Algorithm~\ref{alg:multiacceleratorstrassen}. This workload of $4N$ host-threads
is deadlock-free due to the design of the pipeline---blocking is needed to 
prevent race conditions occurring on the arrays $T_\ell,S_\ell,Q_\ell$,
the accelerator devices, and subarrays of the array $\hat C$, but each pipeline 
will always be flushed eventually. We also observe that the order in which the 
subproblems $(h_1,\ldots,h_{d^(\ho)})$ are processed is insignificant as long 
as the order is well-defined, which is guaranteed by the locks on subarrays of 
$\hat C$. Here we have presented a $4N$-thread design with a single $4$-thread 
pipeline for each of the $N$ accelerators; the design can be easily extended to
multiple pipelines competing for a single accelerator device to better saturate 
the accelerators with data as appropriate.

\begin{algorithm*}
  \small
  \caption{Procedure for coordinating work at a host joined to 
    $N$ accelerators. The input arrays are $\hat A$ and 
    $\hat B$ and the output array is $\hat C$. The workload is formed of
    $4N$ threads. Each thread is associated with one of the $N$ accelerators
    and a specific part of the pipeline for this accelerator. 
    },
  \label{alg:multiacceleratorstrassen}
  \vspace*{-2mm}
  \begin{algorithmic}[1]
    \Procedure{HostCoordinate}{$\hat A,\hat B$}
    \State Let $T_\ell$ for $\ell\in [N]$ be an array of shape
$s^{(\se)}_1t^{(\se)}_1\cdots s^{(\se)}_{d^{(\se)}}t^{(\se)}_{d^{(\se)}}
s^{(\pa)}_1t^{(\pa)}_1\cdots s^{(\pa)}_{d^{(\pa)}}t^{(\pa)}_{d^{(\pa)}}
s^{(\nn)}_1t^{(\nn)}_1\cdots s^{(\nn)}_{d^{(\nn)}}t^{(\nn)}_{d^{(\nn)}}$
    \State Let $S_\ell$ for $\ell\in [N]$ be an array of shape 
$t^{(\se)}_1u^{(\se)}_1\cdots t^{(\se)}_{d^{(\se)}}u^{(\se)}_{d^{(\se)}}
t^{(\pa)}_1u^{(\pa)}_1\cdots t^{(\pa)}_{d^{(\pa)}}u^{(\pa)}_{d^{(\pa)}}
t^{(\nn)}_1u^{(\nn)}_1\cdots t^{(\nn)}_{d^{(\nn)}}u^{(\nn)}_{d^{(\nn)}}$
    \State Let $Q_\ell$ for $\ell\in [N]$ be an array of shape
$s^{(\se)}_1u^{(\se)}_1\cdots s^{(\se)}_{d^{(\se)}}u^{(\se)}_{d^{(\se)}}
s^{(\pa)}_1u^{(\pa)}_1\cdots s^{(\pa)}_{d^{(\pa)}}u^{(\pa)}_{d^{(\pa)}}
s^{(\nn)}_1u^{(\nn)}_1\cdots s^{(\nn)}_{d^{(\nn)}}u^{(\nn)}_{d^{(\nn)}}$\vspace*{-1mm}
    \State Initialize $\hat C$ as zero
%    \State For $(i_1,k_2,\ldots,i_{d^{(\ho)}},k_{d^{(\ho)}})\in[s_1^{(\ho)}]\times[u_1^{(\ho)}]\times\cdots\times [s_{d^{(\ho)}}^{(\ho)}]\times[u_{d^{(\ho)}}^{(\ho)}]$, initialize mutex $M_{i_1k_1\cdots i_{d^{(\ho)}}k_{d^{(\ho)}}}$
    \ParallelFor{{\bf thread} $t\in [4N]$}
    \State $\ell \gets t \bmod N$
    \For{$(h_1,\ldots,h_{d^{(\ho)}})\in[r_1^{(\ho)}]\times\cdots\times [r_{d^{(\ho)}}^{(\ho)}]$ {\bf such that} $\ell\equiv\sum_{i=1}^{d^{(\ho)}} h_i\prod_{j=i+1}^{d^{(\ho)}}r_j^{(\ho)}\pmod N$}\vspace*{-1mm}
    \If{$0\leq t \leq N-1$}
    \State Block until $T_\ell$ is free
    \State Initialize $T_\ell$ as zero
    \For{$(i_1,j_1,\ldots,i_{d^{(\ho)}},j_{d^{(\ho)}})\in [s_1^{(\ho)}]\times[t_1^{(\ho)}]\times\cdots\times [s_{d^{(\ho)}}^{(\ho)}]\times [t_{d^{(\ho)}}^{(\ho)}]$}\vspace*{-1mm} 
        \If{$\alpha^{(\ho,1)}_{h_1i_1j_1}\cdots\alpha^{(\ho,d^{(\ho)})}_{h_{d^{(\ho)}}i_{d^{(\ho)}}j_{d^{(\ho)}}}\neq 0$}
           \State $T_\ell\gets T_\ell + \hat A_{i_1j_1\cdots i_{d^{(\ho)}}j_{d^{(\ho)}}}$
        \EndIf\vspace*{-1mm}
    \EndFor
    \State Mark $T_\ell$ occupied
    \ElsIf{$N \leq t \leq 2N-1$}
    \State Block until $S_\ell$ is free
    \State Initialize $S_\ell$ as zero
    \For{$(j_1,k_1\ldots,j_{d^{(\ho)}},k_{d^{(\ho)}})\in [t_1^{(\ho)}]\times[u_1^{(\ho)}]\times\cdots\times [t_{d^{(\ho)}}^{(\ho)}]\times [u_{d^{(\ho)}}^{(\ho)}]$}
        \If{$\beta^{(\ho,1)}_{h_1j_1k_1}\cdots\beta^{(\ho,d^{(\ho)})}_{h_{d^{(\ho)}}j_{d^{(\ho)}}k_{d^{(\ho)}}}\neq 0$}
           \State $S_\ell\gets S_\ell + \hat B_{j_1k_1\cdots j_{d^{(\ho)}}k_{d^{(\ho)}}}$
        \EndIf\vspace*{-1mm}
    \EndFor
    \State Mark $S_\ell$ occupied
    \ElsIf{$2N \leq t \leq 3N-1$}
    \State Block until $T_\ell$ and $S_\ell$ are occupied and $Q_\ell$ is free
    \State Upload $T_\ell$ and $S_\ell$ to accelerator $\ell$
    \State Mark $T_\ell$ and $S_\ell$ free
    \State Multiply $T_\ell$ and $S_\ell$ on accelerator $\ell$ to obtain the result $Q_\ell$
    \State Download $Q_\ell$ from accelerator $\ell$
    \State Mark $Q_\ell$ occupied
    \ElsIf{$3N \leq t \leq 4N-1$}
     \State Block until $Q_\ell$ is occupied
     \For{$(i_1,k_2,\ldots,i_{d^{(\ho)}},k_{d^{(\ho)}})\in[s_1^{(\ho)}]\times[u_1^{(\ho)}]\times\cdots\times [s_{d^{(\ho)}}^{(\ho)}]\times[u_{d^{(\ho)}}^{(\ho)}]$}\vspace*{-1mm}
       \If{$\gamma^{(\ho,1)}_{i_1k_1h_1}\cdots\gamma^{(\ho,d^{(\ho)})}_{i_{d^{(\ho)}}k_{d^{(\ho)}}h_{d^{(\ho)}}}\neq 0$}
         \State Acquire the lock for $\hat C_{i_1k_1\cdots i_{d^{(\ho)}}k_{d^{(\ho)}}}$
         \State $\hat C_{i_1k_1\cdots i_{d^{(\ho)}}k_{d^{(\ho)}}} \gets \hat C_{i_1k_1\cdots i_{d^{(\ho)}}k_{d^{(\ho)}}} + Q_\ell$
         \State Release the lock for $\hat C_{i_1k_1\cdots i_{d^{(\ho)}}k_{d^{(\ho)}}}$
       \EndIf
     \EndFor
     \State Mark $Q_\ell$ free
    \EndIf
    \EndFor
  \EndParallelFor
  \EndProcedure
  \end{algorithmic}
\end{algorithm*}

\subsection{Parameterizing the Implementation}

Let us now turn to the detailed parameterization of the framework. Our current 
implementation uses the framework in a diagonal setting based on either of 
the two $\bra 2,2,2\ket_7$ alternative-basis decompositions developed 
in \S\ref{sect:mb} and \S\ref{sect:ab}, respectively. Accordingly, we set 
\[
\begin{split}
s_j^{d^{(\ho)}}=s_j^{d^{(\se)}}=s_j^{d^{(\pa)}}=2\,,\\
t_j^{d^{(\ho)}}=t_j^{d^{(\se)}}=t_j^{d^{(\pa)}}=2\,,\\
u_j^{d^{(\ho)}}=u_j^{d^{(\se)}}=u_j^{d^{(\pa)}}=2\,,\\
r_j^{d^{(\ho)}}=r_j^{d^{(\se)}}=r_j^{d^{(\pa)}}=7\,.
\end{split}
\]
When more efficient decompositions are discovered, these can be immediately 
used in the present framework. 

The present accelerator hardware has $V$-length vectorization of threads 
with $V=32$, and the maximum word length supported per thread for memory 
transactions is $W=128$ bits.%
\footnote{The latter implemented with instructions that work with four 
consecutive 32-bit words.}{}
We structure the inner layer to work with $M\times M$ matrices,
$M=64$, so that $M^2=VW$ and the $M\times M$ multiplication can work with 
instructions for shuffling data between threads across each $V$-length 
vector of threads in a thread array. Accordingly, we let the inner layer 
consists of a single level with
\[
d^{(\nn)}=1, \qquad s_1^{(\nn)}=t_1^{(\nn)}=u_1^{(\nn)}=M\,.
\]
This level works in the standard basis and does not use basis changes.

The setting of the parameters $d^{(\ho)}$, $d^{(\se)}$, and $d^{(\pa)}$
for best performance depends on the details of the target platform, such as
the communication bandwidth available between the host and the accelerators,
and the memory capacities in each component. Here we will give a discussion 
of considerations for parameterization that we expect to generalize and 
withstand the test of time. In the present case we have up to $N=8$ 
accelerator devices available in the host. 

The following considerations affect the detailed parameterization:
\begin{enumerate}
\item
The host layer should consist of as few levels $d^{(\ho)}$ as possible
since the $N$ accelerators with their parallel capacity have far superior 
aggregate instruction bandwidth and they should shoulder the bulk of the 
workload. Constraints on decreasing $d^{(\ho)}$ include the memory capacity 
at the host for maintaining the arrays $T_\ell,S_\ell,Q_\ell$ associated with 
each of the $N$ host-side accelerator pipelines, as well as the bandwidth of 
the communication interconnect between the host and the accelerators---large 
subinstances and their solutions take more time to transfer over the 
interconnect, but on the other hand take more time to solve on the accelerator,
leaving the interconnect idle and thus enabling the hiding of transfer 
latency. When working with inputs close to the memory capacity of the host
as per our engineering goal, the memory capacity at the host is the 
primary constraint that determines how many levels need to be performed at
the host layer. 
\item
The accelerator serial layer in its $d^{(\se)}$ levels enables each accelerator
to accept larger inputs than would be otherwise possible by resorting to the accelerator parallel layer only. That is, because of limited memory available at 
the accelerator, the data-expanding/-compressing accelerator parallel layers 
can only be invoked on input sizes that tolerate a constant-factor expansion 
in the size of the data at each level, up to the inner layer. The serial layer 
is more space-efficient by working through the $r_\ell^{(\se)}$ cases at each 
level recursively and serially, reusing space, but becomes time-inefficient 
when the size of a case becomes so small that its parallel processing does not 
saturate the parallel hardware at the accelerator. 
\item
The accelerator parallel layer should consist of sufficiently many levels 
$d^{(\pa)}$ to saturate the parallel hardware of the accelerator and to 
enable utilization of the optimizations in \S\ref{sect:opt}. 
\end{enumerate}
The setting of the parameters $d^{(\ho)}$, $d^{(\se)}$, and $d^{(\pa)}$ that 
gives the fastest overall performance may be done by first identifying 
the memory-capacity-induced constraints on the parameters and then, subject 
to these constraints, empirically finding the best parameter combination that
optimizes performance among the relatively few choices that remain.

For the detailed target platforms considered in what follows, subproblems of 
shape $65536\times 65536$ (512~MiB per matrix) are small enough to be 
accommodated in the host-side arrays $T_\ell,S_\ell,Q_\ell$ for $\ell\in[N]$ 
alongside the input arrays. For example, with subproblems of shape 
$65536\times 65536=2^{16}\times 2^{16}$ delegated to the accelerators, 
an input of shape $1048576\times 1048576=2^{20}\times 2^{20}$ 
(1~Tib; or what is the same, 128~GiB per matrix) will perform 
$d^{(\ho)}=20-16=4$ levels in the host layer using 
a $\bra 2,2,2\ket_7$-decomposition. On an input of shape 
$65536\times 65536=2^{16}\times 2^{16}$, we parameterize the layers on the 
accelerators so that each accelerator performs two levels in the accelerator 
serial layer ($d^{(\se)}=2$) using a $\bra 2,2,2\ket_7$-decomposition, and 
then proceeds with the accelerator parallel layer. That is, the parallel layer
is started with inputs of shape 
$16384\times 16384=2^{14}\times 2^{14}$ (32 MiB per matrix).

\subsection{Boolean Multiplication}

We also provide reference implementations for both Boolean and binary 
multiplication using the elementary cubic algorithm distributed to the
accelerators. For an $s\times t$ by $t\times u$ multiplication to yield 
an $s\times u$ product, the $N$ accelerators each work with subproblems of 
size $\underline s\times \underline t$ by $\underline t\times \underline u$ 
such that $\underline s$ divides $s$, $\underline t$ divides $t$, and 
$\underline u$ divides $u$. That is, the total workload of volume $stu$ 
(in precise terms, $stu+s(t-1)u$ bit operations) gets executed on the 
accelerators so that each unit of work for an accelerator
has volume $\underline{s}\underline{t}\underline{u}$. 
Each accelerator executes a thread array of shape 
$\frac{\underline{s}}{M}\times \frac{\underline{u}}{M}\times \frac{M\cdot M}{W}$
such that the least significant $M^2/W$ threads are responsible for
aggregating one $M\times M$ block of the $\underline{s}\times\underline{u}$
output, with low-level vectorization and vector shuffle operations used for
each work unit of volume $M^3$. 
Empirically, $\underline{s}=\underline{t}=\underline{u}=131072$ and $M=64$
give best parameterization for our target hardware with $W=128$ and 
$V=\frac{M\cdot M}{W}=32$. Coordination at the host is simplified compared with 
Algorithm~\ref{alg:multiacceleratorstrassen}; namely, we partition the 
$s\times u$ result matrix to disjoint segments of size 
$\underline{s}\times\underline{u}$, and set up threads on the host so that 
a single thread is responsible for each segment, which eliminates the need for
synchronization between threads to guard against race conditions. In addition,
we set up each host thread so that it integrates each 
$\underline{s}\times\underline{u}$ subresult it has downloaded from an 
accelerator in parallel when the next subproblem is being solved on the 
accelerator; this effectively hides the latency of integrating each subresult
because the time to solve each subproblem is greater than the time to 
integrate the result. 

\section{Experiments}
\label{sect:experiments}

This section describes the experiments we ran to evaluate the
performance of the framework described in the previous section. 
In particular, our goal is to observe that a careful implementation of 
the framework enables efficient use of current multiple-accelerator systems 
at input sizes that are close to the host memory capacity of the entire system. 

The implementation was written in
C++~\cite{ISO14882}, and the accelerator device routines were written
in CUDA~C~\cite{Nvidia:2018b}. Host-level parallelization was prepared
using the OpenMP API~\cite{OpenMP:2015}. We ran three sets of
experiments: (i) alternative-basis binary matrix multiplication,
(ii) classical binary matrix multiplication, and
(iii) Boolean matrix multiplication. 

The single-accelerator version was evaluated at all powers of two until
the subproblem size for the multiple-accelerator case. 
The multiple-accelerator case was evaluated up to one terabinary-bit (1 Tib)
input size ($n=2^{20}$).

\subsection{Hardware configuration}
The target hardware for our implementation was an NVIDIA DGX-1 system
with eight NVIDIA Tesla V100 accelerators, 512 GiB of memory, and
$2\times 20$ CPU cores. We also ran experiments 
on two less powerful configurations: One with four NVIDIA Tesla P100 
accelerators, 256 GiB of memory, and $2\times 12$ CPU cores, and one with 
four NVIDIA Tesla K80 accelerators, 128~GiB of memory, and $2\times 6$ CPU 
cores. A more detailed description of the hardware configurations is given in
Table~\ref{tbl:hardware}.

Table~\ref{tbl:peakperformance} lists technical details of the
accelerators: the number of cores, boost clock speed, and the peak
performance that presents an upper bound on the number of bit
operations (bops) that could be theoretically achieved by simultaneous use 
of all the accelerators in each configuration. In practice, this 
upper bound is unattainable due to thermal effects and the need to
coordinate data across the storage hierarchy ranging from per-thread 
registers to the global memory of each accelerator. Yet this upper bound
presents an uncompromising benchmark against which to measure the 
performance of the actual implementation. 

The power consumption of the systems used in our experiments is listed
in Table~\ref{tbl:power}. The table lists the nominal power
consumption of the CPUs and GPUs, and the system power intake as indicated
in the basic system documentation of each vendor. While one may perhaps 
expect the CPUs and GPUs to operate essentially at their maximum power, 
the given system power intake is perhaps expected to be higher 
than the actual power usage. While we would like to be able to measure the 
actual energy consumption by each system and its components during the
computation more precisely, such instrumentation has been unavailable to us, 
so in the present experiments we resort to using the tabulated power values 
as the mean power over the duration of each computation, which then gives us 
a crude approximation of the energy used by each computation. In particular,
we would like to highlight the importance of implementation engineering 
for low energy consumption as an important goal beyond the present study, 
with the hope of having in the future available more fine-grained measurement
tools. 

Finally, to record a rough benchmark of the performance of the different
system components in our main target configuration, measurements on the 
bandwith for transferring data within host memory, and between the accelerator 
device and the host system are shown in Tables~\ref{table:memorybenchmark} 
and~\ref{table:pciebenchmark}, respectively.

\begin{table}
  \begin{center}
    \caption{Comparison of the hardware used in our experiments.}
    \label{tbl:hardware}
    {\small 
  \begin{tabular}{p{7em}p{8em}p{8.5em}p{7em}l}
    \hline
    System & CPUs & CPU RAM & GPUs & RAM/GPU \\\hline
    Dell PowerEdge C4130 & $2\times 6$-core Xeon E5 2620v3 2.50~GHz (Haswell) & $8\times 16$~GiB = 128~GiB DDR4-2133 & $4\times 2$ Tesla K80 & 12~GiB GDDR5 \\
    Dell PowerEdge C4130 & $2\times 12$-core Xeon E5-2680v3 2.50~GHz (Haswell) & $16\times 16$~GiB = 256~GiB DDR4-2400 & $4 \times\textrm{Tesla}$ P100 & 16~GiB HBM2\\
    NVIDIA DGX-1   & $2\times 20$-core Xeon E5-2698v4 (Broadwell) & $16\times 32$~GiB = 512~GiB DDR4-2133 & $8\times\textrm{Tesla}$ V100 SXM2 & 16~GiB HBM2\\\hline
  \end{tabular}
}
\end{center}
\end{table}

\begin{table}
  \begin{center}
    \caption{Peak performance of the GPU accelerators used in our experiments. 
      This presents an upper bound on the number of bit operations that could
      theoretically be achieved by the hardware.}
    \label{tbl:peakperformance}
    \begin{tabular}{llll}
      \hline
      GPU       & \#Cores         & Boost clock & Peak performance  \\\hline
      $4\times\textrm{Tesla}$ K80~\cite{Nvidia:2015TESLAK80} & $8\cdot 2496$   & \phantom{1}875~MHz     & $8\cdot 2496 \cdot \phantom{1}875 \cdot 10^6 \cdot 32 \approx 5.60 \cdot 10^{14}$ bops  \\
      $4\times\textrm{Tesla}$ P100~\cite{Nvidia:2016TESLAP100} & $4\cdot 3584$  & 1480~MHz & $4\cdot 3584 \cdot 1480 \cdot 10^6 \cdot 32 \approx 6.79\cdot 10^{14}$ bops \\
      $8\times\textrm{Tesla}$ V100~\cite{Nvidia:2018d} & $8\cdot 5120$ & 1530~MHz & $8\cdot 5120 \cdot 1530 \cdot 10^6 \cdot 32 \approx 2.01\cdot 10^{15}$ bops  \\\hline
    \end{tabular}
    \end{center}
\end{table}

\begin{table}
  \begin{center}
    \caption{Power consumption of the system and the GPUs.}
    \label{tbl:power}
    \begin{tabular}{llll}
      \hline
      System     & CPU watts       & GPU watts       & System watts \\\hline
      C4130/K80  & $2\times \phantom{1}85$~W  & $4\times 300$~W & 2000~W \\
      C4130/P100 & $2\times 120$~W & $4\times 300$~W & 2000~W \\
      DGX-1/V100 & $2\times 135$~W & $8\times 300$~W & 3500~W \\\hline
    \end{tabular}
    \end{center}
\end{table}

\begin{table}
  \caption{Measured host-memory bandwiths of the DGX-1
    node. The values shown are averages of five consecutive
    repetitions.}
  \label{table:memorybenchmark}
  \begin{center}
    \begin{tabular}{lrr}
      \hline
      Benchmark & Single core & All cores \\\hline
      Read from linear addresses (consecutive 64-bit words) & 9.08 GiB/s & 37.28 GiB/s\\
      Write to linear addresses (consecutive 64-bit words) & 6.92 GiB/s & 19.93 GiB/s\\
      Read from random addresses (individual 64-bit words) & 0.17 GiB/s & 3.13 GiB/s\\
      Read from random addresses (full cache lines) & 0.78 GiB/s & 14.09 GiB/s\\\hline
    \end{tabular}
  \end{center}
\end{table}

\begin{table}
  \caption{Measured memory bandwiths of
    transferring data within a single GPU device or between the host and the
    GPU device. Measurements on the DGX-1 node.}
  \label{table:pciebenchmark}
  \begin{center}
    \begin{tabular}{lr}
      \hline
      Benchmark & Bandwith \\\hline
      Host to device & 10.5 GiB/s \\ 
      Device to host & 11.8 GiB/s \\ 
      Device to device & 704.9 GiB/s \\ 
      \hline
    \end{tabular}
  \end{center}
\end{table}

\begin{table}
  \caption{Running times for the routine that
    transposes each of the $64\times 64$ submatrices. The procedure is
    parallelized and works using 32-bit words as the base data
    type. All runtimes reported are the median of five repeats on the 
    DGX-1 node.}
  \label{tbl:transpose}
  \begin{center}
    {
      \begin{tabular}{rr}
        \hline
        $n$ & $64\times 64$ submatrix transpose time\\\hline
        1024 & $1.56 \cdot 10^{-5}$~s\\
        2048 & $3.69 \cdot 10^{-5}$~s\\
        4096 & $1.23 \cdot 10^{-4}$~s\\
        8192 & $4.63 \cdot 10^{-4}$~s\\
        16384 & $1.84 \cdot 10^{-3}$~s\\
        32768 & $7.45 \cdot 10^{-3}$~s\\
        65536 & $2.98 \cdot 10^{-2}$~s\\
        131072 & $1.29 \cdot 10^{-1}$~s\\
        262144 & $4.94 \cdot 10^{-1}$~s\\
        524288 & $2.03 \cdot 10^{0\phantom{-}}$~s\\
        1048576 & $7.99 \cdot 10^{0\phantom{-}}$~s\\
        \hline\end{tabular}
    }
  \end{center}
\end{table}

\begin{table}
  \caption{Scalability of running times for the various
    change-of-basis routines. The forward columns show the running times
    for the forward transformations (that is, corresponding to $\psi$
    and $\phi$) whereas the inverse columns show the inverse
    transformations (corresponding to $\chi$). For the chaining
    variant, measurements were made to confirm that accounting for
    transposed data in the right-hand operand has no effect on the
    running time. The procedures have been crafted using AVX2
    intrinsics and to work with 512-bit cache lines on the DGX-1
    host hardware. All runtimes reported are the median of five
    repeats on the DGX-1 node.}
  \label{tbl:cob}
  \begin{center}
    {
      \begin{tabular}{rrrrrr}
        \hline
        $n$ & Forward self-inverse & Inverse self-inverse & Forward chain left & Forward chain right & Inverse chain\\\hline
        1024 & $3.28 \cdot 10^{-5}$~s & $3.42 \cdot 10^{-5}$~s & $3.41 \cdot 10^{-5}$~s & $3.44 \cdot 10^{-5}$~s & $3.41 \cdot 10^{-5}$~s\\
        2048 & $4.70 \cdot 10^{-5}$~s & $4.88 \cdot 10^{-5}$~s & $5.02 \cdot 10^{-5}$~s & $5.13 \cdot 10^{-5}$~s & $5.00 \cdot 10^{-5}$~s\\
        4096 & $8.10 \cdot 10^{-5}$~s & $8.80 \cdot 10^{-5}$~s & $9.42 \cdot 10^{-5}$~s & $9.33 \cdot 10^{-5}$~s & $9.35 \cdot 10^{-5}$~s\\
        8192 & $1.80 \cdot 10^{-4}$~s & $2.09 \cdot 10^{-4}$~s & $2.04 \cdot 10^{-4}$~s & $2.07 \cdot 10^{-4}$~s & $2.05 \cdot 10^{-4}$~s\\
        16384 & $7.15 \cdot 10^{-4}$~s & $8.78 \cdot 10^{-4}$~s & $8.55 \cdot 10^{-4}$~s & $8.68 \cdot 10^{-4}$~s & $8.67 \cdot 10^{-4}$~s\\
        32768 & $2.06 \cdot 10^{-2}$~s & $2.56 \cdot 10^{-2}$~s & $2.59 \cdot 10^{-2}$~s & $2.61 \cdot 10^{-2}$~s & $2.59 \cdot 10^{-2}$~s\\
        65536 & $9.36 \cdot 10^{-2}$~s & $1.20 \cdot 10^{-1}$~s & $1.20 \cdot 10^{-1}$~s & $1.20 \cdot 10^{-1}$~s & $1.20 \cdot 10^{-1}$~s\\
        131072 & $4.10 \cdot 10^{-1}$~s & $5.30 \cdot 10^{-1}$~s & $5.27 \cdot 10^{-1}$~s & $5.30 \cdot 10^{-1}$~s & $5.26 \cdot 10^{-1}$~s\\
        262144 & $1.81 \cdot 10^{0\phantom{-}}$~s & $2.31 \cdot 10^{0\phantom{-}}$~s & $2.31 \cdot 10^{0\phantom{-}}$~s & $2.33 \cdot 10^{0\phantom{-}}$~s & $2.32 \cdot 10^{0\phantom{-}}$~s\\
        524288 & $7.82 \cdot 10^{0\phantom{-}}$~s & $1.01 \cdot 10^{1\phantom{-}}$~s & $1.01 \cdot 10^{1\phantom{-}}$~s & $1.01 \cdot 10^{1\phantom{-}}$~s & $1.01 \cdot 10^{1\phantom{-}}$~s\\
        1048576 & $3.45 \cdot 10^{1\phantom{-}}$~s & $4.41 \cdot 10^{1\phantom{-}}$~s & $4.46 \cdot 10^{1\phantom{-}}$~s & $4.46 \cdot 10^{1\phantom{-}}$~s & $4.40 \cdot 10^{1\phantom{-}}$~s\\
        \hline\end{tabular}
    }
  \end{center}
\end{table}

\subsection{Results}
Runtimes, effective bit operations per second, and energy consumption
at different sizes of input are shown in
Tables~\ref{tbl:binaryresults}, \ref{tbl:booleanresults},
\ref{tbl:strassenwinogradresults}, \ref{tbl:absinvresults},
and~\ref{tbl:abchainresults}.  All values shown are medians over five
consecutive runs.  Since the instrumentation does not permit us to
measure the actual power use, two different values are shown for the
energy consumption: one is computed assuming full CPU utilization with
nominal CPU wattage and the maximum wattage of a single GPU
accelerator times the number of GPUs in use, and the other by simply
taking the maximum power intake of the entire system in account. We
expect the actual energy consumption to lie somewhere between these
two values, since the former ignores macroscopic factors such as
memory and cooling, and the latter is simply the maximum amount of
power that the entire system can supply.

In Tables~\ref{tbl:binaryresults} and~\ref{tbl:booleanresults}, given
runtime $T$, the exact bit operation count of $2n^3-n^2$ is used to
compute $(2n^3-n^2)/T$, the number of individual bit operations
performed in a second on average. The same number is also used in
Tables~\ref{tbl:strassenwinogradresults}, \ref{tbl:absinvresults},
and~\ref{tbl:abchainresults} to compute the \emph{effective} bit
operation count; that is, to highlight the relative difference in
performance, we show how many elementary bit operations per second a
classical implementation operating on individual bits would have to
perform to achieve the same wall-clock runtime as the
alternative-basis design.

The times reported here do not include the time required for data
permutation or change of basis operations at the host; it is assumed
that the data is already in the desired format. Table~\ref{tbl:cob}
shows the runtime of an implementation of the change of basis with
AVX2 vector-intrinsics on the DGX-1 host. It can be seen that, in the
case of the best-performing algorithm (with self-inversion), the
forward-transformation takes approximately 34.5 seconds per 1-Tib
input matrix, and 44.1 seconds for the inverse transformation of the
result. The favorable difference is due to the structure of the
transformation which means that only 1 of the 4 subarrays needs to be
modified at any level of recursion. Also, to preserve linearity of
access in memory, we assume that the right-hand-side operand is
transposed. For Strassen-like algorithms, this needs to be only done
for the $64\times 64$ submatrices; we implemented a parallelized,
recursive transpose function using word-operations that can perform
the transpose for all $2^{28}$ submatrices of a 1-TiB input matrix in
7.99~s (median of five consecutive runs, see Table~\ref{tbl:transpose}
for scalability).

With two 1-Tib input matrices, on the DGX-1, the classical cubic algorithm runs 
for 1880 and 2110~seconds, and performs over 1.23 and 1.09~peta-bit-operations
per second, in the binary and Boolean cases, respectively. This means
we are able to achieve over 50~\% of the theoretical maximum peak
performance of 2.01~Pbops per second. At the assumed level of power
used, this means that one such multiplication takes at most 1.83 and
2.05~kWh of electricity, or 2.85 and 3.20~pJ/bop, in the binary and
Boolean cases, respectively. In both cases, a single V100 accelerator
processes a $131072\times 131072$-subproblem in just under 30~seconds,
meaning that the runtime obtained here is accelerator-bound. The time
required for data upload and download, and result integration of
individual subproblems is in the order of seconds, making it negligible 
in comparison and effectively hidden by the host-side pipeline design.

Due to their limited memory, the P100 and K80 systems can only be
evaluated up to $n=524288$. However, otherwise the results are very
similar to those of the DGX-1, apart from worse energy consumption and
slower processing power, owing to the older hardware in use.

As an interesting artifact of the low-level implementation, we see in the
Tables~\ref{tbl:binaryresults} and~\ref{tbl:booleanresults} that, with
the V100 and P100 accelerators, the Boolean version is slightly slower
than the binary version of the algorithm, even though the only
differences in code are with the replacement of the popcount-parity
check with the nonzero test, and the replacement of XOR with
OR. Furthermore, the difference is inverse with the K80 accelerator. This 
artifact appears to be due to the optimizations performed by
the NVIDIA compiler: On the P100 and the V100, the compiler compresses
several bitwise operations into one LOP3.LUT instruction, or
arbitrary trinary logical instructions, and for one reason or another,
the sequence is slightly more efficient in the XOR case. However,
these instructions are not available on the K80.

For binary multiplication, the Strassen-Winograd design in standard
basis executes in 2610~seconds, the alternative basis with self-inverse
in 821~seconds, and the alternative basis with chaining in
972~seconds, on 1-Tib inputs on the DGX-1, and achieve 0.8, 2.8, and
2.3~effective Pbops per second, respectively. The best alternative-basis 
variant exceeds the theoretical peak hardware performance of the elementary
algorithm by more than 30 \%. One alternative-basis multiplication
takes less than 0.8~kWh of electricity, or less than 1.26
pJ/bop. This is less than one half of that required by the cubic
multiplication algorithm. Furthermore, considering that a subproblem
of size $n=65536$ can be solved by an accelerator in 1.87~seconds,
this means that the available host-side bandwidth forms the bottleneck
(cf.~Tables~\ref{table:memorybenchmark} and
\ref{table:pciebenchmark}).  Indeed, if subproblem construction and
aggregation as well as the data transfer over the PCIe link were
completely free, the multiplication could be executed in
$\frac{7^4}{8}\times 1.87$~s $\approx 560$~s. This in particular
highlights the need to carefully engineer workloads and their
pipelining both at the host and at the accelerators to obtain balanced
overall performance from a multi-accelerator
system. Tables~\ref{tbl:absinvresults} and~\ref{tbl:abchainresults}
also show that the present approach generalizes to alternative 
multiple-accelerator configurations, even though the more limited memory 
on the K80 accelerators requires a smaller subproblem size.

\begin{table}
  \caption{Scalability of the running times, the bit operations per second, and the energy requirements for the classical (cubic) binary (AND/XOR) multiplication procedure. The subproblem size is 131072 for the V100 (DGX-1) and P100 accelerators, and 65536 for the K80. The number of bit operations is $2n^3-n^2$. The first two energy columns are computed assuming full usage of CPU Watts and the Wattage of a single GPU times the number GPUs in use. The latter two columns are computed using the full system power intake.}
  \label{tbl:binaryresults}
  \begin{center}
    {
      \small
      \begin{tabular}{rrrrrrrrr}
        \hline
        GPU & $n$ & $d$ & Runtime & Bop/s & Energy/bop & Total energy & System~en./bop & Total~system~en.\\\hline
        V100 & 1024 & 1 & $1.07 \cdot 10^{-4}$~s & 20.22~Tbop/s & 132.06~pJ/bop & $7.88 \cdot 10^{-8}$~kWh & 173.12~pJ/bop & $1.04 \cdot 10^{-7}$~kWh\\
        V100 & 2048 & 1 & $2.15 \cdot 10^{-4}$~s & 79.94~Tbop/s & 33.40~pJ/bop & $1.60 \cdot 10^{-7}$~kWh & 43.79~pJ/bop & $2.09 \cdot 10^{-7}$~kWh\\
        V100 & 4096 & 1 & $1.16 \cdot 10^{-3}$~s & 118.78~Tbop/s & 22.48~pJ/bop & $8.59 \cdot 10^{-7}$~kWh & 29.47~pJ/bop & $1.13 \cdot 10^{-6}$~kWh\\
        V100 & 8192 & 1 & $8.28 \cdot 10^{-3}$~s & 132.83~Tbop/s & 20.10~pJ/bop & $6.14 \cdot 10^{-6}$~kWh & 26.35~pJ/bop & $8.05 \cdot 10^{-6}$~kWh\\
        V100 & 16384 & 1 & $5.21 \cdot 10^{-2}$~s & 169.11~Tbop/s & 15.79~pJ/bop & $3.86 \cdot 10^{-5}$~kWh & 20.70~pJ/bop & $5.06 \cdot 10^{-5}$~kWh\\
        V100 & 32768 & 1 & $4.09 \cdot 10^{-1}$~s & 172.46~Tbop/s & 15.48~pJ/bop & $3.03 \cdot 10^{-4}$~kWh & 20.29~pJ/bop & $3.97 \cdot 10^{-4}$~kWh\\
        V100 & 65536 & 1 & $3.28 \cdot 10^{0\phantom{-}}$~s & 171.74~Tbop/s & 15.55~pJ/bop & $2.44 \cdot 10^{-3}$~kWh & 20.38~pJ/bop & $3.19 \cdot 10^{-3}$~kWh\\
        V100 & 131072 & 1 & $2.65 \cdot 10^{1\phantom{-}}$~s & 170.14~Tbop/s & 15.69~pJ/bop & $1.97 \cdot 10^{-2}$~kWh & 20.57~pJ/bop & $2.58 \cdot 10^{-2}$~kWh\\
        V100 & 262144 & 8 & $6.76 \cdot 10^{1\phantom{-}}$~s & 533.46~Tbop/s & 5.01~pJ/bop & $5.01 \cdot 10^{-2}$~kWh & 6.56~pJ/bop & $6.57 \cdot 10^{-2}$~kWh\\
        V100 & 524288 & 8 & $2.40 \cdot 10^{2\phantom{-}}$~s & 1204.30~Tbop/s & 2.22~pJ/bop & $1.78 \cdot 10^{-1}$~kWh & 2.91~pJ/bop & $2.33 \cdot 10^{-1}$~kWh\\
        V100 & 1048576 & 8 & $1.88 \cdot 10^{3\phantom{-}}$~s & 1230.18~Tbop/s & 2.17~pJ/bop & $1.40 \cdot 10^{0\phantom{-}}$~kWh & 2.85~pJ/bop & $1.83 \cdot 10^{0\phantom{-}}$~kWh\\
        P100 & 1024 & 1 & $6.22 \cdot 10^{-5}$~s & 34.56~Tbop/s & 41.67~pJ/bop & $2.49 \cdot 10^{-8}$~kWh & 57.87~pJ/bop & $3.46 \cdot 10^{-8}$~kWh\\
        P100 & 2048 & 1 & $2.96 \cdot 10^{-4}$~s & 58.21~Tbop/s & 24.74~pJ/bop & $1.19 \cdot 10^{-7}$~kWh & 34.36~pJ/bop & $1.64 \cdot 10^{-7}$~kWh\\
        P100 & 4096 & 1 & $1.88 \cdot 10^{-3}$~s & 73.31~Tbop/s & 19.64~pJ/bop & $7.50 \cdot 10^{-7}$~kWh & 27.28~pJ/bop & $1.05 \cdot 10^{-6}$~kWh\\
        P100 & 8192 & 1 & $1.25 \cdot 10^{-2}$~s & 88.56~Tbop/s & 16.26~pJ/bop & $4.97 \cdot 10^{-6}$~kWh & 22.58~pJ/bop & $6.90 \cdot 10^{-6}$~kWh\\
        P100 & 16384 & 1 & $8.64 \cdot 10^{-2}$~s & 101.90~Tbop/s & 14.13~pJ/bop & $3.46 \cdot 10^{-5}$~kWh & 19.63~pJ/bop & $4.80 \cdot 10^{-5}$~kWh\\
        P100 & 32768 & 1 & $6.83 \cdot 10^{-1}$~s & 103.03~Tbop/s & 13.98~pJ/bop & $2.74 \cdot 10^{-4}$~kWh & 19.41~pJ/bop & $3.80 \cdot 10^{-4}$~kWh\\
        P100 & 65536 & 1 & $5.46 \cdot 10^{0\phantom{-}}$~s & 103.15~Tbop/s & 13.96~pJ/bop & $2.19 \cdot 10^{-3}$~kWh & 19.39~pJ/bop & $3.04 \cdot 10^{-3}$~kWh\\
        P100 & 131072 & 1 & $4.37 \cdot 10^{1\phantom{-}}$~s & 103.21~Tbop/s & 13.95~pJ/bop & $1.75 \cdot 10^{-2}$~kWh & 19.38~pJ/bop & $2.43 \cdot 10^{-2}$~kWh\\
        P100 & 262144 & 4 & $9.57 \cdot 10^{1\phantom{-}}$~s & 376.85~Tbop/s & 3.82~pJ/bop & $3.83 \cdot 10^{-2}$~kWh & 5.31~pJ/bop & $5.32 \cdot 10^{-2}$~kWh\\
        P100 & 524288 & 4 & $7.24 \cdot 10^{2\phantom{-}}$~s & 398.11~Tbop/s & 3.62~pJ/bop & $2.90 \cdot 10^{-1}$~kWh & 5.02~pJ/bop & $4.03 \cdot 10^{-1}$~kWh\\
        K80 & 1024 & 1 & $2.27 \cdot 10^{-4}$~s & 9.47~Tbop/s & 144.69~pJ/bop & $8.63 \cdot 10^{-8}$~kWh & 211.23~pJ/bop & $1.26 \cdot 10^{-7}$~kWh\\
        K80 & 2048 & 1 & $1.95 \cdot 10^{-3}$~s & 8.83~Tbop/s & 155.16~pJ/bop & $7.41 \cdot 10^{-7}$~kWh & 226.50~pJ/bop & $1.09 \cdot 10^{-6}$~kWh\\
        K80 & 4096 & 1 & $9.94 \cdot 10^{-3}$~s & 13.84~Tbop/s & 99.01~pJ/bop & $3.78 \cdot 10^{-6}$~kWh & 144.54~pJ/bop & $5.52 \cdot 10^{-6}$~kWh\\
        K80 & 8192 & 1 & $6.44 \cdot 10^{-2}$~s & 17.09~Tbop/s & 80.15~pJ/bop & $2.45 \cdot 10^{-5}$~kWh & 117.00~pJ/bop & $3.58 \cdot 10^{-5}$~kWh\\
        K80 & 16384 & 1 & $4.28 \cdot 10^{-1}$~s & 20.59~Tbop/s & 66.52~pJ/bop & $1.63 \cdot 10^{-4}$~kWh & 97.11~pJ/bop & $2.38 \cdot 10^{-4}$~kWh\\
        K80 & 32768 & 1 & $3.43 \cdot 10^{0\phantom{-}}$~s & 20.55~Tbop/s & 66.68~pJ/bop & $1.31 \cdot 10^{-3}$~kWh & 97.34~pJ/bop & $1.91 \cdot 10^{-3}$~kWh\\
        K80 & 65536 & 1 & $2.86 \cdot 10^{1\phantom{-}}$~s & 19.70~Tbop/s & 69.54~pJ/bop & $1.09 \cdot 10^{-2}$~kWh & 101.52~pJ/bop & $1.59 \cdot 10^{-2}$~kWh\\
        K80 & 131072 & 8 & $6.34 \cdot 10^{1\phantom{-}}$~s & 71.11~Tbop/s & 19.27~pJ/bop & $2.42 \cdot 10^{-2}$~kWh & 28.13~pJ/bop & $3.52 \cdot 10^{-2}$~kWh\\
        K80 & 262144 & 8 & $2.51 \cdot 10^{2\phantom{-}}$~s & 143.73~Tbop/s & 9.53~pJ/bop & $9.54 \cdot 10^{-2}$~kWh & 13.91~pJ/bop & $1.40 \cdot 10^{-1}$~kWh\\
        K80 & 524288 & 8 & $1.98 \cdot 10^{3\phantom{-}}$~s & 146.29~Tbop/s & 9.36~pJ/bop & $7.50 \cdot 10^{-1}$~kWh & 13.67~pJ/bop & $1.10 \cdot 10^{0\phantom{-}}$~kWh\\
        \hline\end{tabular}
    }
  \end{center}
\end{table}

\begin{table}
  \caption{Scalability of the running times, the bit operations per second, and the energy requirements for the classic (cubic) Boolean (AND/OR) multiplication procedure. The subproblem size is 131072 for the V100 (DGX-1) and P100 accelerators, and 65536 for the K80. The number of bit operations is $2n^3-n^2$. The first two energy columns are computed assuming full usage of CPU Watts and the Wattage of a single GPU times the number GPUs in use. The latter two columns are computed using the full system power intake.}
  \label{tbl:booleanresults}
  \begin{center}
    {\small
      \begin{tabular}{rrrrrrrrr}
        \hline
        GPU & $n$ & $d$ & Runtime & Bop/s & Energy/bop & Total energy & System~en./bop & Total~system~en.\\\hline
        V100 & 1024 & 1 & $9.84 \cdot 10^{-5}$~s & 21.83~Tbop/s & 122.33~pJ/bop & $7.30 \cdot 10^{-8}$~kWh & 160.36~pJ/bop & $9.57 \cdot 10^{-8}$~kWh\\
        V100 & 2048 & 1 & $1.93 \cdot 10^{-4}$~s & 89.25~Tbop/s & 29.92~pJ/bop & $1.43 \cdot 10^{-7}$~kWh & 39.22~pJ/bop & $1.88 \cdot 10^{-7}$~kWh\\
        V100 & 4096 & 1 & $1.30 \cdot 10^{-3}$~s & 106.45~Tbop/s & 25.08~pJ/bop & $9.58 \cdot 10^{-7}$~kWh & 32.88~pJ/bop & $1.26 \cdot 10^{-6}$~kWh\\
        V100 & 8192 & 1 & $7.86 \cdot 10^{-3}$~s & 140.00~Tbop/s & 19.07~pJ/bop & $5.83 \cdot 10^{-6}$~kWh & 25.00~pJ/bop & $7.64 \cdot 10^{-6}$~kWh\\
        V100 & 16384 & 1 & $5.94 \cdot 10^{-2}$~s & 148.16~Tbop/s & 18.02~pJ/bop & $4.41 \cdot 10^{-5}$~kWh & 23.62~pJ/bop & $5.78 \cdot 10^{-5}$~kWh\\
        V100 & 32768 & 1 & $4.69 \cdot 10^{-1}$~s & 150.33~Tbop/s & 17.76~pJ/bop & $3.48 \cdot 10^{-4}$~kWh & 23.28~pJ/bop & $4.56 \cdot 10^{-4}$~kWh\\
        V100 & 65536 & 1 & $3.73 \cdot 10^{0\phantom{-}}$~s & 151.07~Tbop/s & 17.67~pJ/bop & $2.77 \cdot 10^{-3}$~kWh & 23.17~pJ/bop & $3.63 \cdot 10^{-3}$~kWh\\
        V100 & 131072 & 1 & $2.98 \cdot 10^{1\phantom{-}}$~s & 151.27~Tbop/s & 17.65~pJ/bop & $2.21 \cdot 10^{-2}$~kWh & 23.14~pJ/bop & $2.90 \cdot 10^{-2}$~kWh\\
        V100 & 262144 & 8 & $7.51 \cdot 10^{1\phantom{-}}$~s & 479.81~Tbop/s & 5.56~pJ/bop & $5.57 \cdot 10^{-2}$~kWh & 7.29~pJ/bop & $7.31 \cdot 10^{-2}$~kWh\\
        V100 & 524288 & 8 & $2.71 \cdot 10^{2\phantom{-}}$~s & 1065.59~Tbop/s & 2.51~pJ/bop & $2.01 \cdot 10^{-1}$~kWh & 3.28~pJ/bop & $2.63 \cdot 10^{-1}$~kWh\\
        V100 & 1048576 & 8 & $2.11 \cdot 10^{3\phantom{-}}$~s & 1094.98~Tbop/s & 2.44~pJ/bop & $1.57 \cdot 10^{0\phantom{-}}$~kWh & 3.20~pJ/bop & $2.05 \cdot 10^{0\phantom{-}}$~kWh\\
        P100 & 1024 & 1 & $5.09 \cdot 10^{-5}$~s & 42.24~Tbop/s & 34.09~pJ/bop & $2.04 \cdot 10^{-8}$~kWh & 47.35~pJ/bop & $2.83 \cdot 10^{-8}$~kWh\\
        P100 & 2048 & 1 & $2.72 \cdot 10^{-4}$~s & 63.35~Tbop/s & 22.73~pJ/bop & $1.09 \cdot 10^{-7}$~kWh & 31.57~pJ/bop & $1.51 \cdot 10^{-7}$~kWh\\
        P100 & 4096 & 1 & $1.62 \cdot 10^{-3}$~s & 85.13~Tbop/s & 16.91~pJ/bop & $6.46 \cdot 10^{-7}$~kWh & 23.49~pJ/bop & $8.97 \cdot 10^{-7}$~kWh\\
        P100 & 8192 & 1 & $1.16 \cdot 10^{-2}$~s & 95.10~Tbop/s & 15.14~pJ/bop & $4.63 \cdot 10^{-6}$~kWh & 21.03~pJ/bop & $6.43 \cdot 10^{-6}$~kWh\\
        P100 & 16384 & 1 & $8.90 \cdot 10^{-2}$~s & 98.89~Tbop/s & 14.56~pJ/bop & $3.56 \cdot 10^{-5}$~kWh & 20.22~pJ/bop & $4.95 \cdot 10^{-5}$~kWh\\
        P100 & 32768 & 1 & $7.09 \cdot 10^{-1}$~s & 99.29~Tbop/s & 14.50~pJ/bop & $2.84 \cdot 10^{-4}$~kWh & 20.14~pJ/bop & $3.94 \cdot 10^{-4}$~kWh\\
        P100 & 65536 & 1 & $5.67 \cdot 10^{0\phantom{-}}$~s & 99.40~Tbop/s & 14.49~pJ/bop & $2.27 \cdot 10^{-3}$~kWh & 20.12~pJ/bop & $3.15 \cdot 10^{-3}$~kWh\\
        P100 & 131072 & 1 & $4.53 \cdot 10^{1\phantom{-}}$~s & 99.45~Tbop/s & 14.48~pJ/bop & $1.82 \cdot 10^{-2}$~kWh & 20.11~pJ/bop & $2.52 \cdot 10^{-2}$~kWh\\
        P100 & 262144 & 4 & $9.89 \cdot 10^{1\phantom{-}}$~s & 364.55~Tbop/s & 3.95~pJ/bop & $3.96 \cdot 10^{-2}$~kWh & 5.49~pJ/bop & $5.50 \cdot 10^{-2}$~kWh\\
        P100 & 524288 & 4 & $7.50 \cdot 10^{2\phantom{-}}$~s & 384.46~Tbop/s & 3.75~pJ/bop & $3.00 \cdot 10^{-1}$~kWh & 5.20~pJ/bop & $4.17 \cdot 10^{-1}$~kWh\\
        K80 & 1024 & 1 & $2.15 \cdot 10^{-4}$~s & 10.03~Tbop/s & 136.60~pJ/bop & $8.15 \cdot 10^{-8}$~kWh & 199.41~pJ/bop & $1.19 \cdot 10^{-7}$~kWh\\
        K80 & 2048 & 1 & $1.21 \cdot 10^{-3}$~s & 14.22~Tbop/s & 96.35~pJ/bop & $4.60 \cdot 10^{-7}$~kWh & 140.66~pJ/bop & $6.72 \cdot 10^{-7}$~kWh\\
        K80 & 4096 & 1 & $8.08 \cdot 10^{-3}$~s & 17.02~Tbop/s & 80.48~pJ/bop & $3.08 \cdot 10^{-6}$~kWh & 117.49~pJ/bop & $4.49 \cdot 10^{-6}$~kWh\\
        K80 & 8192 & 1 & $5.52 \cdot 10^{-2}$~s & 19.92~Tbop/s & 68.77~pJ/bop & $2.11 \cdot 10^{-5}$~kWh & 100.40~pJ/bop & $3.07 \cdot 10^{-5}$~kWh\\
        K80 & 16384 & 1 & $3.26 \cdot 10^{-1}$~s & 27.06~Tbop/s & 50.62~pJ/bop & $1.24 \cdot 10^{-4}$~kWh & 73.90~pJ/bop & $1.81 \cdot 10^{-4}$~kWh\\
        K80 & 32768 & 1 & $2.53 \cdot 10^{0\phantom{-}}$~s & 27.82~Tbop/s & 49.24~pJ/bop & $9.63 \cdot 10^{-4}$~kWh & 71.88~pJ/bop & $1.41 \cdot 10^{-3}$~kWh\\
        K80 & 65536 & 1 & $2.11 \cdot 10^{1\phantom{-}}$~s & 26.72~Tbop/s & 51.27~pJ/bop & $8.02 \cdot 10^{-3}$~kWh & 74.84~pJ/bop & $1.18 \cdot 10^{-2}$~kWh\\
        K80 & 131072 & 8 & $4.63 \cdot 10^{1\phantom{-}}$~s & 97.47~Tbop/s & 14.06~pJ/bop & $1.76 \cdot 10^{-2}$~kWh & 20.52~pJ/bop & $2.57 \cdot 10^{-2}$~kWh\\
        K80 & 262144 & 8 & $1.80 \cdot 10^{2\phantom{-}}$~s & 200.24~Tbop/s & 6.84~pJ/bop & $6.85 \cdot 10^{-2}$~kWh & 9.99~pJ/bop & $10.00 \cdot 10^{-2}$~kWh\\
        K80 & 524288 & 8 & $1.41 \cdot 10^{3\phantom{-}}$~s & 205.47~Tbop/s & 6.67~pJ/bop & $5.34 \cdot 10^{-1}$~kWh & 9.73~pJ/bop & $7.80 \cdot 10^{-1}$~kWh\\
        \hline\end{tabular}
    }
  \end{center}
\end{table}

\begin{table}
  \caption{Scalability the running times, the effective bit operations
    per second, and the energy requirements for the standard-basis
    binary Strassen-Winograd multiplication procedure. The subproblem
    size is 65536 for the V100 (DGX-1) and P100 accelerators, and
    32768 for the K80. The effective number of bit operations is
    computed with the same value of $2n^3-n^2$ as in the cubic case to
    highlight the relative difference in performance. The first two
    energy columns are computed assuming full usage of CPU Watts and
    the Wattage of a single GPU times the number GPUs in use. The
    latter two columns are computed using the full system power
    intake. The right-hand-side input matrix is assumed to be
    pre-transposed; the times reported here do not include the
    transposition of submatrices.}
  \label{tbl:strassenwinogradresults}
  \begin{center}
    {\small
      \begin{tabular}{rrrrrrrrr}
        \hline
        GPU & $n$ & $d$ & Runtime & Effective bop/s & Energy/bop & Total energy & System~en./bop & Total~system~en.\\\hline
        V100 & 1024 & 1 & $1.67 \cdot 10^{-4}$~s & 12.92~Tbop/s & 206.67~pJ/bop & $1.24 \cdot 10^{-7}$~kWh & 270.92~pJ/bop & $1.62 \cdot 10^{-7}$~kWh\\
        V100 & 2048 & 1 & $2.72 \cdot 10^{-4}$~s & 63.24~Tbop/s & 42.22~pJ/bop & $2.02 \cdot 10^{-7}$~kWh & 55.34~pJ/bop & $2.65 \cdot 10^{-7}$~kWh\\
        V100 & 4096 & 1 & $1.05 \cdot 10^{-3}$~s & 132.07~Tbop/s & 20.22~pJ/bop & $7.72 \cdot 10^{-7}$~kWh & 26.50~pJ/bop & $1.02 \cdot 10^{-6}$~kWh\\
        V100 & 8192 & 1 & $5.86 \cdot 10^{-3}$~s & 187.69~Tbop/s & 14.23~pJ/bop & $4.35 \cdot 10^{-6}$~kWh & 18.65~pJ/bop & $5.70 \cdot 10^{-6}$~kWh\\
        V100 & 16384 & 1 & $3.97 \cdot 10^{-2}$~s & 222.03~Tbop/s & 12.03~pJ/bop & $2.94 \cdot 10^{-5}$~kWh & 15.76~pJ/bop & $3.86 \cdot 10^{-5}$~kWh\\
        V100 & 32768 & 1 & $2.81 \cdot 10^{-1}$~s & 251.04~Tbop/s & 10.64~pJ/bop & $2.08 \cdot 10^{-4}$~kWh & 13.94~pJ/bop & $2.73 \cdot 10^{-4}$~kWh\\
        V100 & 65536 & 1 & $1.97 \cdot 10^{0\phantom{-}}$~s & 285.95~Tbop/s & 9.34~pJ/bop & $1.47 \cdot 10^{-3}$~kWh & 12.24~pJ/bop & $1.92 \cdot 10^{-3}$~kWh\\
        V100 & 131072 & 8 & $3.74 \cdot 10^{0\phantom{-}}$~s & 1205.38~Tbop/s & 2.22~pJ/bop & $2.78 \cdot 10^{-3}$~kWh & 2.90~pJ/bop & $3.64 \cdot 10^{-3}$~kWh\\
        V100 & 262144 & 8 & $2.34 \cdot 10^{1\phantom{-}}$~s & 1543.54~Tbop/s & 1.73~pJ/bop & $1.74 \cdot 10^{-2}$~kWh & 2.27~pJ/bop & $2.27 \cdot 10^{-2}$~kWh\\
        V100 & 524288 & 8 & $2.66 \cdot 10^{2\phantom{-}}$~s & 1086.54~Tbop/s & 2.46~pJ/bop & $1.97 \cdot 10^{-1}$~kWh & 3.22~pJ/bop & $2.58 \cdot 10^{-1}$~kWh\\
        V100 & 1048576 & 8 & $2.61 \cdot 10^{3\phantom{-}}$~s & 884.18~Tbop/s & 3.02~pJ/bop & $1.94 \cdot 10^{0\phantom{-}}$~kWh & 3.96~pJ/bop & $2.54 \cdot 10^{0\phantom{-}}$~kWh\\
        P100 & 1024 & 1 & $1.39 \cdot 10^{-4}$~s & 15.54~Tbop/s & 92.64~pJ/bop & $5.53 \cdot 10^{-8}$~kWh & 128.67~pJ/bop & $7.68 \cdot 10^{-8}$~kWh\\
        P100 & 2048 & 1 & $3.34 \cdot 10^{-4}$~s & 51.50~Tbop/s & 27.96~pJ/bop & $1.34 \cdot 10^{-7}$~kWh & 38.84~pJ/bop & $1.86 \cdot 10^{-7}$~kWh\\
        P100 & 4096 & 1 & $1.53 \cdot 10^{-3}$~s & 90.17~Tbop/s & 15.97~pJ/bop & $6.10 \cdot 10^{-7}$~kWh & 22.18~pJ/bop & $8.47 \cdot 10^{-7}$~kWh\\
        P100 & 8192 & 1 & $9.12 \cdot 10^{-3}$~s & 120.55~Tbop/s & 11.94~pJ/bop & $3.65 \cdot 10^{-6}$~kWh & 16.59~pJ/bop & $5.07 \cdot 10^{-6}$~kWh\\
        P100 & 16384 & 1 & $5.76 \cdot 10^{-2}$~s & 152.77~Tbop/s & 9.43~pJ/bop & $2.31 \cdot 10^{-5}$~kWh & 13.09~pJ/bop & $3.20 \cdot 10^{-5}$~kWh\\
        P100 & 32768 & 1 & $4.08 \cdot 10^{-1}$~s & 172.84~Tbop/s & 8.33~pJ/bop & $1.63 \cdot 10^{-4}$~kWh & 11.57~pJ/bop & $2.27 \cdot 10^{-4}$~kWh\\
        P100 & 65536 & 1 & $2.87 \cdot 10^{0\phantom{-}}$~s & 196.61~Tbop/s & 7.32~pJ/bop & $1.15 \cdot 10^{-3}$~kWh & 10.17~pJ/bop & $1.60 \cdot 10^{-3}$~kWh\\
        P100 & 131072 & 4 & $7.37 \cdot 10^{0\phantom{-}}$~s & 611.79~Tbop/s & 2.35~pJ/bop & $2.95 \cdot 10^{-3}$~kWh & 3.27~pJ/bop & $4.09 \cdot 10^{-3}$~kWh\\
        P100 & 262144 & 4 & $4.37 \cdot 10^{1\phantom{-}}$~s & 824.68~Tbop/s & 1.75~pJ/bop & $1.75 \cdot 10^{-2}$~kWh & 2.43~pJ/bop & $2.43 \cdot 10^{-2}$~kWh\\
        P100 & 524288 & 4 & $3.02 \cdot 10^{2\phantom{-}}$~s & 956.67~Tbop/s & 1.51~pJ/bop & $1.21 \cdot 10^{-1}$~kWh & 2.09~pJ/bop & $1.68 \cdot 10^{-1}$~kWh\\
        K80 & 1024 & 1 & $3.07 \cdot 10^{-4}$~s & 7.00~Tbop/s & 195.67~pJ/bop & $1.17 \cdot 10^{-7}$~kWh & 285.64~pJ/bop & $1.71 \cdot 10^{-7}$~kWh\\
        K80 & 2048 & 1 & $1.19 \cdot 10^{-3}$~s & 14.50~Tbop/s & 94.47~pJ/bop & $4.51 \cdot 10^{-7}$~kWh & 137.91~pJ/bop & $6.58 \cdot 10^{-7}$~kWh\\
        K80 & 4096 & 1 & $6.96 \cdot 10^{-3}$~s & 19.77~Tbop/s & 69.31~pJ/bop & $2.65 \cdot 10^{-6}$~kWh & 101.18~pJ/bop & $3.87 \cdot 10^{-6}$~kWh\\
        K80 & 8192 & 1 & $4.08 \cdot 10^{-2}$~s & 26.99~Tbop/s & 50.76~pJ/bop & $1.56 \cdot 10^{-5}$~kWh & 74.10~pJ/bop & $2.27 \cdot 10^{-5}$~kWh\\
        K80 & 16384 & 1 & $2.35 \cdot 10^{-1}$~s & 37.44~Tbop/s & 36.59~pJ/bop & $8.95 \cdot 10^{-5}$~kWh & 53.42~pJ/bop & $1.31 \cdot 10^{-4}$~kWh\\
        K80 & 32768 & 1 & $1.62 \cdot 10^{0\phantom{-}}$~s & 43.51~Tbop/s & 31.49~pJ/bop & $6.16 \cdot 10^{-4}$~kWh & 45.97~pJ/bop & $8.99 \cdot 10^{-4}$~kWh\\
        K80 & 65536 & 8 & $2.34 \cdot 10^{0\phantom{-}}$~s & 241.49~Tbop/s & 5.67~pJ/bop & $8.88 \cdot 10^{-4}$~kWh & 8.28~pJ/bop & $1.30 \cdot 10^{-3}$~kWh\\
        K80 & 131072 & 8 & $1.36 \cdot 10^{1\phantom{-}}$~s & 331.26~Tbop/s & 4.14~pJ/bop & $5.18 \cdot 10^{-3}$~kWh & 6.04~pJ/bop & $7.56 \cdot 10^{-3}$~kWh\\
        K80 & 262144 & 8 & $1.33 \cdot 10^{2\phantom{-}}$~s & 272.80~Tbop/s & 5.02~pJ/bop & $5.03 \cdot 10^{-2}$~kWh & 7.33~pJ/bop & $7.34 \cdot 10^{-2}$~kWh\\
        K80 & 524288 & 8 & $1.60 \cdot 10^{3\phantom{-}}$~s & 180.20~Tbop/s & 7.60~pJ/bop & $6.09 \cdot 10^{-1}$~kWh & 11.10~pJ/bop & $8.89 \cdot 10^{-1}$~kWh\\
        \hline\end{tabular}
    }
  \end{center}
\end{table}

\begin{table}
  \caption{Scalability the running times, the effective bit operations
    per second, and the energy requirements for the alternative-basis
    (with self-inversion) binary multiplication procedure. The
    subproblem size is 65536 for the V100 (DGX-1) and P100
    accelerators, and 32768 for the K80. The effective number of bit
    operations is computed with the same value of $2n^3-n^2$ as in the
    cubic case to highlight the relative difference in
    performance. The first two energy columns are computed assuming
    full usage of CPU Watts and the Wattage of a single GPU times the
    number GPUs in use. The latter two columns are computed using the
    full system power intake. The right-hand-side input matrix is
    assumed to be pre-transposed and in the alternate basis; the times
    reported here do not include the transposition of submatrices or
    the change of basis.}
  \label{tbl:absinvresults}
  \begin{center}
    {\small
      \begin{tabular}{rrrrrrrrr}
        \hline
        GPU & $n$ & $d$ & Runtime & Effective bop/s & Energy/bop & Total energy & System~en./bop & Total~system~en.\\\hline
        V100 & 1024 & 1 & $1.51 \cdot 10^{-4}$~s & 14.26~Tbop/s & 187.30~pJ/bop & $1.12 \cdot 10^{-7}$~kWh & 245.53~pJ/bop & $1.47 \cdot 10^{-7}$~kWh\\
        V100 & 2048 & 1 & $2.45 \cdot 10^{-4}$~s & 70.36~Tbop/s & 37.95~pJ/bop & $1.82 \cdot 10^{-7}$~kWh & 49.74~pJ/bop & $2.38 \cdot 10^{-7}$~kWh\\
        V100 & 4096 & 1 & $9.97 \cdot 10^{-4}$~s & 137.96~Tbop/s & 19.35~pJ/bop & $7.39 \cdot 10^{-7}$~kWh & 25.37~pJ/bop & $9.69 \cdot 10^{-7}$~kWh\\
        V100 & 8192 & 1 & $5.55 \cdot 10^{-3}$~s & 198.32~Tbop/s & 13.46~pJ/bop & $4.12 \cdot 10^{-6}$~kWh & 17.65~pJ/bop & $5.39 \cdot 10^{-6}$~kWh\\
        V100 & 16384 & 1 & $3.76 \cdot 10^{-2}$~s & 234.14~Tbop/s & 11.40~pJ/bop & $2.79 \cdot 10^{-5}$~kWh & 14.95~pJ/bop & $3.66 \cdot 10^{-5}$~kWh\\
        V100 & 32768 & 1 & $2.66 \cdot 10^{-1}$~s & 265.39~Tbop/s & 10.06~pJ/bop & $1.97 \cdot 10^{-4}$~kWh & 13.19~pJ/bop & $2.58 \cdot 10^{-4}$~kWh\\
        V100 & 65536 & 1 & $1.87 \cdot 10^{0\phantom{-}}$~s & 301.77~Tbop/s & 8.85~pJ/bop & $1.39 \cdot 10^{-3}$~kWh & 11.60~pJ/bop & $1.82 \cdot 10^{-3}$~kWh\\
        V100 & 131072 & 8 & $3.33 \cdot 10^{0\phantom{-}}$~s & 1356.04~Tbop/s & 1.97~pJ/bop & $2.47 \cdot 10^{-3}$~kWh & 2.58~pJ/bop & $3.23 \cdot 10^{-3}$~kWh\\
        V100 & 262144 & 8 & $1.83 \cdot 10^{1\phantom{-}}$~s & 1970.12~Tbop/s & 1.36~pJ/bop & $1.36 \cdot 10^{-2}$~kWh & 1.78~pJ/bop & $1.78 \cdot 10^{-2}$~kWh\\
        V100 & 524288 & 8 & $1.19 \cdot 10^{2\phantom{-}}$~s & 2440.74~Tbop/s & 1.09~pJ/bop & $8.76 \cdot 10^{-2}$~kWh & 1.43~pJ/bop & $1.15 \cdot 10^{-1}$~kWh\\
        V100 & 1048576 & 8 & $8.21 \cdot 10^{2\phantom{-}}$~s & 2809.41~Tbop/s & 0.95~pJ/bop & $6.09 \cdot 10^{-1}$~kWh & 1.25~pJ/bop & $7.98 \cdot 10^{-1}$~kWh\\
        P100 & 1024 & 1 & $1.18 \cdot 10^{-4}$~s & 18.22~Tbop/s & 79.02~pJ/bop & $4.72 \cdot 10^{-8}$~kWh & 109.75~pJ/bop & $6.55 \cdot 10^{-8}$~kWh\\
        P100 & 2048 & 1 & $2.94 \cdot 10^{-4}$~s & 58.49~Tbop/s & 24.62~pJ/bop & $1.18 \cdot 10^{-7}$~kWh & 34.19~pJ/bop & $1.64 \cdot 10^{-7}$~kWh\\
        P100 & 4096 & 1 & $1.38 \cdot 10^{-3}$~s & 100.27~Tbop/s & 14.36~pJ/bop & $5.49 \cdot 10^{-7}$~kWh & 19.95~pJ/bop & $7.62 \cdot 10^{-7}$~kWh\\
        P100 & 8192 & 1 & $8.31 \cdot 10^{-3}$~s & 132.34~Tbop/s & 10.88~pJ/bop & $3.33 \cdot 10^{-6}$~kWh & 15.11~pJ/bop & $4.62 \cdot 10^{-6}$~kWh\\
        P100 & 16384 & 1 & $5.37 \cdot 10^{-2}$~s & 163.82~Tbop/s & 8.79~pJ/bop & $2.15 \cdot 10^{-5}$~kWh & 12.21~pJ/bop & $2.99 \cdot 10^{-5}$~kWh\\
        P100 & 32768 & 1 & $3.79 \cdot 10^{-1}$~s & 185.81~Tbop/s & 7.75~pJ/bop & $1.52 \cdot 10^{-4}$~kWh & 10.76~pJ/bop & $2.11 \cdot 10^{-4}$~kWh\\
        P100 & 65536 & 1 & $2.67 \cdot 10^{0\phantom{-}}$~s & 211.34~Tbop/s & 6.81~pJ/bop & $1.07 \cdot 10^{-3}$~kWh & 9.46~pJ/bop & $1.48 \cdot 10^{-3}$~kWh\\
        P100 & 131072 & 4 & $7.12 \cdot 10^{0\phantom{-}}$~s & 633.00~Tbop/s & 2.27~pJ/bop & $2.85 \cdot 10^{-3}$~kWh & 3.16~pJ/bop & $3.96 \cdot 10^{-3}$~kWh\\
        P100 & 262144 & 4 & $4.12 \cdot 10^{1\phantom{-}}$~s & 876.27~Tbop/s & 1.64~pJ/bop & $1.65 \cdot 10^{-2}$~kWh & 2.28~pJ/bop & $2.29 \cdot 10^{-2}$~kWh\\
        P100 & 524288 & 4 & $2.60 \cdot 10^{2\phantom{-}}$~s & 1110.41~Tbop/s & 1.30~pJ/bop & $1.04 \cdot 10^{-1}$~kWh & 1.80~pJ/bop & $1.45 \cdot 10^{-1}$~kWh\\
        K80 & 1024 & 1 & $2.87 \cdot 10^{-4}$~s & 7.50~Tbop/s & 182.63~pJ/bop & $1.09 \cdot 10^{-7}$~kWh & 266.62~pJ/bop & $1.59 \cdot 10^{-7}$~kWh\\
        K80 & 2048 & 1 & $1.12 \cdot 10^{-3}$~s & 15.44~Tbop/s & 88.74~pJ/bop & $4.24 \cdot 10^{-7}$~kWh & 129.54~pJ/bop & $6.19 \cdot 10^{-7}$~kWh\\
        K80 & 4096 & 1 & $6.56 \cdot 10^{-3}$~s & 20.95~Tbop/s & 65.38~pJ/bop & $2.50 \cdot 10^{-6}$~kWh & 95.45~pJ/bop & $3.65 \cdot 10^{-6}$~kWh\\
        K80 & 8192 & 1 & $4.17 \cdot 10^{-2}$~s & 26.40~Tbop/s & 51.89~pJ/bop & $1.59 \cdot 10^{-5}$~kWh & 75.75~pJ/bop & $2.32 \cdot 10^{-5}$~kWh\\
        K80 & 16384 & 1 & $2.24 \cdot 10^{-1}$~s & 39.43~Tbop/s & 34.75~pJ/bop & $8.50 \cdot 10^{-5}$~kWh & 50.73~pJ/bop & $1.24 \cdot 10^{-4}$~kWh\\
        K80 & 32768 & 1 & $1.52 \cdot 10^{0\phantom{-}}$~s & 46.45~Tbop/s & 29.49~pJ/bop & $5.77 \cdot 10^{-4}$~kWh & 43.05~pJ/bop & $8.42 \cdot 10^{-4}$~kWh\\
        K80 & 65536 & 8 & $2.04 \cdot 10^{0\phantom{-}}$~s & 276.04~Tbop/s & 4.96~pJ/bop & $7.77 \cdot 10^{-4}$~kWh & 7.25~pJ/bop & $1.14 \cdot 10^{-3}$~kWh\\
        K80 & 131072 & 8 & $1.24 \cdot 10^{1\phantom{-}}$~s & 365.09~Tbop/s & 3.75~pJ/bop & $4.70 \cdot 10^{-3}$~kWh & 5.48~pJ/bop & $6.86 \cdot 10^{-3}$~kWh\\
        K80 & 262144 & 8 & $7.79 \cdot 10^{1\phantom{-}}$~s & 462.55~Tbop/s & 2.96~pJ/bop & $2.97 \cdot 10^{-2}$~kWh & 4.32~pJ/bop & $4.33 \cdot 10^{-2}$~kWh\\
        K80 & 524288 & 8 & $5.64 \cdot 10^{2\phantom{-}}$~s & 511.71~Tbop/s & 2.68~pJ/bop & $2.15 \cdot 10^{-1}$~kWh & 3.91~pJ/bop & $3.13 \cdot 10^{-1}$~kWh\\
        \hline\end{tabular}
    }
  \end{center}
\end{table}

\begin{table}
  \caption{Scalability the running times, the effective bit operations
    per second, and the energy requirements for the alternative-basis
    (with chaining) binary multiplication procedure. The subproblem
    size is 65536 for the V100 (DGX-1) and P100 accelerators, and
    32768 for the K80. The effective number of bit operations is
    computed with the same value of $2n^3-n^2$ as in the cubic case to
    highlight the relative difference in performance. The first two
    energy columns are computed assuming full usage of CPU Watts and
    the Wattage of a single GPU times the number GPUs in use. The
    latter two columns are computed using the full system power
    intake. The right-hand-side input matrix is assumed to be
    pre-transposed and in the alternate basis; the times reported here
    do not include the transposition of submatrices or the change of
    basis.}
  \label{tbl:abchainresults}
  \begin{center}
    {\small
      \begin{tabular}{rrrrrrrrr}
        \hline
        GPU & $n$ & $d$ & Runtime & Effective bop/s & Energy/bop & Total energy & System~en./bop & Total~system~en.\\\hline
        V100 & 1024 & 1 & $1.63 \cdot 10^{-4}$~s & 13.20~Tbop/s & 202.35~pJ/bop & $1.21 \cdot 10^{-7}$~kWh & 265.25~pJ/bop & $1.59 \cdot 10^{-7}$~kWh\\
        V100 & 2048 & 1 & $2.53 \cdot 10^{-4}$~s & 68.02~Tbop/s & 39.25~pJ/bop & $1.88 \cdot 10^{-7}$~kWh & 51.45~pJ/bop & $2.46 \cdot 10^{-7}$~kWh\\
        V100 & 4096 & 1 & $1.01 \cdot 10^{-3}$~s & 136.65~Tbop/s & 19.54~pJ/bop & $7.46 \cdot 10^{-7}$~kWh & 25.61~pJ/bop & $9.78 \cdot 10^{-7}$~kWh\\
        V100 & 8192 & 1 & $5.56 \cdot 10^{-3}$~s & 197.78~Tbop/s & 13.50~pJ/bop & $4.13 \cdot 10^{-6}$~kWh & 17.70~pJ/bop & $5.41 \cdot 10^{-6}$~kWh\\
        V100 & 16384 & 1 & $3.77 \cdot 10^{-2}$~s & 233.70~Tbop/s & 11.42~pJ/bop & $2.80 \cdot 10^{-5}$~kWh & 14.98~pJ/bop & $3.66 \cdot 10^{-5}$~kWh\\
        V100 & 32768 & 1 & $2.66 \cdot 10^{-1}$~s & 265.02~Tbop/s & 10.07~pJ/bop & $1.97 \cdot 10^{-4}$~kWh & 13.21~pJ/bop & $2.59 \cdot 10^{-4}$~kWh\\
        V100 & 65536 & 1 & $1.87 \cdot 10^{0\phantom{-}}$~s & 301.20~Tbop/s & 8.86~pJ/bop & $1.39 \cdot 10^{-3}$~kWh & 11.62~pJ/bop & $1.82 \cdot 10^{-3}$~kWh\\
        V100 & 131072 & 8 & $3.50 \cdot 10^{0\phantom{-}}$~s & 1287.65~Tbop/s & 2.07~pJ/bop & $2.60 \cdot 10^{-3}$~kWh & 2.72~pJ/bop & $3.41 \cdot 10^{-3}$~kWh\\
        V100 & 262144 & 8 & $1.83 \cdot 10^{1\phantom{-}}$~s & 1976.47~Tbop/s & 1.35~pJ/bop & $1.36 \cdot 10^{-2}$~kWh & 1.77~pJ/bop & $1.78 \cdot 10^{-2}$~kWh\\
        V100 & 524288 & 8 & $1.37 \cdot 10^{2\phantom{-}}$~s & 2105.28~Tbop/s & 1.27~pJ/bop & $1.02 \cdot 10^{-1}$~kWh & 1.66~pJ/bop & $1.34 \cdot 10^{-1}$~kWh\\
        V100 & 1048576 & 8 & $9.72 \cdot 10^{2\phantom{-}}$~s & 2374.71~Tbop/s & 1.12~pJ/bop & $7.21 \cdot 10^{-1}$~kWh & 1.47~pJ/bop & $9.45 \cdot 10^{-1}$~kWh\\
        P100 & 1024 & 1 & $1.26 \cdot 10^{-4}$~s & 17.17~Tbop/s & 83.89~pJ/bop & $5.01 \cdot 10^{-8}$~kWh & 116.51~pJ/bop & $6.95 \cdot 10^{-8}$~kWh\\
        P100 & 2048 & 1 & $3.04 \cdot 10^{-4}$~s & 56.63~Tbop/s & 25.43~pJ/bop & $1.22 \cdot 10^{-7}$~kWh & 35.32~pJ/bop & $1.69 \cdot 10^{-7}$~kWh\\
        P100 & 4096 & 1 & $1.38 \cdot 10^{-3}$~s & 99.65~Tbop/s & 14.45~pJ/bop & $5.52 \cdot 10^{-7}$~kWh & 20.07~pJ/bop & $7.67 \cdot 10^{-7}$~kWh\\
        P100 & 8192 & 1 & $8.38 \cdot 10^{-3}$~s & 131.30~Tbop/s & 10.97~pJ/bop & $3.35 \cdot 10^{-6}$~kWh & 15.23~pJ/bop & $4.66 \cdot 10^{-6}$~kWh\\
        P100 & 16384 & 1 & $5.40 \cdot 10^{-2}$~s & 163.10~Tbop/s & 8.83~pJ/bop & $2.16 \cdot 10^{-5}$~kWh & 12.26~pJ/bop & $3.00 \cdot 10^{-5}$~kWh\\
        P100 & 32768 & 1 & $3.82 \cdot 10^{-1}$~s & 184.64~Tbop/s & 7.80~pJ/bop & $1.53 \cdot 10^{-4}$~kWh & 10.83~pJ/bop & $2.12 \cdot 10^{-4}$~kWh\\
        P100 & 65536 & 1 & $2.69 \cdot 10^{0\phantom{-}}$~s & 210.00~Tbop/s & 6.86~pJ/bop & $1.08 \cdot 10^{-3}$~kWh & 9.52~pJ/bop & $1.49 \cdot 10^{-3}$~kWh\\
        P100 & 131072 & 4 & $7.08 \cdot 10^{0\phantom{-}}$~s & 636.99~Tbop/s & 2.26~pJ/bop & $2.83 \cdot 10^{-3}$~kWh & 3.14~pJ/bop & $3.93 \cdot 10^{-3}$~kWh\\
        P100 & 262144 & 4 & $4.12 \cdot 10^{1\phantom{-}}$~s & 875.64~Tbop/s & 1.64~pJ/bop & $1.65 \cdot 10^{-2}$~kWh & 2.28~pJ/bop & $2.29 \cdot 10^{-2}$~kWh\\
        P100 & 524288 & 4 & $2.63 \cdot 10^{2\phantom{-}}$~s & 1096.61~Tbop/s & 1.31~pJ/bop & $1.06 \cdot 10^{-1}$~kWh & 1.82~pJ/bop & $1.47 \cdot 10^{-1}$~kWh\\
        K80 & 1024 & 1 & $2.97 \cdot 10^{-4}$~s & 7.25~Tbop/s & 189.02~pJ/bop & $1.13 \cdot 10^{-7}$~kWh & 275.94~pJ/bop & $1.65 \cdot 10^{-7}$~kWh\\
        K80 & 2048 & 1 & $1.13 \cdot 10^{-3}$~s & 15.22~Tbop/s & 90.03~pJ/bop & $4.30 \cdot 10^{-7}$~kWh & 131.43~pJ/bop & $6.28 \cdot 10^{-7}$~kWh\\
        K80 & 4096 & 1 & $6.70 \cdot 10^{-3}$~s & 20.53~Tbop/s & 66.73~pJ/bop & $2.55 \cdot 10^{-6}$~kWh & 97.42~pJ/bop & $3.72 \cdot 10^{-6}$~kWh\\
        K80 & 8192 & 1 & $3.97 \cdot 10^{-2}$~s & 27.71~Tbop/s & 49.44~pJ/bop & $1.51 \cdot 10^{-5}$~kWh & 72.17~pJ/bop & $2.21 \cdot 10^{-5}$~kWh\\
        K80 & 16384 & 1 & $2.24 \cdot 10^{-1}$~s & 39.33~Tbop/s & 34.84~pJ/bop & $8.52 \cdot 10^{-5}$~kWh & 50.86~pJ/bop & $1.25 \cdot 10^{-4}$~kWh\\
        K80 & 32768 & 1 & $1.58 \cdot 10^{0\phantom{-}}$~s & 44.61~Tbop/s & 30.71~pJ/bop & $6.01 \cdot 10^{-4}$~kWh & 44.84~pJ/bop & $8.77 \cdot 10^{-4}$~kWh\\
        K80 & 65536 & 8 & $2.15 \cdot 10^{0\phantom{-}}$~s & 262.51~Tbop/s & 5.22~pJ/bop & $8.17 \cdot 10^{-4}$~kWh & 7.62~pJ/bop & $1.20 \cdot 10^{-3}$~kWh\\
        K80 & 131072 & 8 & $1.31 \cdot 10^{1\phantom{-}}$~s & 344.31~Tbop/s & 3.98~pJ/bop & $4.98 \cdot 10^{-3}$~kWh & 5.81~pJ/bop & $7.27 \cdot 10^{-3}$~kWh\\
        K80 & 262144 & 8 & $8.05 \cdot 10^{1\phantom{-}}$~s & 447.61~Tbop/s & 3.06~pJ/bop & $3.07 \cdot 10^{-2}$~kWh & 4.47~pJ/bop & $4.48 \cdot 10^{-2}$~kWh\\
        K80 & 524288 & 8 & $6.92 \cdot 10^{2\phantom{-}}$~s & 416.80~Tbop/s & 3.29~pJ/bop & $2.64 \cdot 10^{-1}$~kWh & 4.80~pJ/bop & $3.85 \cdot 10^{-1}$~kWh\\
        \hline\end{tabular}
    }
  \end{center}
\end{table}

\begin{table}
  \caption{This table presents a best-case comparison between the
    runtimes of the different algorithms as run on the DGX-1 with 8
    GPUs and subproblem size of 131072 for the cubic binary and
    Boolean algorithms and 65536 for Strassen-like algorithms. The
    right-hand-side operand is assumed to be pre-transposed and in the
    desired basis; the change of basis or the transpose of the
    submatrices is not included in the reported times.}
  \label{tbl:dgxcomparison}
  \begin{center}
    {\small
      \begin{tabular}{rrrrrr}
        \hline
        $n$ & Cubic & Boolean & Strassen-Winograd & Alt.-basis self-inverse & Alt.-basis chaining\\\hline
        1024 & $1.07 \cdot 10^{-4}$~s & $9.84 \cdot 10^{-5}$~s & $1.67 \cdot 10^{-4}$~s & $1.51 \cdot 10^{-4}$~s & $1.63 \cdot 10^{-4}$~s\\
        2048 & $2.15 \cdot 10^{-4}$~s & $1.93 \cdot 10^{-4}$~s & $2.72 \cdot 10^{-4}$~s & $2.45 \cdot 10^{-4}$~s & $2.53 \cdot 10^{-4}$~s\\
        4096 & $1.16 \cdot 10^{-3}$~s & $1.30 \cdot 10^{-3}$~s & $1.05 \cdot 10^{-3}$~s & $9.97 \cdot 10^{-4}$~s & $1.01 \cdot 10^{-3}$~s\\
        8192 & $8.28 \cdot 10^{-3}$~s & $7.86 \cdot 10^{-3}$~s & $5.86 \cdot 10^{-3}$~s & $5.55 \cdot 10^{-3}$~s & $5.56 \cdot 10^{-3}$~s\\
        16384 & $5.21 \cdot 10^{-2}$~s & $5.94 \cdot 10^{-2}$~s & $3.97 \cdot 10^{-2}$~s & $3.76 \cdot 10^{-2}$~s & $3.77 \cdot 10^{-2}$~s\\
        32768 & $4.09 \cdot 10^{-1}$~s & $4.69 \cdot 10^{-1}$~s & $2.81 \cdot 10^{-1}$~s & $2.66 \cdot 10^{-1}$~s & $2.66 \cdot 10^{-1}$~s\\
        65536 & $3.28 \cdot 10^{0\phantom{-}}$~s & $3.73 \cdot 10^{0\phantom{-}}$~s & $1.97 \cdot 10^{0\phantom{-}}$~s & $1.87 \cdot 10^{0\phantom{-}}$~s & $1.87 \cdot 10^{0\phantom{-}}$~s\\
        131072 & $2.65 \cdot 10^{1\phantom{-}}$~s & $2.98 \cdot 10^{1\phantom{-}}$~s & $3.74 \cdot 10^{0\phantom{-}}$~s & $3.33 \cdot 10^{0\phantom{-}}$~s & $3.50 \cdot 10^{0\phantom{-}}$~s\\
        262144 & $6.76 \cdot 10^{1\phantom{-}}$~s & $7.51 \cdot 10^{1\phantom{-}}$~s & $2.34 \cdot 10^{1\phantom{-}}$~s & $1.83 \cdot 10^{1\phantom{-}}$~s & $1.83 \cdot 10^{1\phantom{-}}$~s\\
        524288 & $2.40 \cdot 10^{2\phantom{-}}$~s & $2.71 \cdot 10^{2\phantom{-}}$~s & $2.66 \cdot 10^{2\phantom{-}}$~s & $1.19 \cdot 10^{2\phantom{-}}$~s & $1.37 \cdot 10^{2\phantom{-}}$~s\\
        1048576 & $1.88 \cdot 10^{3\phantom{-}}$~s & $2.11 \cdot 10^{3\phantom{-}}$~s & $2.61 \cdot 10^{3\phantom{-}}$~s & $8.21 \cdot 10^{2\phantom{-}}$~s & $9.72 \cdot 10^{2\phantom{-}}$~s\\
        \hline\end{tabular}
    }
  \end{center}
\end{table}

\section*{Acknowledgments.} 
The research leading to these results has received funding from the European Research Council under the European Union's Seventh Framework Programme (FP/2007-2013) / ERC Grant Agreement 338077 ``Theory and Practice of Advanced Search and Enumeration''. We gratefully acknowledge the use of computational resources provided by the Aalto Science-IT project at Aalto University.

%%%%%%%%%%%%%%%%%%%%%%%%%%%%%%%%%%%%%%%%%%%%%%%%%%%%%%%%%%%%%%%% References %%%

\bibliographystyle{ACM-Reference-Format}
\bibliography{../../bib/refs.bib}

%%%%%%%%%%%%%%%%%%%%%%%%%%%%%%%%%%%%%%%%%%%%%%%%%%%%%%%%%%%%%%%%%% Appendix %%%

\end{document}